\documentclass[fleqn,usenatbib]{mnras}

\usepackage{newtxtext,newtxmath}

\usepackage[T1]{fontenc}
\usepackage{ae,aecompl}

\usepackage{graphicx}	
\usepackage{amsmath}	
\usepackage{amssymb}	
\usepackage[figure,figure*]{hypcap}
\usepackage{float}

\title[The HERON Project]{The Halos and Environments of Nearby Galaxies (HERON) I: Imaging, Sample Characteristics, and Envelope Diameters}

\author[R. M. Rich et al.]{
R. Michael Rich,$^{1}$\thanks{E-mail: rmr@astro.ucla.edu}
Aleksandr Mosenkov,$^{2}$
Henry Lee-Saunders,$^{1}$
Andreas Koch,$^{3}$
\newauthor
John Kormendy,$^{4}$
Julia Kennefick,$^{5}$
Noah Brosch,$^{6}$
Laura Sales,$^{7}$
James Bullock,$^{8}$
\newauthor
Andreas Burkert,$^{9}$
Michelle Collins,$^{10}$
Michael Cooper,$^{8}$
Michael Fusco,$^{5}$
\newauthor
David Reitzel,$^{11}$
David Thilker,$^{13}$
Dave G. Milewski,$^{14}$ $^{16}$
Lydia Elias,$^{7}$
\newauthor
M. L. Saade,$^{1}$
and Laura De Groot$^{15}$
\\
$^{1}$Department of Physics \& Astronomy, Univ. of California Los Angeles, 430 Portola Plaza, Los Angeles, CA 90095-1547, USA\\
$^{2}$Central Astronomical Observatory, Russian Academy of Sciences, 65/1 Pulkovskoye chaussee, St. Petersburg, 196140 Russia\\
$^{3}$Zentrum f\"ur Astronomie der Universit\"at Heidelberg, Astronomisches Rechen-Institut, 69120 Heidelberg, Germany\\
$^{4}$Department of Astronomy, Univ. of Texas at Austin, 2515 Speedway, Mail Stop C1400, Austin, TX 78712-1205, USA\\
$^{5}$Department of Physics, Univ. of Arkansas, 825 West Dickson Street, Fayetteville, AR 72701, USA\\
$^{6}$Wise Observatory, Tel Aviv University, 69978 Tel Aviv, Israel\\
$^{7}$Department of Physics \& Astronomy, Univ. of California Riverside, Riverside 92521, CA, USA\\
$^{8}$Department of Physics \& Astronomy, Univ. of California, Irvine, 4129 Frederick Reines Hall, Irvine 92697-4575, CA, USA\\
$^{9}$Department of Physics, Ludwig-Maximillians Universitat M\"unchen, Schellingstrasse 4, 80799 Munich, Germany\\
$^{10}$Department of Physics, Univ. of Surrey, Surrey GU2 7XH, United Kingdom\\
$^{11}$Griffith Observatory, 2800 East Observatory Road, Los Angeles 90027, CA, USA\\
$^{13}$Department of Physics and Astronomy, Johns Hopkins University, 3400 N. Charles Street, Baltimore, MD 21218, USA\\
$^{14}$Department of Earth, Planetary, \& Space Sciences, 595 Charles Young Drive, Univ. of California Los Angeles, CA 90095-1567, USA\\
$^{15}$Department of Physics, The College of Wooster, 308 E. University Street, Wooster, OH 44691, USA \\
$^{16}$NASA Jet Propulsion Laboratory, 4800 Oak Grove Drive, Pasadena, CA 91109-8001, USA
}

\date{Accepted XXX. Received YYY; in original form ZZZ}

\pubyear{2018}

\begin{document}
\label{firstpage}
\pagerange{\pageref{firstpage}--\pageref{lastpage}}
\maketitle

\begin{abstract}
We use a dedicated 0.7-m telescope to image the halos of 119 galaxies in the Local Volume to $\mu_r \sim 28-30$~mag/arcsec$^2$. The sample is primarily from the 2MASS Large Galaxy Atlas and extended to include nearby dwarf galaxies and more distant giant ellipticals, and spans fully the galaxy colour--magnitude diagram including the blue cloud and red sequence. We present an initial overview, including deep images of our galaxies. Our observations reproduce previously reported low surface brightness structures, including extended plumes in M\,51, and a newly discovered tidally extended dwarf galaxy in NGC\,7331. Low surface brightness structures, or ``envelopes'', exceeding 50~kpc in diameter are found mostly in galaxies with $M_V<-20.5$, and classic interaction signatures are infrequent. Defining a halo diameter at the surface brightness 28~mag/arcsec$^2$, we find that halo diameter is correlated with total galaxy luminosity. Extended signatures of interaction are found throughout the galaxy colour--magnitude diagram without preference for the red or blue sequences, or the green valley.  Large envelopes may be found throughout the colour--magnitude diagram with some preference for the bright end of the red sequence.  Spiral and S0 galaxies have broadly similar sizes, but ellipticals extend to notably greater diameters, reaching 150~kpc. 
We propose that the extended envelopes of disk galaxies are dominated by an extension of the disk population rather than by a classical population II halo.  
\end{abstract}

\begin{keywords}
Galaxies: evolution - formation - halos - interactions - photometry - structure
\end{keywords}



\section{Introduction}
The stellar outskirts or envelopes of galaxies are frequently described using the the term ``halo'', even though their component stellar populations are unlikely to resemble the Population II stellar population normally associated with the Milky Way halo. These envelopes may range from classical old Population II stars to accreted low luminosity galaxies that may still host star formation and gas. The situation is further complicated by evidence that galactic halos consist of a mixture of accreted (outer) and in-situ (inner) components (e.g. \citealt{Cooper13}; \citealt{Pillepich15}), or metal rich stars related to the spheroid  (e.g. \citealt{Bellazzini03}). As such, these extended stellar envelopes might consist of ancient halo stars formed in-situ, debris from an ingested galaxy, relics of stages in the formation of the disk, or stars ejected from the disk of the host galaxy experiencing accretion.

By way of example, deep {\it Hubble Space Telescope (HST)} imaging on the minor axis of M\,31 (\citealt{Brown03}; \citealt{Brown06}) revealed a substantial intermediate-age population. The halo of M\,31 is mostly metal rich (\citealt{Rich96}; \citealt{durrell01}; \citealt{Bellazzini03}) while also hosting both globular clusters, RR Lyrae stars, and a host of substructures in the form of satellites and streams (e.g. \citealt{Mcconnachie09}). Simulations like those of e.g. \citet{Mori08} show that a minor merger with a galaxy of mass $M \sim 10^9 M_\odot$ can reproduce the many complex structures similar to those seen in the M\,31 halo as well as ejecting stars from the disk. 

Ideally, this study would separate out the different populations of halo stars (in-situ, accreted, and disturbed disk) using the colours and magnitudes of the stars. However, the age--metallicity degeneracy that affects interpretation of colours and spectra also would affect the stars in the resolved stellar population. As the origins of these structures might be so varied, we must use the term ``halos'' to mean a more complicated type of stellar population than the conventional e.g. Population II. Recent extensive surveys of interaction signatures of galaxies bear this out; studies by \citet{Atkinson13}, \citet{hood18}, and \citet{morales18} report a wide variety of interaction signatures across the galaxy color--magnitude diagram.  These range from organized shells observed around luminous red galaxies, to a huge array of disturbances associated with disks. An examination of the range of interactions raises the question of what the extended structures of galaxies are comprised of, likely a mix of debris from the host and the merger galaxy.  For the purposes of our investigation, these low surface brightness structures will be referred to as envelopes, as the term ``halo'' might imply a more specific stellar population. 

Studies have also shown that the type of envelope a galaxy possesses may relate to its mass. A small survey of galaxy envelope fields from photometric metallicities of resolved stars imaged using {\it HST} finds an interesting dichotomy, in which galaxies with $M_V < -21$ have relatively metal rich envelopes (e.g. $[Fe/H] = -0.7$) and lower luminosity envelopes have lower metallicity, closer to $-1.5$ dex. (\citet{Mouhcine05}; \citet{Monachesi16}; \citet{Harmsen17}). 
Although this kind of investigation requires more development, it is important to reflect that more luminous galaxies may have envelopes more related to their central bulges and spheroids, while disk galaxies lacking bulges (e.g. NGC\,4244) have a more metal-poor envelope.  The field of low surface brightness imaging has recently been associated with the discovery of shell structures around elliptical galaxies, spectacular interaction streams, and a zoo of peculiar extensions and structures (e.g. \citealt{Martinez10}; \citealt{Duc17}). Relatively less attention has been paid to investigating the systematics of the quotidian envelopes extending to low surface brightness surrounding galaxies, that represent the potentially very long-lived structures. Our investigation explores the correlation of this relaxed envelope diameter with galaxy absolute magnitude, and in this initial study, we explore how envelope diameter varies across the galaxy colour-- magnitude diagram. 

This paper formally introduces the {\sl HERON} project, an international collaboration of observers and theorists working to motivate observations of the low surface brightness extensions of galaxies and to compare those results with theory.   
Future papers will report analysis of quantitative surface brightness profiles, outer envelope morphologies, and other properties including comparisons of extended structures to imaging data at other wavelengths. We will also publish catalogues and luminosities for all low surface brightness candidate companions in the survey. In this project, we build on the heritage of \citet{Kormendy74} and report envelope diameters to 28~mag/arcsec$^2$ for 119 galaxies. Starting in 2019 October, we will begin populating the {\sl HERON} archive at IRSA/IPAC in the community data archives; \footnote{\url{www.irsa.ipac.caltech.edu/data/HERON/}}  we will provide data tables, JPEG, and FITS images for the full galaxy sample in the {\sl HERON} survey.

This paper is organized as follows:
Sect.~2  presents our instrumentation, observing strategy, and data reduction.
The sample of {\sl HERON} galaxies is introduced in Sect.~3, along with the measurement of their envelope diameters. In Sect.~4 we discuss our results, while in Sect.~5 we correlate our results with general galaxy properties, before summarizing in Sect.~6.

\subsection{A brief history of low-surface brightness features}

The study of low surface brightness galaxy envelopes and extensions has a 
long history with many interesting subjects; \citet{Zwicky56} used apertures from the 18-inch Schmidt to the Hale 200-inch telescope, and called attention to tidal tails and extensions of galaxies including NGC\,3628.  He also emphasized the value of studying such systems by noting the much earlier works of, e.g., \citet{Pease20} and \citet{Lundmark20}. Other early contributions include the career-long work of \citet{Karachentsev65}, which continues to the present day (\citealt{Karachentsev17}).  Photoelectric photometry exploring the extent of M\,87 to a full degree (\citealt{Vaucouleurs69}; \citealt{Arp69}) was also a remarkably early application of technology to the problem. The analysis of scanned photographic plates of \citet{Kormendy74} was the first to report a photographic image of the giant stream of NGC\,3628, and to report the diameter measurements for a large number of envelopes,  some as large as 100 kpc.  

Low surface brightness studies returned to the spotlight with the pioneering work of e.g. \citet{Malin78}, \citet{Malin79}, and \citet{Malin80}. \citet{Malin97} illustrate the remarkable low surface brightness envelopes and streams of spiral galaxies, including a giant arc near M\,83. These efforts included the development of `unsharp masking' and the discovery of low surface brightness shells around elliptical galaxies. \citet{Binggeli88} catalogued the Virgo cluster including dwarf galaxies, continued by \citet{Ferguson89}. The subject has seen contemporary vitality with the now decades-old explosion of modest aperture telescopes with CCD detectors that have revealed surprisingly extended envelopes of nearby galaxies (e.g. \citealt{Tal09}; \citealt{Martinez10}; \citealt{Rich12}; \citealt{Martinez12}; \citealt{vanDokkum15}; \citealt{Trujillo16}) and significant work using the Burrell Schmidt Telescope 
(\citealt{Mihos05}; \citealt{Watkins15}; \citealt{Mihos17}). \citet{Duc15} undertook low surface brightness imaging of the environs of ATLAS3D \footnote{\url{www-astro.physics.ox.ac.uk/atlas3d/}} ellipticals, updating the \citet{Tal09} sample, and finding numerous cases of streams, shells, and extended disk star formation. The Pan-Andromeda Archaeological Survey (PandAS) map of the resolved stellar halo of the nearby Andromeda-M\,33 complex reveals significant halo structures, demonstrating that the entire extent of the M\,31 envelope may stretch halfway to the Milky Way and significantly overlaps with that of the neighbouring spiral M\,33 (\citealt{Mcconnachie09}; \citealt{Ibata13}; see also \citealt{Koch08}).  
The interesting science questions raised by these analyses have inspired a range of investigations that include networks of small telescopes \citep{Martinez12}, 
an array of 8 (now 2$\times$24) Canon telephoto camera lenses (the Dragonfly project; \citealt{Abraham14}; \citealt{vanDokkum14}; \citealt{Merritt16}) and our project -- the {\em Halos and Environments of Nearby Galaxies (HERON)} survey \citealt{Rich17}.

Any new entry into this subject area faces the long history of research and also serious challenges such as the correct treatment of scattered light \citep{Sandin14,Sandin15}.  Furthermore, the instrumentation required to enter the field, for the most part, is modest in cost, enabling numerous individuals and teams to participate. A successful philosophy has been to aggregate very long exposures obtained by citizen scientists with state of the art, commercial off the shelf equipment (e.g. \citealt{Martinez10}). The Dragonfly array employs commercial technology to minimize scattered light, and builds a powerful instrument from multiple focal planes.  An additional relatively recent project is the Purple Mountain $1.0$ -m Schmidt Near Earth Object Survey Telescope \citep{Shi17}; this facility has been used to investigate ultra-diffuse galaxies. Special purpose telescopes are under construction and a small space mission has been proposed to undertake imaging at low surface brightness \citep{Muslimov17}. Serendipitous imaging of galaxy outskirts into resolved stars will also occur as part of the WFIRST mission.  
In fact, the systematics of Galactic cirrus and scattered light probably, in all likelihood, limit quantitative investigations to the $30-32$~mag/arcsec$^2$ (in the $r$ band) level (for an extreme case of the effects of infrared cirrus, as e.g. for NGC\,7743). Pushing fainter than this will require  space-based  mission that is capable of resolving the low surface brightness structures into stars.

\subsection {Theoretical motivation}

The low-surface brightness envelopes of nearby galaxies offer a unique window into galaxy formation and evolution, and possibly cosmology.  These regions hold clues to the hierarchical build-up of structure formation on sub-galactic scales, the very scales where the dominant $\Lambda$CDM paradigm is facing its most difficult challenges.  Some of the most profound concerns about the $\Lambda$CDM theory have arisen in comparison to dwarf satellites and low-surface brightness features seen around just two galaxies: the Milky Way and M\,31 \citep{Boylankolchin11,Boylankolchin12}.  But by relying on just two galaxies of similar luminosity and type we are potentially biasing ourselves significantly.  
Despite its marked successes in reproducing the large-scale properties of the Universe, the $\Lambda$CDM cosmological model faces some significant problems on the mass scales of dwarf galaxies $(M_{\ast}\sim M_{vir}= 10^{5-9} M_{\odot})$.  
The overall count of dwarfs throughout the Local Group is significantly lower than might naively be expected in $\Lambda$CDM-based models of galaxy formation (the ``missing satellites'' problem; \citealt{Klypin99}; \citealt{Moore99}). Moreover, the measured internal mass densities of
dwarf satellites are significantly lower than predicted for the $\sim$10 most massive dark matter halos near galaxies similar to the Milky Way (the ``too big to fail'' problem; \citealt{Boylankolchin11,Boylankolchin12}; \citealt{Agertz16}).  

These anomalies do not necessarily mean that our cosmology needs to be revised, as plausible astrophysical solutions have been proposed (e.g. \citealt{Bullock00}; \citealt{Governato10}; \citealt{Dicintio14}; \citealt{Wetzel16}), but they do strongly motivate the need for alternative tests of the paradigm on the mass scales of dwarf galaxies. One particularly robust test involves looking for tell-tale signs of past dwarf-size merger events around local galaxies: low-surface brightness streams, diffuse halo light, and faint heated disk material (\citealt{Johnston98}; \citealt{Bullock01}, \citealt{Bullock05}; \citealt{Kazantzidis08}; \citealt{Purcell10}; \citealt{Cooper10}; \citealt{Amorisco17}). While theoretically well motivated, searches of this kind have been limited largely to resolved-star studies around the Milky Way and M\,31 (\citealt{Mcconnachie09}; \citealt{Belokurov06}; \citealt{Belokurov07a}; \citealt{Bechtol15}; \citealt{Cunningham16}; \citealt{Mackey16}). The reason is that the predicted features are extremely low surface brightness, $\sim29-30$~mag/arcsec$^2$.  Until recently, faint features of this kind were prohibitively difficult to detect without resolving them into stars. 

Why are low-surface brightness features so important?  Primarily because the predictions are robust: they rely on the assumption that structure formation is hierarchical down to small scales -- 
one of the fundamental predictions of cold dark matter cosmologies.  
More specifically, if $\Lambda$CDM is correct, then galaxy-size dark matter halos should be built by the steady accretion of smaller, dwarf-size dark matter clumps (e.g. \citealt{Cole00}; \citealt{Stewart08}; \citealt{Fakhouri10}). The rate and timing of these dark-matter halo mergers are robustly predicted, at least statistically speaking. Moreover, the stellar content of these dark matter mergers is also well constrained, as the stellar mass of dwarf galaxies needs to drop steeply with decreasing halo mass in a well-defined way to solve the missing satellites problem and (more generally) to explain the observed faint-end of the luminosity function (e.g. \citealt{Behroozi13}; \citealt{GarrisonKimmel17a}; \citealt{Read17}, loosely called ``abundance matching''). 

The implication is that it should be relatively straightforward to predict the fraction of a galaxy's light contained in diffuse, low-surface brightness material as a function of galaxy stellar mass. 
\citet{Purcell10} made this point explicitly using a semi-analytic model, and showed that the diffuse light fraction (stellar halo light fraction) should vary strongly with galaxy mass over the galaxy scales of interest here.

\section{Observations and data reduction}
Our approach to the {\sl HERON} project has been to use and maintain two dedicated telescopes that are proven to reach low levels of surface brightness rapidly, allowing us to upgrade the focal planes, and experiment with different filters and observing modes, all at reasonable cost.  

\subsection{Instrumentation}

We employ the 0.7-m Jeanne Rich Telescope Centurion 28 (C28) at the Polaris Observatory Association site, a dedicated f/3.2 telescope with a prime focus imager behind a Ross doublet corrector, consisting of FS2 and BK7 glass. A conical baffle with ring stops is placed in front of the corrector group to control scattered light.   
The primary mirror is honeycomb light weighted, and the truss consists of graphite epoxy rods in tension; the optical telescope assembly is supported on an equatorial yoke mount. The control system employs the FS2 astro-electronic telescope control system, by Michael Koch \footnote{\url{www.astro-electronic.de/fs2.htm}}. 
Focus is achieved using a motor by Robofocus, which focuses by moving the corrector group. The telescope is illustrated in \citet{Brosch15a} and \citet{Rich17}. 

The observatory is located in Lockwood Valley, near Frazier Park, CA at an elevation of 1615~m.  Although the greater Los Angeles area creates a light dome in the Southeast, the site is $21.7-22.0$~mag/arcsec$^2$ at the zenith and very dark in the West.  The manufacturer, James Riffle of AstroWorks Corp in Arizona, produced a number of 0.5-m telescopes of similar design (\citealt{Brosch08}) and an identical telescope (the 0.7-m Jay Baum Rich Telescope) operates at Wise Observatory, Israel near Mitzpe Ramon (\citealt{Brosch15a}; \citealt{Brosch15b}).  
A companion project using the Wise 0.7-m Jay Baum Rich Telescope of Tel Aviv University addresses the deep imaging of edge-on disk galaxies and Hickson compact groups and is underway (P.I. N. Brosch).   

The present image quality is $2.5-3^{\prime\prime}$ and poses no impediment to imaging low surface brightness features typically $>60^{\prime\prime}$ in size. Remote operations are now routine for the Jeanne Rich C28 telescope.  
Most of the dataset we report on here was obtained using an SBIG STL 11000 CCD which includes a guide CCD alongside the main detector. This camera also has an internal filter wheel that holds 5 round 50mm filters. The detector is thermoelectrically cooled to typically $-25\degr$C; the detector is a KAI1100M interline transfer CCD with 9 $\mu \rm m$ pixels in a $4008\times2672$~pixel format. The scale is 0.83~arcsec/pix with a field of view of 0.57~sq.~degree. The STL11000m camera has a readout noise of 13$e^-$, and the A/D conversion (single binned) is $0.8e^-$/ADU, and double-binned (a minority of our images) is $1.6e^-$/ADU.  The FLI09000 used for a subset of our data has 11$e^-$ read noise in the 1MHz readout mode and close to 1.0$e^-$/ADU.  About 20\% of our data are double-binned and so indicated.  The compact design of this camera with the internal filter wheel results in images with only modest vignetting near the corners. The noise in these images is dominated by the sky background, usually $\sim 2000$ counts, on a given frame.

Since early 2015, we have used an FLI09000 detector on loan from Arizona State University; it is nearly identical to the STL11000m, with a $3056\times 3056$~pixel format, lower read noise and $12 \mu \rm m$~pix. Guiding for the FLI camera is accomplished via a Lodestar X2 guide CCD mounted on an Astrodon Mega MOAG off-axis guider, in front of the filter wheel. We use a Finger Lakes Instruments CFW2 five position filter wheel that holds 2~inch square filters (currently supplied by Astrodon). In order to address some low level image persistence issues, data taken after mid-2016 employs the slow (1~MHz) readout mode using an RBI (Residual Background Image) flood with 400~ms flood time, 4 flushes, and a bin factor of 4, accessed via ``advanced camera settings''.

Although persistence is an issue with the FLI camera, it can be eliminated by using the RBI Flood read mode, which preflashes the imager at the expense of a slight increase in read noise; in any case nearly all of the data we report here were obtained using the SBIG camera. The FLI Camera operates behind the off-axis guider and filter wheel; this results in greater vignetting than experienced for the STL11000m, however, we are able to flat field our images successfully and our exposure times overcome the modest loss of light due to vignetting.  Our telescope operation and acquisition software is The Sky 6 by Software Bisque and is used for telescope control, Maxim DL version 6 for CCD camera control, and the commercial software Focusmax controlling a robofocus unit, for focus control. Temperature, humidity, sky darkness, and clarity are monitored using a Boltwood cloud sensor from Cyanogen, Ltd.

\begin{figure}
	\includegraphics[width=\columnwidth]{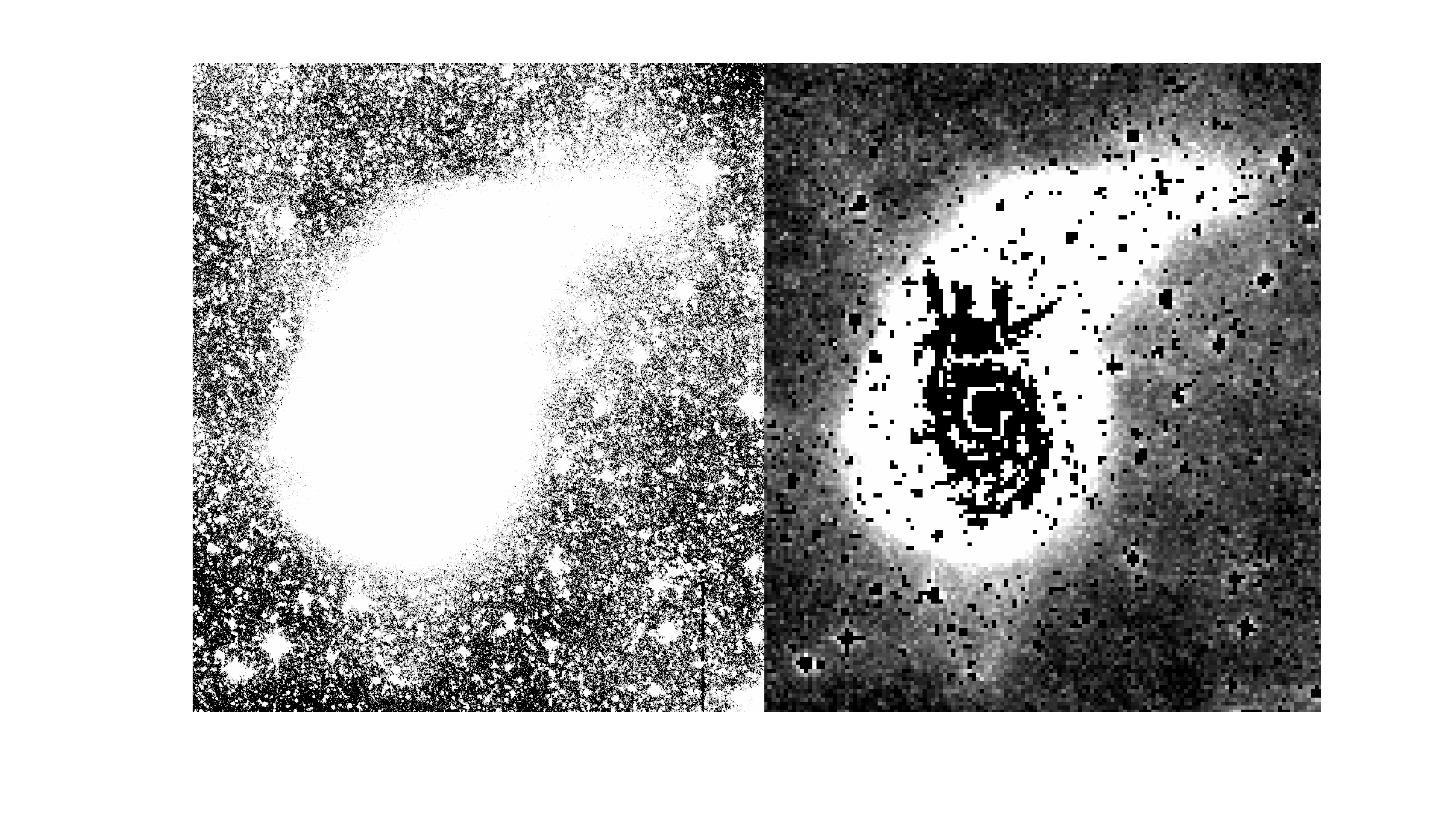}
    \caption{A comparison between our Luminance image of M\,51 using the C28 and SBIG11000m imager for $12\times 300$~sec (left) and $V$ band exposure courtesy of A. Watkins \citep{Watkins15} $31 \times 1200$~sec (roughly 10~hours) using the 24/36-inch Burrell Schmidt at Kitt Peak (right). Two completely independent approaches to deep imaging are reaching similar depths and revealing similar low surface brightness morphology. This result also addresses concerns that wide angle, faint scattered light might affect our measurements. Both figures show a limiting surface brightness of $\sim$ 30~mag/arcsec$^2$. The dark inserted feature shows the approximate 
    extent of the higher surface brightness portions of M51. The Burrell Schmidt image has has had bright stars subtracted while ours has not.}
    \label{watkinsM51}
\end{figure}

\begin{figure*}
\centering
\includegraphics[scale=0.33]{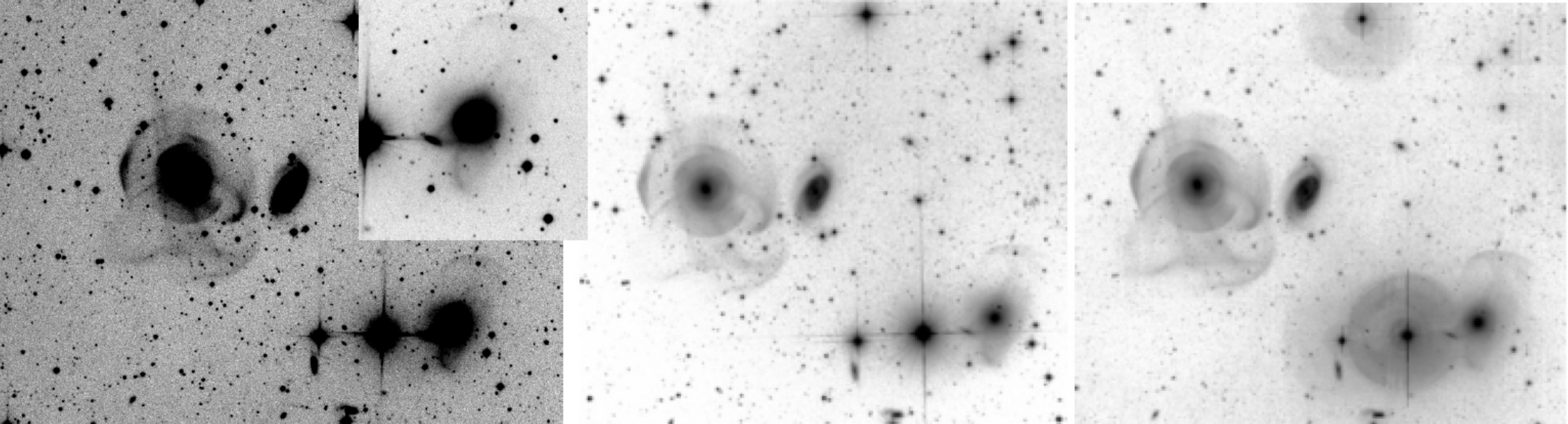}
\caption{Images of NGC\,474 and 470 using different platforms, each showing the faintest features at $\sim 29$~mag/arcsec$^2$.  This illustrates our excellent scattered light control; see also the inset of arcs associated with a galaxy interacting with the elliptical galaxy NGC 467 (Left): $25\times300$~sec exposure (SDSS $g$) using the Jeanne Rich 28-inch telescope.  (Middle) A 21.5~hour exposure using the Irida Observatory 12-inch astrograph; (Right) 0.7~hour exposure in SDSS $g$ using the Canada-France-Hawaii Telescope (middle and right-hand panels from \citealt{Duc15}, fig.~7). Notice the scattered light in the 3 bright images at lower right, and our clean detection of the faint arc.}
\label{n474}
\end{figure*}

\begin{figure}
\centering
\includegraphics[width=\columnwidth]{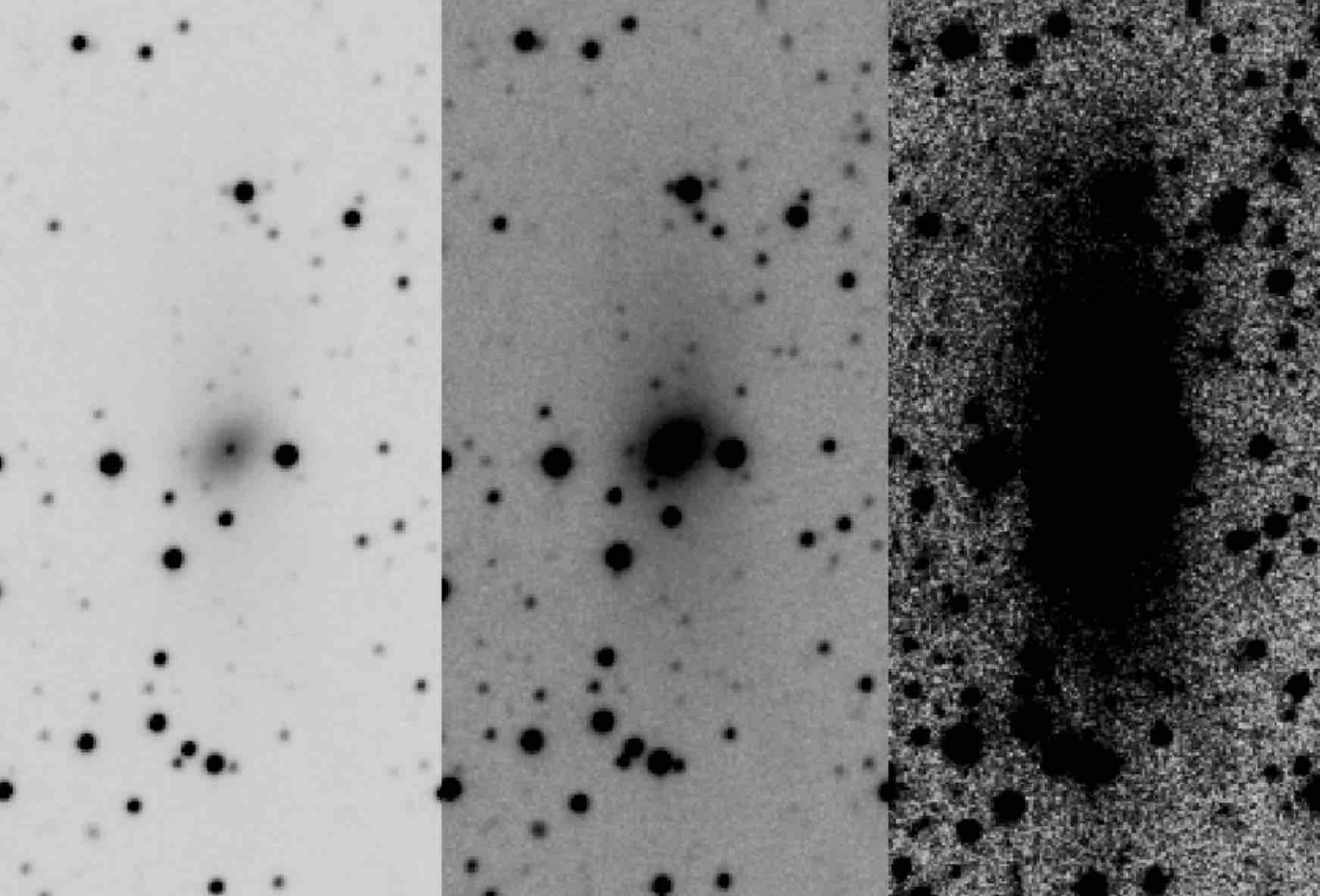}
\caption{The nucleated dwarf galaxy associated with NGC\,7331 at RA 2000: 22h~37m~12.4s, DEC 2000: +34$^{\circ}$~37'~12.1". The tidal tails were first reported in \citet{Paudel14} where the galaxy is HdE6. N is up and E to the left. The full frame height is 6~arcmin; the suspected tidal tail feature appears as nearly vertical wings. In this 12$\times$300~s exposure, the dwarf is illustrated at 3 different stretches (compare with fig.~2 of \citealt{Blauensteiner17} from a 26~hour exposure). The full extent of the dwarf is $1\times 4^\prime$, corresponding to $4\times 16$~kpc at the 14~Mpc distance of NGC\,7331. \citet{Duc18} (their figs~2 and 3) also illustrate this galaxy and show the same extended tails, but do not explicitly mention the galaxy or this structure.}
\label{n7331}
\end{figure}

\begin{figure*}
\centering
\includegraphics[width=4.3cm]{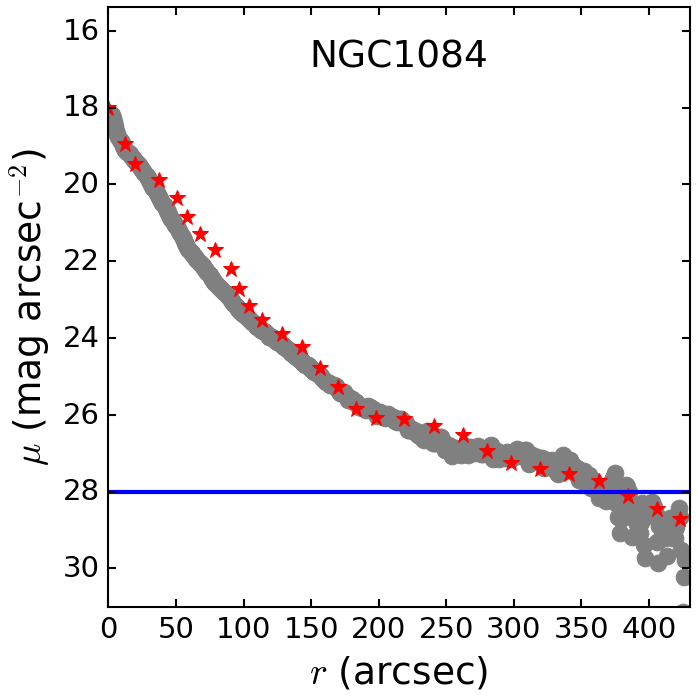}
\includegraphics[width=4.3cm]{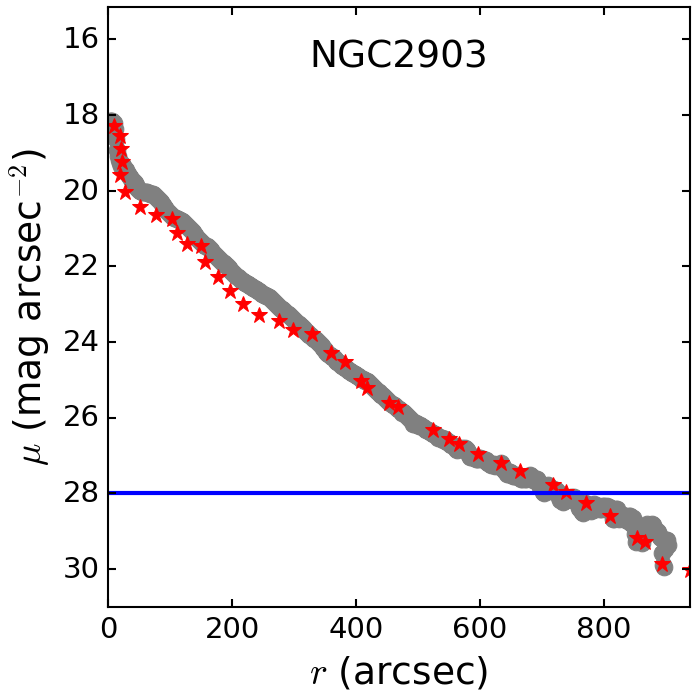}
\includegraphics[width=4.3cm]{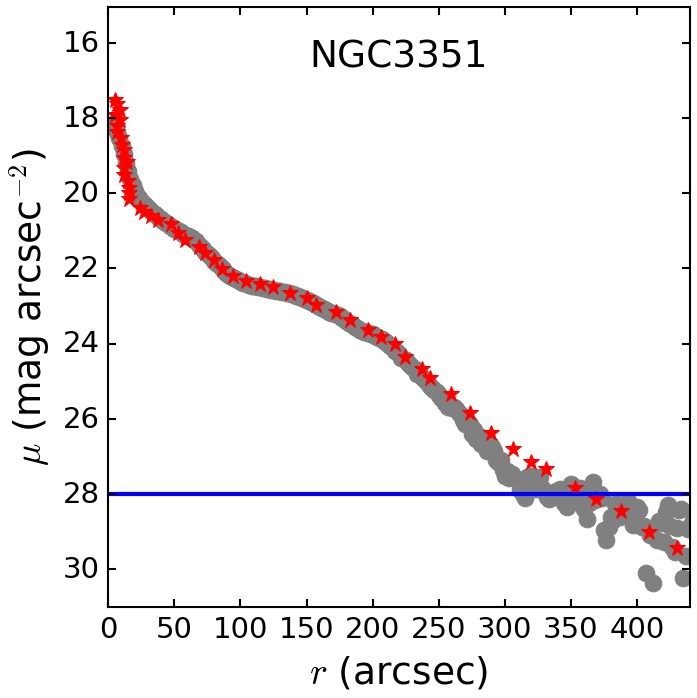}
\includegraphics[width=4.3cm]{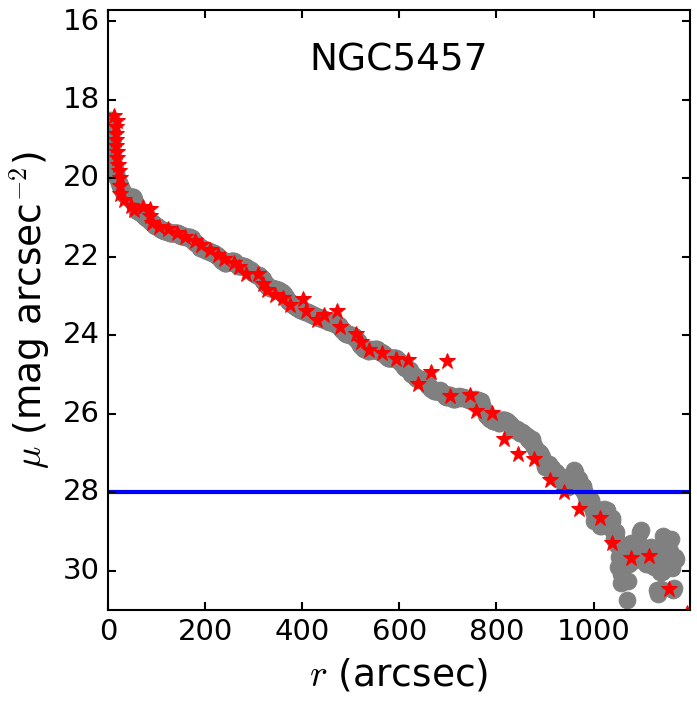}
\caption{Comparison of our measured surface brightness profiles (grey) for four common galaxies with those observed by \citet{Merritt16} (extracted from their paper using \url{https://apps.automeris.io}). There is excellent agreement between the surface photometry of {\sl HERON} and those of \citet{Merritt16}. Our data reach to $\mu\sim 29-30$ mag/arcsec$^2$. The blue line shows the 28 mag/arcsec$^2$ isophote, at which we measure the envelope diameter.}
\label{comparesb}
\end{figure*}

\begin{figure}
\centering\includegraphics[width=\columnwidth]{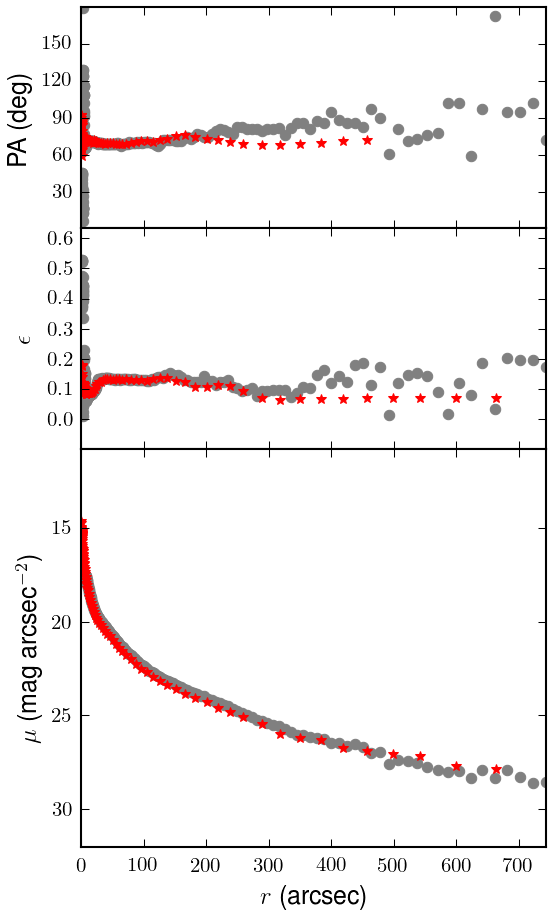}
\caption{Comparison of our surface brightness profile of NGC\,3379 with that of \citet{Kormendy09} (red stars). {\sl HERON} are data illustrated in grey. Our data reach to $\mu\sim 28.5$ mag/arcsec$^2$.}
\label{n3379}
\end{figure}

\subsection{Observations and Data Reduction}

The imaging reported in this paper uses a luminance filter, a square bandpass filter from $4000-7000$~\AA\  with an actual transmission equivalent very approximately to the full SDSS (Sloan Digital Sky Survey, \citealt{2014ApJS..211...17A}) $g$ and $r$ passbands.
Its speed enables us to exploit good conditions and to build a large data set rapidly.
We calibrate this filter using photometry of stars from the SDSS in the field of our target galaxies; this is described in \citet{Rich12}.
We have concluded that the time needed for two-colour integrations is worthwhile only in exceptional cases; the errors in $g-r$ become significant at 25--26~mag/arcsec$^2$ \citep{vanDokkum14, Merritt16}.
Our objective is to build a large sample of systems with well-measured stellar envelope luminosities, diameters, morphologies, and satellite properties.  A single passband enables these science goals and facilitates follow-up multi-wavelength studies if such follow-up appears to be compelling. 

We obtain typically $12-36$ images of 300s duration for each target. For the case of NGC\,128, we measured its diameter for a coadd of 3 and 9 exposures respectively. We found that the 3 exposure coadd resulted in a 0.1 arcmin smaller diameter than the 9 exposure coadd, which is less than a 3\% change, which is smaller than our other sources of error. These are randomly dithered during acquisition by $\sim 20$~arcsec on each move.  We attempt to image galaxies near the meridian and avoid results with image quality poorer than 4~arcsec. The MaximDL acquisition software (http://diffractionlimited.com/product/maxim-dl/) includes a guiding control package, and this is employed using the Starlight express off-axis guider.  
As the CCD camera is cooled thermo-electrically only to a temperature ranging from $-20$ to $-40 \degr C$, there is still dark current, and we must also acquire dark frames of equal length to the images, at the same temperature as the image frames. These calibrations, along with bias and flat fields, are acquired every few nights. We obtain flat fields frequently, usually every few nights, using the 76~cm square Alnitak XL electroluminescent panel, finding no difference between using the panel and twilight flats. We use an SBIG 340C all sky camera to monitor climate and transparency, and we do not take data in moon or even light cirrus. We verify that the sky is essentially photometric before and during imaging and observations during any moonlight is avoided.

All data are reduced using standard IRAF routines including the subtraction of 300s darks obtained at the same operating temperature as the science frames. In the rare cases where the flat fielding is not optimal, 
we use the IRAF routine IMSURFIT to arrive at a final flattened background. IMSURFIT fits two-dimensional low order polynomials to the sky background; we limit the polynomial to order $<5$ and confirm that the application of IMSURFIT does not affect our derived surface brightness profile or measurement of envelope diameter.  It was also very rarely necessary to add a small constant to the frame to avoid oversubtraction.  We conclude that our agreement with other studies as evidenced by Figs. \ref{watkinsM51}-\ref{n3379} reassures that our instrumentation and reduction procedures reach the standards of other similar studies.

\subsection{Surface Photometry}
\label{sec:surf_phot}

The image preparation was done employing a procedure developed by one of the authors (A. Mosenkov) as follows.

An initial step was including astrometry using the website Astrometry.net\footnote{\url{http://nova.astrometry.net/}}.

Then we performed photometric calibration using a range of photometric sources. In each frame, isolated, non-saturated stars with a high signal-to-noise ratio were automatically selected (an average number of such stars for all fields was 59). Then we cross-correlated the selected stars with photometric databases of SDSS, Pan-STARRS \citep{2016arXiv161205560C,2016arXiv161205243F}, and {\it Gaia} DR2 \citep{2016A&A...595A...1G,2018A&A...616A...1G} and transformed their measured magnitudes to the SDSS $r$ band using \citet{2016ApJ...822...66F} and sec. 5.3.7 from the {\it Gaia} data release documentation\footnote{\url{https://gea.esac.esa.int/archive/documentation/GDR2/}} and taking into account Galactic extinction using a 3D dust map from \citet{2019arXiv190502734G}\footnote{\url{http://argonaut.skymaps.info/}}. The average calibration error is 0.046 mag.
Third, sky was carefully estimated for a sub-field, which includes the target galaxy and some empty background (typically, we used a box with a side 5 times larger than the galaxy diameter). We used {\sc sextractor} \citep{1996A&AS..117..393B} to create a segmentation map for this sub-field and mask out all detected objects. To minimise the impact of the scattered light from the masked objects on the sky fitting, the {\sc sextractor} mask for each object was increased by a factor of 1.5. We then used this mask to fit the sky with a polynomial of some degree, starting from 0 (constant sky level) to 5 (significantly non-linear sky background), increasing this value after each iteration if the sky-subtracted image still had a gradient. Additionally, after subtracting the best-fit polynomial from the original image, we re-estimated the sky within an elliptical annulus, built around the galaxy on the basis of the preliminary 1D azimuthally-averaged profile where the profile flattens (typically, at a radius of the double optical radius $R_{25}$ with an annulus width of 32 pixels). Also, to estimate background variations within the annulus, which can be left after flat-fielding or caused by scattered light from stars, extended objects (satellite galaxies), Galactic cirrus, or low surface brightness features (tidal streams, stellar flows etc.) we determined the variations of the median (using 3$\sigma$ clipping) inside uniformly located boxes with a side of 32 pixels placed along the annulus. We define the standard deviation of the measured median values within these boxes as the background estimation error. Note that it is different from the standard deviation of the sky level, which is measured for the background within the whole elliptical annulus.

Finally, galaxy images were cropped to encompass the outermost galaxy isophotes plus some space beyond them (1.5 times larger than the diameter of the outermost isophotes). 

To generate surface brightness profiles for a galaxy, we used the following technique. We masked out all foreground stars and other galaxies detected in the final frame. For this, we used the masks, which had been produced earlier, and revisited them by eye. Also, to them we added masks created for the space inside the galaxy by searching for local maxima above the 2D galaxy intensity profile.

For each frame we estimated the Point Spread Function (PSF) for those stars which we used for photometric calibration. Then we performed {\sc galfit} \citep{2002AJ....124..266P,2010AJ....139.2097P} fitting of each galaxy image using a single S\'ersic model \citep{1968adga.book.....S}, convolved with the corresponding PSF. From this modelling, we were able to estimate general parameters of the galaxy position and orientation (position angle) and its ellipticity. We used these values as an initial guess for the {\sc iraf/ellipse} routine. We performed isophote fitting starting from the center and extending up to the outermost isophote which can be detected in the galaxy. The galaxy center, position angle and ellipticity in each fit were set free. From the output results of the ellipse fitting we created azimuthally averaged profiles which were corrected for Galactic extinction using \citet{2011ApJ...737..103S}. The profiles will be presented in detail and discussed in a future work, while here we only use them to estimate the galaxy diameter (see Sec.~\ref{sec:diam_measure}).

\subsection{Scattered light issues and reproducibility of {\sl HERON}}

Amateur operated telescopes of professional quality contribute data that reach 29--30~mag/arcsec$^2$ in tens of hours of integration \citep{Javanmardi16}; the 8 lens Dragonfly array attained 32~mag/arcsec$^2$ in 35~hours and it is reasonable to assume that a 24 lens array reaches these levels in half the time.  Our single detector at the f/3.2 prime focus with a 0.7-m primary is able to detect all of the faint companions reported by  \citet{Javanmardi16}, all except the faintest details in M\,101 reported in \citet{vanDokkum14}, and all faint extensions detected in 10 hr by the 24/36-inch Burrell Schmidt in M\,51 (\citealt{Watkins15}; Fig. \ref{watkinsM51}). We also detect all of the streams reported in \citet{Miskolczi11}. At the faint levels we work (29--30~mag/arcsec$^2$) it is reasonable to be concerned that wide-angle scattered light might compromise our ability to image and measure faint structures. We have compared our deep images and surface brightness profiles with other work (e.g. \citealt{Watkins15}; Fig. \ref{watkinsM51}) and \citet{Merritt16}.  These two programs are respectively from the Burrell Schmidt and Dragonfly array, and they report surface photometry to 30~mag/arcsec$^2$. We also note the excellent agreement between low surface brightness details for NGC\,4449 and its tidally-disrupting dwarf galaxy companion NGC\,4449B illustrated in \citet{Rich12}  and \citet{Martinez12}. Fig.~\ref{n474} compares our $12\times 300$s exposure of NGC\,474 with a 21~hour exposure from the 30-cm astrograph of Irida Observatory \footnote{\url{www.irida-observatory.org/}} and the CFHT (Canada-France-Hawaii Telescope; 0.7~hours $g$-band exposure; both of these illustrated in fig.~7 of \citealt{Duc15}). An example of the performance attained by our system, reaching $\sim 29$~mag/arcsec$^2$, is shown by the tidal structures revealed in \citet{muller19}. 

\citet{Duc15} emphasized that the CFHT has serious scattered light halos not present in the long exposure using the astrograph at Irida Observatory. Our 0.7-m C28 telescope shows exceptionally low scattered light, as good as that of the astrograph and substantially better than the raw CFHT image. The comparison illustrated in our Fig.~\ref{n474} with that of fig.~11 in \citet{Duc15} shows that our images can reach at least 29~mag/arcsec$^2$ in surface brightness. While not illustrated, our {\sl HERON} imaging of Stephens Quintet reproduces all of the faint structure detected by \citet{Duc18} also using the CFHT telescope. Significantly, however, Fig.~\ref{n474} convinces us that our control for scattered light is excellent, reaching or exceeding that of the Irida Observatory astrograph \citep{Duc15}. Additional comparisons with $\sim 10$~hour long images obtained using an 8-inch doublet refractor confirm our excellent scattered light control (B. Megdal, private communication). Example images and comparisons are available on the {\sl HERON} website at IRSA. We will not further discuss the scattered light issue in this paper as it has no impact on our measurements or conclusions.

Fig.~\ref{n7331} illustrates our image of a dwarf galaxy near NGC\,7331. \citet{Blauensteiner17} published images ranging from 6 to 21 hours in depth, however, the tidal tail was actually discovered from SDSS3 images by \citet{Paudel14}. Here, we are able to discern the full extent of the tidal structure (right panel) reaching 4~arcmin = 16~kpc in total length, making it one of the largest known tidally affected galaxies; our measured size is over a factor of 2 larger than that of \citet{Paudel14}.  We also recover the large tidal tails that can be noted in the CFHT images of \citet{Duc18} (their figs.~2 and 3), but those authors did not explicitly note the object. This tidal feature is twice the extent of NGC\,4449B, the tidal dwarf near NGC\,4449 (\citealt{Rich12}). {\it We have confidence in our data quality because our low surface brightness features are detected by other investigators using different telescopes and reduction methods. We have not failed to measure any features reported by others in the literature.}  

In Fig.~\ref{comparesb} we illustrate the exceptionally good comparison of our surface brightness profiles with those of 4 galaxies in \citet{Merritt16}. One galaxy, NGC\,4258, is excluded from the comparison as our image does not cover its outskirts, and, hence, our profile is cropped. Table \ref{tab:table1} presents our agreement in diameter measurement with \citet{Merritt16}. Fig.~\ref{n3379} compares our surface brightness profile of NGC\,3379 with \citet{Kormendy09} and we attain very good agreement to 28~mag/arcsec$^2$. Note that in Figs.~\ref{comparesb} and \ref{n3379} the original profiles from \citet{Merritt16} and \citet{Kormendy09} were shifted upwards to match the inner part of our profiles (excluding the very center where our observations can be saturated), as their observations were done in the $g$ and $V$ band, respectively.   

\begin{table}
	\centering
	\caption{Envelope diameter comparisons between the {\sl HERON} measurements (at the level of 28~mag/arcsec$^2$) and those from \citet{Merritt16}. We only compared galaxies that were in both samples, except for NGC\,4258 (see text). The radii were measured on the basis of Fig.~\ref{comparesb}.}
	\label{tab:table1}
	\begin{tabular}{ccc} 
		\hline
		NGC & {\em HERON} Diameter & Merritt et al. Diameter \\
         & kpc & kpc \\
		\hline
		1084 & 27.2 & 28.0  \\
		2903 & 32.6 & 33.5  \\
		3351 & 16.9 & 17.4  \\
		M\,101 & 33.8 & 32.5  \\
		\hline
	\end{tabular}
\end{table}

In Sec.~\ref{sec:psf_impact} we consider the impact of the extended PSF on the measured diameters.

\section{Analysis }
\subsection{Sample}
Most of our galaxies were selected from the Two Micron All Sky Survey (2MASS) Large Galaxy Atlas (\citealt{Jarrett03}). We also observed a smaller sample of nearby low luminosity galaxies selected to have large angular diameters from the \citet{Karachentsev17} catalog, and with more distant early type galaxies from the ATLAS3D survey.  

While {\sl HERON} is by no means intended to be a volume-complete survey, we note that our sample encompasses $\sim$1\% of the galaxies in the Local Volume within 50~Mpc, given in \citet{White11}. 

\begin{figure}
\centering
\includegraphics[width=\columnwidth]{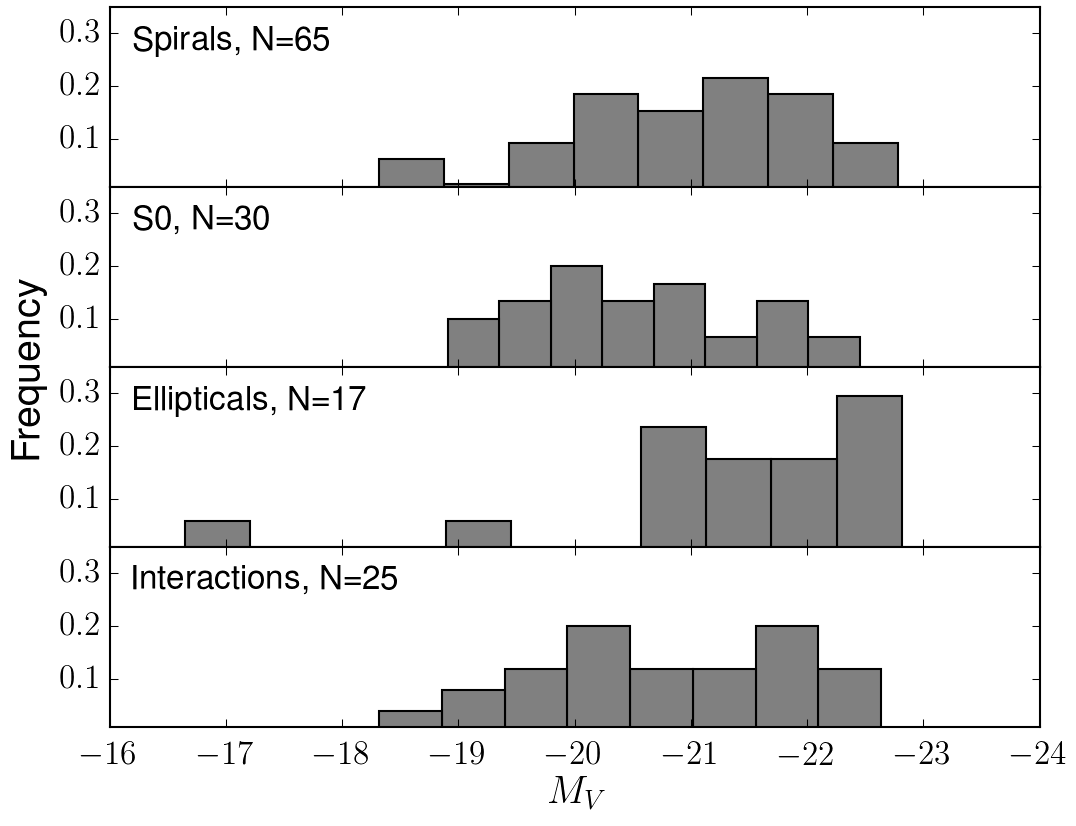}
\caption{The $V$-band absolute magnitude ($M_V$) distribution for the {\sl HERON} sample, divided by Hubble type. The sample labelled as ``interactions'' represent the subset of our sample displaying interaction signatures.}
\label{maghistos}
\end{figure}

Fig.~\ref{maghistos} shows the distribution of absolute magnitudes of our final sample, sorted by Hubble type. Although our sample has a large number of relatively nearby galaxies, it is dominated by galaxies with $M_V<-20$.  We are addressing that shortcoming by observing low luminosity galaxies selected from the \citet{Karachentsev17} catalog.  

\subsection{Depth of the images}
\label{sec:deepness}

Based on the created averaged surface brightness profiles (see Sec.~\ref{sec:surf_phot}), we determined a limiting surface brightness level for each sample galaxy where the error of the intensity is comparable to its value. These values are listed in Table~B2.

\begin{figure}
\centering
\includegraphics[width=\columnwidth]{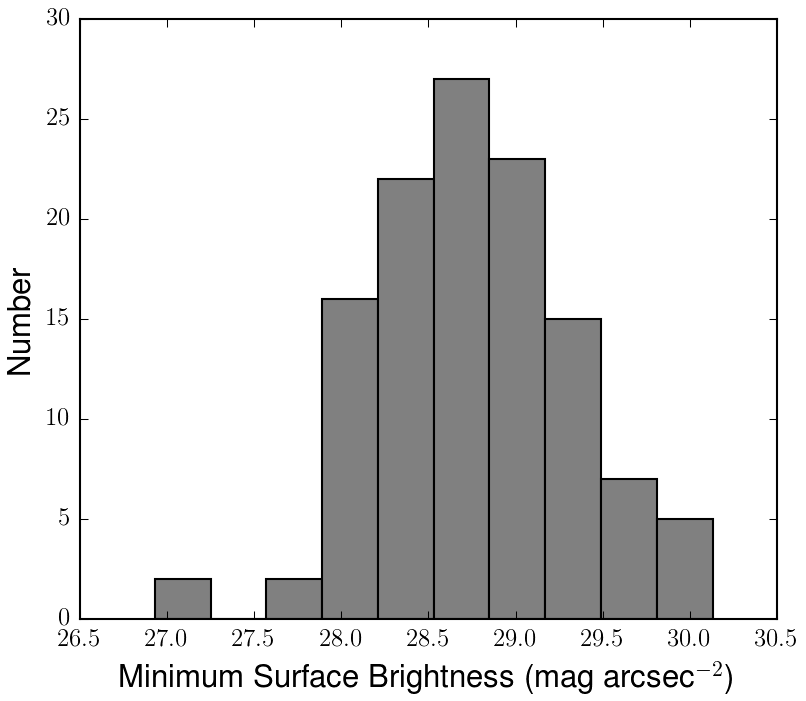}
\caption{Histogram displaying the corresponding surface brightness that our diameter measurements reached. The histogram displays a mean value of 28.8~mag/arcsec$^2$ and a standard deviation of 0.5~mag/arcsec$^2$.}
\label{surfbright}
\end{figure}

Fig.~\ref{surfbright} shows a histogram of the limiting surface brightness levels. The distribution looks normal with a mean surface brightness of $\sim 28.8\pm 0.5$~mag/arcsec$^2$. 115 galaxies (97\%) have limiting diameters at surface brightness levels deeper than 28~mag/arcsec$^2$ and 37 (31\%) deeper than 29~mag/arcsec$^2$. It appeared that four galaxy images in our sample (NGC\,525, NGC\,4258, NGC\,7465, and UGC\,4872) do not demonstrate deep profiles (their limiting surface brightness is higher than 28~mag/arcsec$^2$). Therefore, we do not consider them in our further analysis.

As the vast majority of our sample galaxies have surface brightness profiles that extend to 28~mag/arcsec$^2$ and deeper, we decided to define the diameter of the envelope/halo at this level. Further, we will refer to it as the envelope diameter. Alternatively, we will use the envelope radius as half the envelope diameter.

\subsection{Diameter measurements and magnitudes}
\label{sec:diam_measure}

We find that one of the most reliable and reproducible features is the measurement of the diameter as determined at a fixed surface brightness level. As defined above (see Sec.~\ref{sec:deepness}), the envelope diameter (radius) is measured at the 28~mag/arcsec$^2$ isophote, as almost all galaxies in our sample have surface brightness deepness up to this isophote. As some systematic errors may potentially influence the measured diameter, such as the uncertainty of the background measurement, the calibration error, the local error of the {\sc iraf/ellipse} model, we used Monte Carlo simulations to model galaxy profiles with taking into account all these errors. This allowed us to estimate an error on the diameter. We calculated that an average error of the envelope diameter for the whole sample is $3.4\pm1.6$~\% of the diameter value. In Table~B3 we provide the diameter values and its errors, along with the envelope shape which was visually estimated for each galaxy. We are about to discuss these envelope shapes in our subsequent study.

\begin{figure}
\centering
\includegraphics[width=\columnwidth]{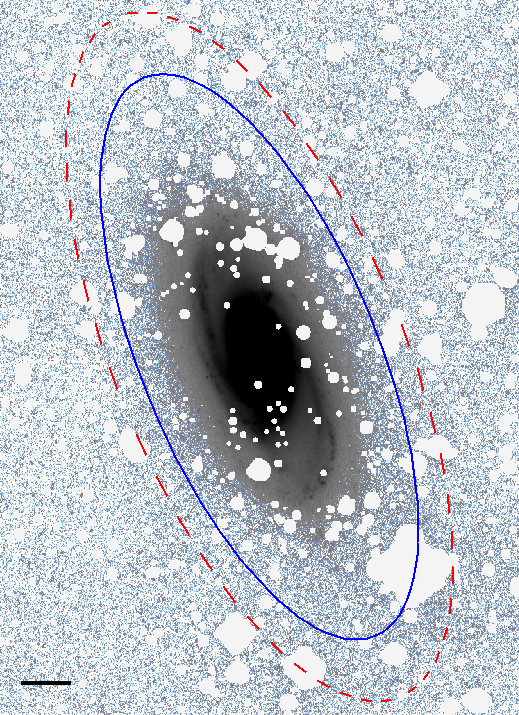}
\caption{This figure displays two envelope diameters for the 
SABbc galaxy NGC\,2903 (see also \citealt{Merritt16}). The blue solid ellipse is related to the isophote 28~mag/arcsec$^2$, whereas the red dashed ellipse to the 29~mag/arcsec$^2$.
 NGC\,2903 has a diameter of $24.1\pm0.6$~kpc at $M_V=-21.57$.
This figure is intended to display the lowest surface brightness we reached; other images with scale bars are shown in the Appendix \ref{Appendix:figs}, Fig.~\ref{invim1}. The scale bar in the left bottom corner shows $2\arcmin$.}
\label{2903d}
\end{figure}

\begin{figure}
\centering
\includegraphics[width=\columnwidth]{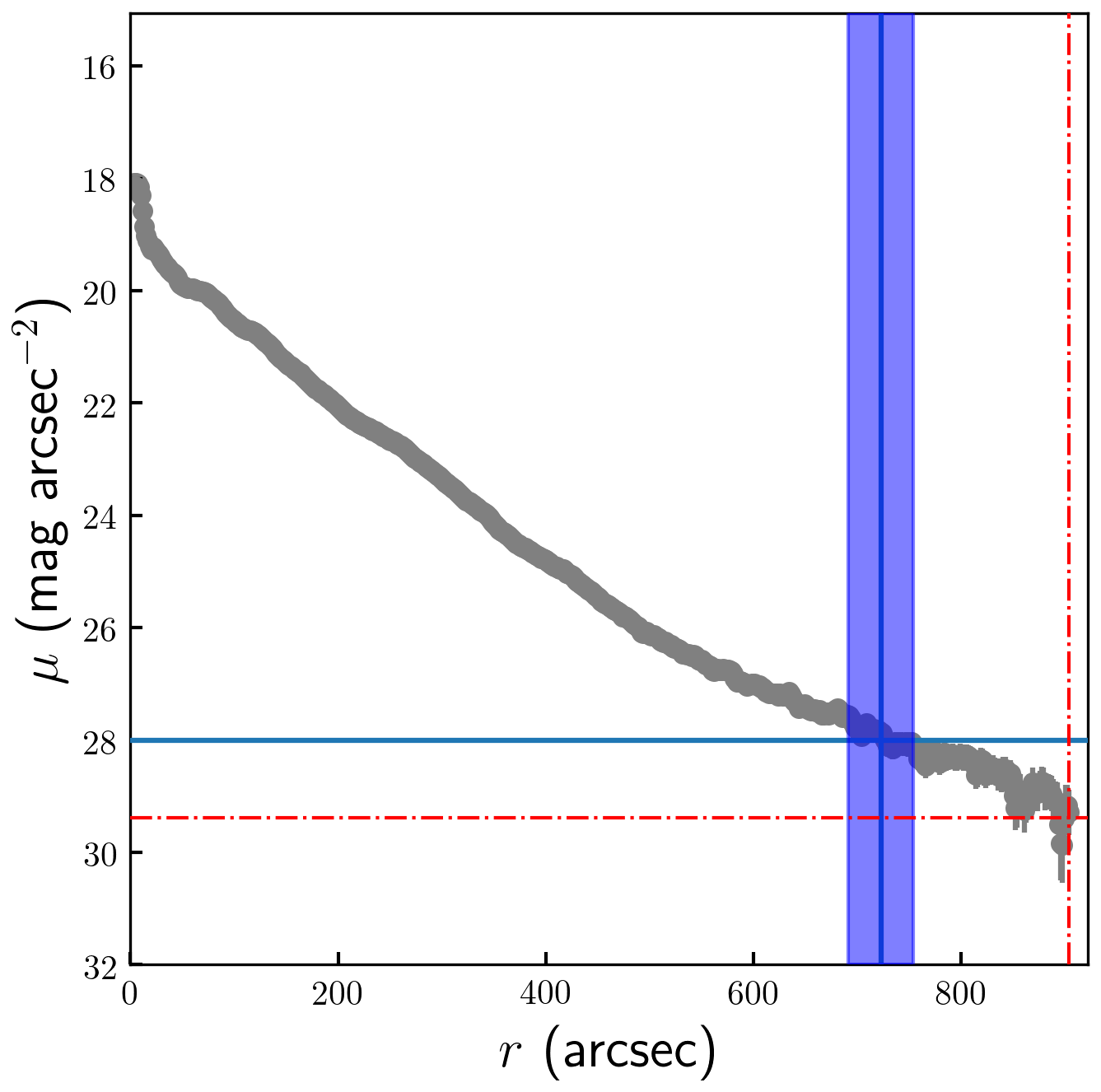}
\caption{Example of the envelope radius measurement for NGC\,2903. The blue solid horizontal line shows the surface brightness level 28~mag/arcsec$^2$, whereas the red dot-dashed horizontal line shows the limiting surface brightness level for this galaxy (29.4~mag/arcsec$^2$). The vertical lines show a measure of the envelope radii at these levels. The blue stripe for the radius of the 28~mag/arcsec$^2$ isophote includes different systematic errors (see text).}
\label{2903_prof}
\end{figure}

An example of a diameter measurement at two surface brightness levels, 28~mag/arcsec$^2$ and 29.4~mag/arcsec$^2$ (the limiting surface brightness level for this galaxy), are illustrated in Fig.~\ref{2903d} and Fig.~\ref{2903_prof}. 

In order to report a physical diameter and absolute magnitude, we must adopt a distance. In this work, we use the redshift-independent distances provided by the HyperLEDA database\footnote{\url{http://leda.univ-lyon1.fr/}} \citep{2014A&A...570A..13M}. For galaxies without redshift-independent HyperLEDA distances, we use the redshift-independent distances provided by the Nasa/ipac Extragalactic Database (NED)\footnote{\url{https://ned.ipac.caltech.edu/}} or the flow-corrected redshift-derived values provided by NED (if no redshift-independent distances are provided).

As some of our galaxies may have saturated nuclear regions, we do not use our photometry to measure galaxy fluxes. Instead, we adopt from HyperLeda the magnitudes $m_V$ and the colours $B-V$, denoted as \textit{btc} and \textit{bvtc}. These have been corrected for Galactic extinction using \citet{2011ApJ...737..103S} and internal absorption (see comments in HyperLeda).

\subsubsection{Impact of the PSF on our measurements of the diameter}
\label{sec:psf_impact}

As the PSF can significantly affect the real galaxy profile \citep[see e.g.][]{Sandin14,Sandin15}, we need to ensure that our measurements of the diameters do not suffer from this effect. For this purpose we should use an extended PSF, where extended wings should be larger than 1.5 times the distance between the outermost galaxy isophote (in our case it is 28~mag/arcsec$^2$) and the center, where we usually observe the maximum intensity. To take into account the PSF effect, two approaches are used. The first method includes a multi-component modelling of the galaxy using the convolution with an extended PSF \citep{Trujillo16}. The second approach uses deconvolution techniques \citep[see e.g.][]{2017A&A...601A..86K}. As our sample consists of a rather large number of objects of different orientation, morphology and angular size, the first approach is extremely time consuming. The second approach is more promising, taking into account a constantly increasing computational power to solve such kind of problems. This work has to be done in the future.

In this study we decided to investigate the effect of an extended PSF on our galaxy profiles by performing simulations of galaxy images and then convolving them with the extended PSF.

\begin{figure}
\centering
\includegraphics[width=8cm]{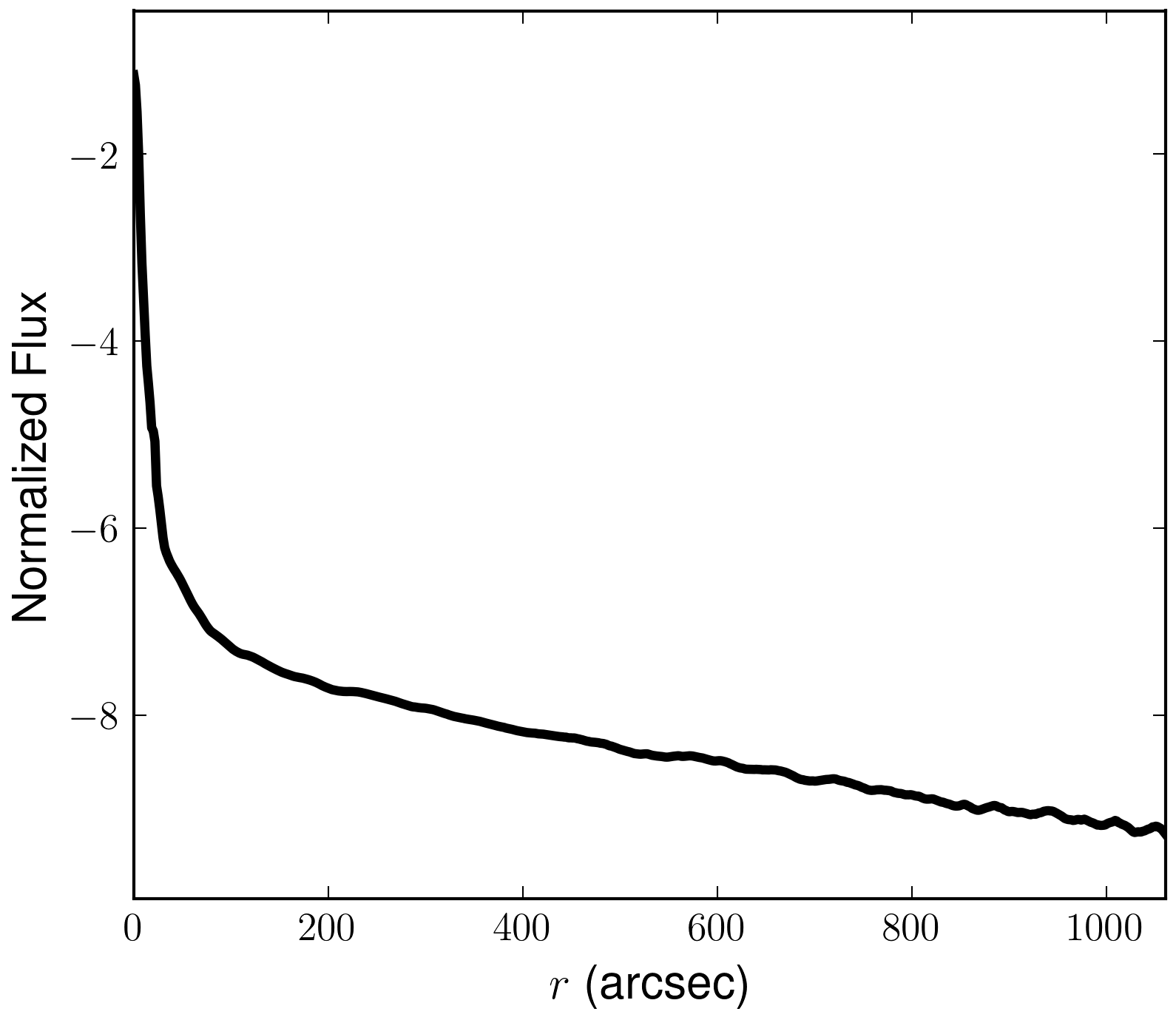}
\caption{Normalized azimuthally-averaged profile of the PSF model profile.}
\label{ext_psf}
\end{figure}

To create an extended PSF for the {\sl HERON}, we used an observation of the bright star HD\,9562. Using the {\sc iraf/ellipse} routine, we created its azimuthally averaged profile up to a radius of $\approx1000\arcsec$. However, as its central part is saturated, we replaced the core of the extracted profile by a non-saturated star (the normalization was done at the intersection of the profiles, see \citealt{2017A&A...601A..86K} for details). The synthesized profile is shown in Fig.~\ref{ext_psf}.

\begin{figure*}
\centering
\includegraphics[width=5.5cm]{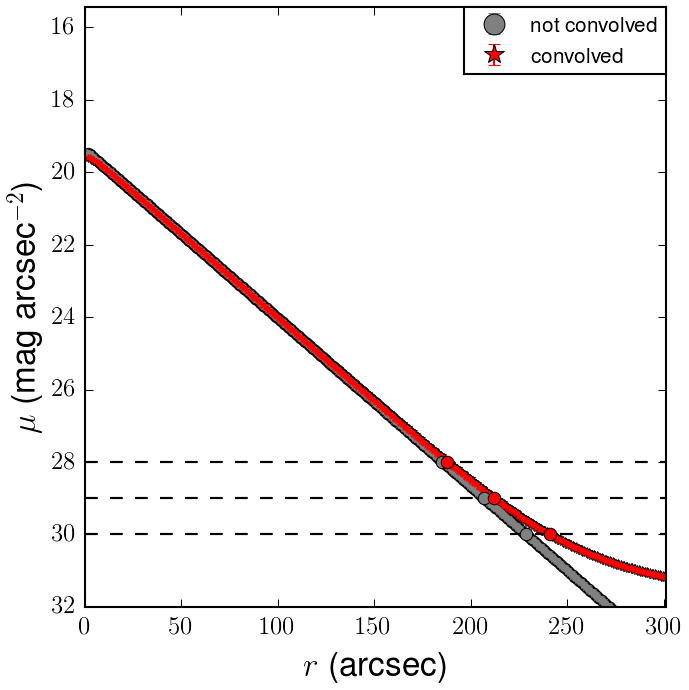}
\includegraphics[width=5.5cm]{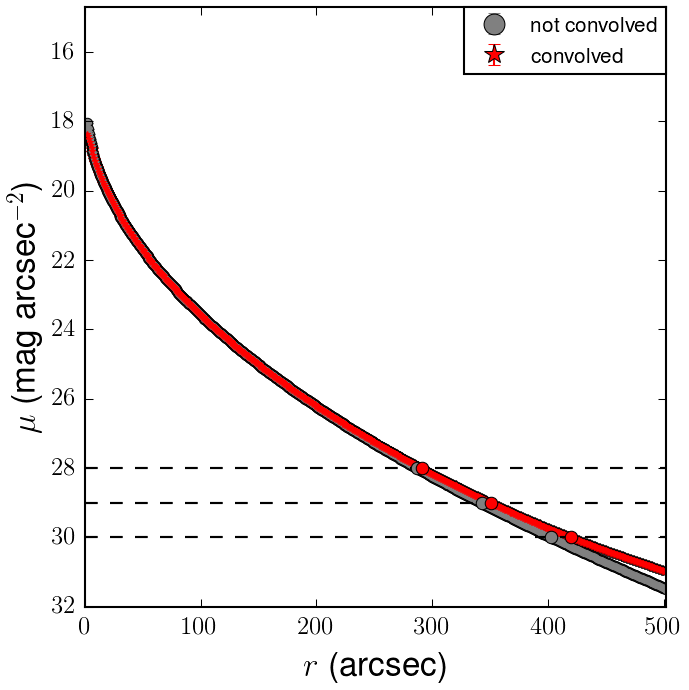}
\includegraphics[width=5.5cm]{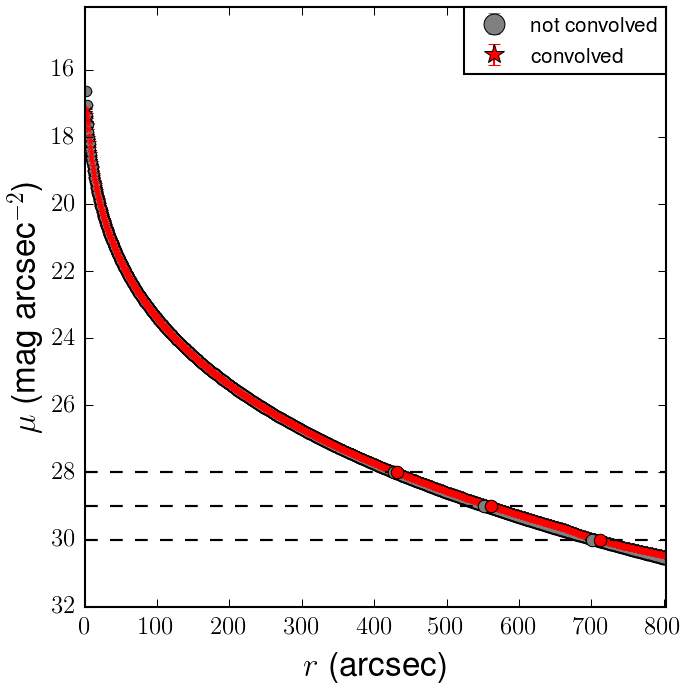}
\caption{The comparison of the original (non-convolved) and convolved profiles for three S\'ersic models with $n=1$ (left), $n=2$ (middle) and $n=4$ (left). The dashed lines show some characteristic surface brightness levels of 28, 29 and 30~mag/arcsec$^2$.}
\label{psf_impact}
\end{figure*}

As in our sample there is a large variety of morphologies, let us consider three simple models of galaxies: with the S\'ersic index $n=1$ (late-type galaxy), 2 (galaxy with a compact bulge) and 4 (elliptical galaxy). Other parameters of the S\'ersic model are: the effective radius is $47\arcsec$ and the effective surface brightness is 21.2~mag/arcsec$^2$ (these are the average values for our sample determined from our {\sc galfit} decomposition, see Sec.~\ref{sec:surf_phot}). The 2D images were simulated with {\sc galfit} and then convolved with the extended {\sl HERON} PSF. The results of the envelope radius estimation are shown in Fig.~\ref{psf_impact}. As one can see, the overestimation of the real radius at the 28~mag/arcsec$^2$ isophote is very small (less than 1.5\% for all $n$). However, it is getting larger for lower surface brightnesses: for the 30~mag/arcsec$^2$ isophote an overestimation increases up to 5.5 \% for $n=1$. 
From this comparison it is obvious, that our measurements of the envelope radius are within the typical errors of the envelope radius, and cannot affect our results. We should notice, however, that the presence of an AGN may change this conclusion as the bright compact source will produce extended wings which can be mis-interpreted as a halo. However, only 12 galaxies in our sample exhibit some activity at the center, therefore, we do not consider this case.    

\subsubsection{Tests of our diameter measurement metric}

In order to reassure us that our measurement of diameters is robust and does not arise due to a selection effect, we have performed a number of tests on the data. 

\begin{figure}
\centering
\includegraphics[width=\columnwidth]{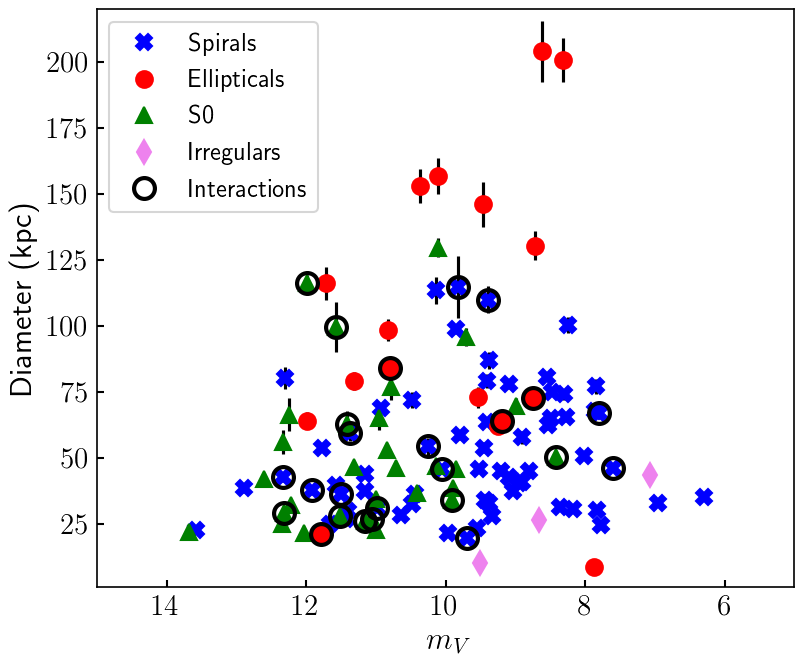}
\caption{Envelope diameter (kpc) versus apparent $V$ magnitude. We do not see any trend, as expected if scattered light does not contribute to the measured envelope diameter. The apparently brightest galaxies might in principle have spuriously larger diameters caused by scattered light from their e.g. bulge or disk components.   
\label{apparmag}}
\end{figure}

\begin{figure}
\centering
\includegraphics[width=\columnwidth]{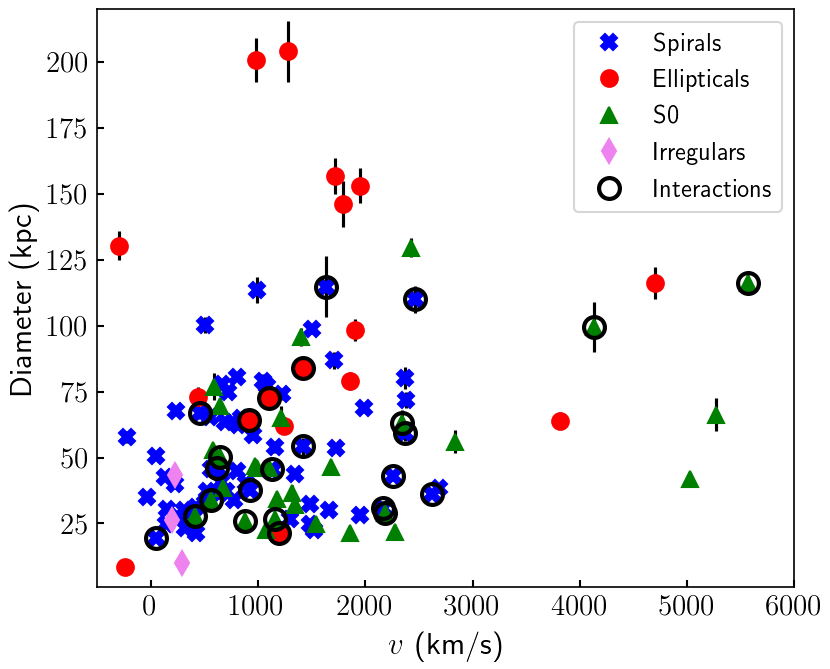}
\caption{Envelope diameter (kpc) versus radial velocity with galaxy classification indicated. No trend is evident; the distant galaxies with large halos correspond to luminous systems that are more rare in our lower redshift sample. The lack of any strong correlation of halo diameter with velocity is expected.
\label{redshiftdiam} }
\end{figure}

\begin{figure}
\centering
\includegraphics[width=\columnwidth]{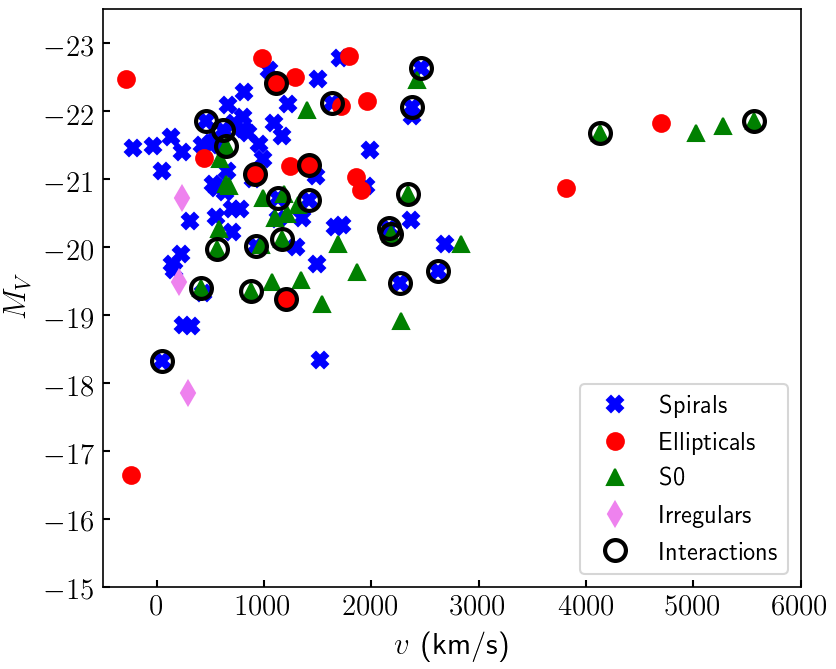}
\caption{A plot of absolute $V$ magnitude ($M_V$) versus radial velocity. The sampling in 
absolute magnitude and radial velocity shows no apparent bias. The present sample is relatively sparse at the low luminosity end; this will be supplemented using the \citet{Karachentsev17} catalogue, as described in the text.}
\label{redshiftlum}
\end{figure}

Fig.~\ref{apparmag} shows that there is no correlation between diameter and apparent $V$ magnitude. If scattered light from bright central bulges or disks were a significant contributor to our measurement of diameter, one might expect to see a correlation between diameter and apparent magnitude. Fig.~\ref{redshiftdiam} presents our correlation between diameter and radial velocity. Again, no strong correlation is expected. However, our most distant galaxies include some of our most luminous. Fig.~\ref{redshiftlum} considers absolute magnitude vs. radial velocity. Our sample is under represented at the faint end; as mentioned previously we are taking steps to address this.  
We conclude from these tests that our measurements of diameter do not suffer from any systematic trends, and they do not behave as expected if the diameter measurements were seriously affected by scattered light or systematic error.

\section{Discussion}

\begin{figure*}
\centering
\includegraphics[scale=1.]{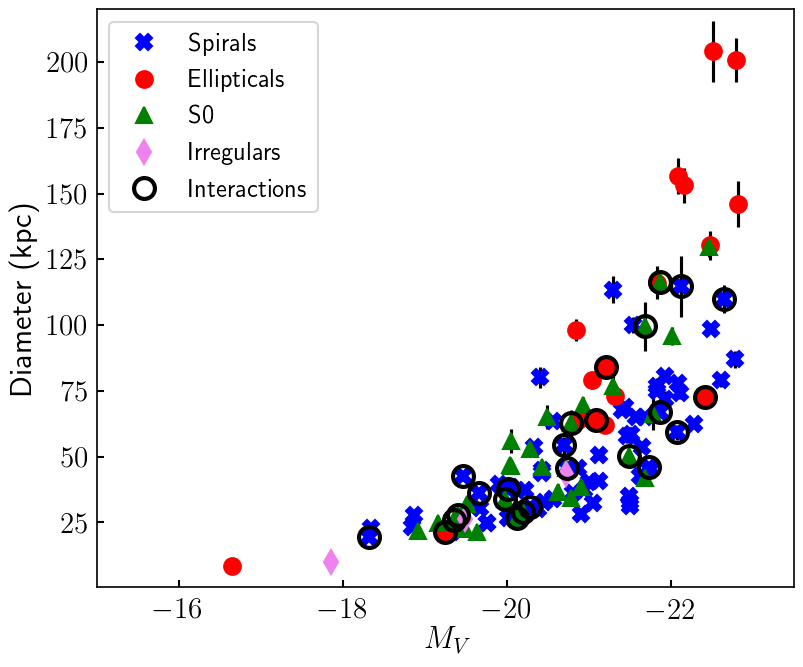}
\caption{A plot of envelope diameter (kpc) versus absolute magnitude $M_V$. Galaxy classification is according the symbols in the legend. Circled symbols indicate galaxies with a stream, extended shell, or otherwise strongly asymmetric interaction. The largest envelopes are found in the most luminous galaxies. 
\label{envelopeskpc} }
\end{figure*}

We now address how our measurement of envelope diameter correlates with various physical properties of galaxies in our sample. We first consider the characteristics of our sample as a whole in Fig.~\ref{envelopeskpc}. S0 and spiral galaxies span the widest range in luminosity. Although our most luminous galaxies are elliptical, the absolute magnitude distribution of galaxies by Hubble type and presence of interaction signatures is remarkably similar.

Fig.~\ref{envelopeskpc} shows a strong primary correlation between the measured physical diameter of the envelope, and absolute $V$ magnitude, agreeing with our work earlier reported in \citet{Rich17}. We have confirmed this trend based on diameters at 25~mag/arcsec$^2$ (D25) from the NED Database. The D25 data confirm the general trend but as expected, do not reach the largest diameters.  Similarly, our work confirms the work of \citet{munoz15} that finds similar trends in galaxy diameter and stellar mass from Spitzer data.
Our data show an apparent transition near $M_V \sim -20.5$ or $L^*$ at which point a subset of galaxies, mostly E and S0, begin to display very large diameters, reaching 150~kpc. However, a few spirals also have large envelopes, as previously noted by \citet{Kormendy74}. We find no tendency for interactions to be detected in a particular magnitude range, except for their relative paucity for $M_V>-20.5$.  This trend is not a simple correlation between galaxy diameter and total mass. For example, at $-20 < M_V < -24$, galaxies of all morphological types exhibit a total range in measured diameter from 20 to 170~kpc and within this most luminous category there is only weak dependence of diameter on luminosity.   

\begin{figure}
\centering
\includegraphics[width=\columnwidth]{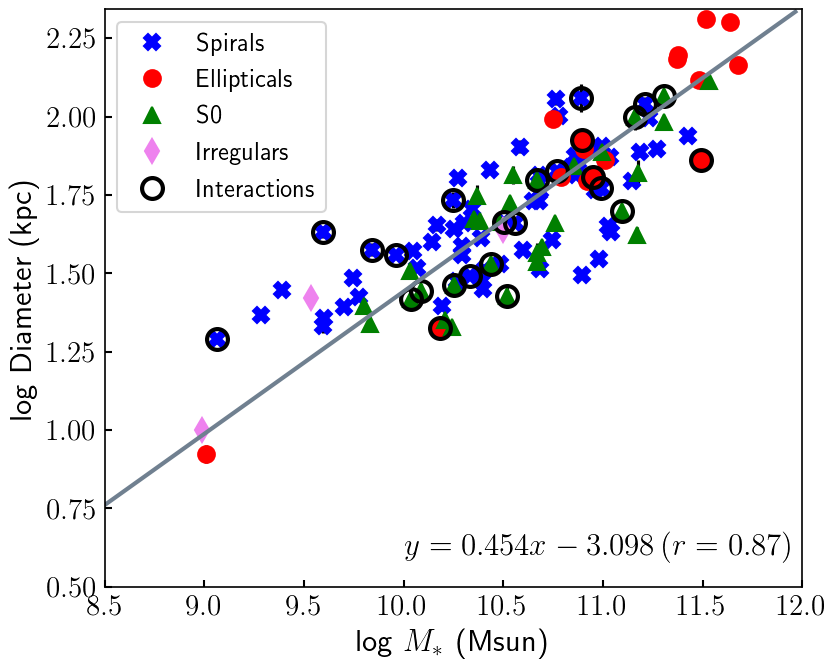}
\caption{Log diameter in kpc vs stellar mass using the relationship of \citet{Bell03}.  The trend extends to lower surface brightness that reported in \citet{munoz15}} 
\label{logenvelope} 
\end{figure}

When plotted as log Diameter vs. stellar mass in Fig.~\ref{logenvelope}, the apparent break at $L^*$ is not as evident; we also observe a trend similar to that found by \citet{munoz15}. This plot also suggests that $\sim200$~kpc may represent an upper limit to envelope diameter, although more observations are required to confirm this. One very large envelope with diameter $\sim 170$~kpc has been found surrounding Hickson group HCG~98 and is reported in \citet{2019MNRAS.482.2284B}. It will be important to explore whether these correlations extend to the total luminosities encompassed in small galaxy groups.

\subsection{The Galaxy Colour--magnitude diagram}

\begin{figure}
\centering
\includegraphics[width=\columnwidth]{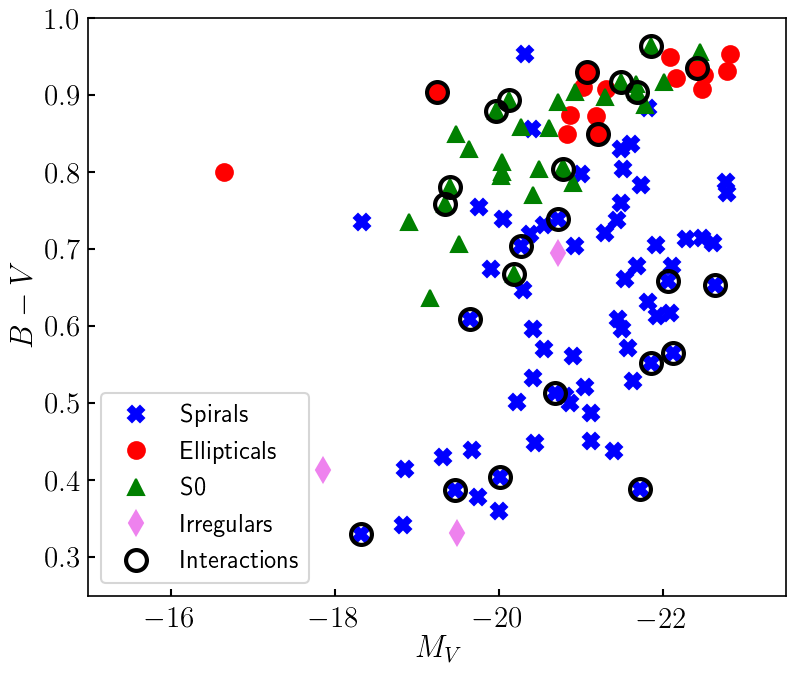}
\caption{A plot of $B-V$ colour versus $M_V$. Circled symbols indicate interactions. The diagram divides cleanly into the luminous red sequence, green valley, and blue cloud. Interactions appear to occur throughout this plot and do not favour a particular location.\label{colormag} }
\end{figure}

We turn next to explore correlations across the galaxy colour--magnitude diagram. Our motivation is to explore whether the signatures of active interactions preferentially
populate any part of the CMD. Fig.~\ref{colormag} shows that galaxies hosting interaction signatures \citep{Duc17} appear to show no preference for populating the blue or red sequence, or the green valley. Interactions may be a significant factor in driving quenching, as the significant infall of baryons and dark matter might be expected to be a factor that induces star formation.  

\begin{figure}
\centering
\includegraphics[width=\columnwidth]{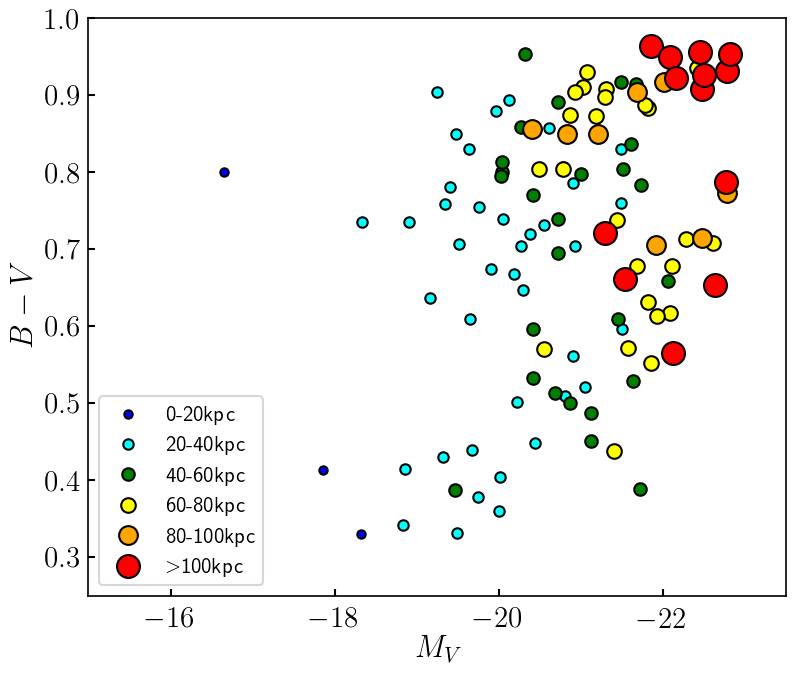}
\caption{The same plot as Fig.~\ref{colormag}, $B-V$ colour magnitude versus $M_V$. Here, the different coloured symbols represent different envelope diameters according to the legend. The bright end of the red sequence hosts the largest envelopes, those with diameter $>100$kpc, but the faint end of the red sequence has halo sizes comparable to that of the blue cloud. The largest envelopes are seen in the most luminous galaxies regardless of whether they occupy the blue cloud, green valley, or red sequence. Envelopes of median diameter can be found across the plot.}
\label{colormagsize}
\end{figure}

Fig.~\ref{colormagsize} shows a new result: the largest envelopes appear to preferentially populate the luminous end of the red sequence and include both S0 and E galaxies.  However, large halos are also seen in the blue cloud.  We previously found that the envelope size is correlated with luminosity and recall that Fig.~\ref{histomags} shows that elliptical galaxies host the largest halos; hence the largest envelopes are found in $M_V<-21$ galaxies on the red sequence.

However, the largest envelopes are not {\it confined} to the red sequence, with the most luminous blue sequence members and at least one green valley galaxy exhibiting large envelopes as well. It will be interesting to consider the role of environment in future work, however, it is noteworthy that only the bright end of the red sequence hosts the largest envelopes. There appears to be no clear preference for larger envelopes on the faint end of the red sequence compared to the blue cloud. The strong primary correlation between intrinsic luminosity and envelope size is of greatest importance, but for galaxies with $M_V<-21$, the envelopes of greatest diameter are found at the bright end of the red sequence and are notably less common in the blue cloud and green valley. Simulations also predict, at fixed stellar mass, more massive stellar halos in red galaxies than in blue \citep{Elias18}. However, as noted originally by \citet{Kormendy74}, the largest envelopes can be found in both spirals and ellipticals. 

The cases of very large envelopes not on the red sequence are unusual. NGC\,474 is an elliptical galaxy in the green valley, but it is involved with a significant (likely recent disk) merger event. The merger shells are bluish on false colour images (see e.g. \citealt{Duc15}); this galaxy will likely migrate to the red sequence after the merger event settles. NGC\,5746 shows one of the largest envelopes found for a galaxy in the blue cloud. This edge-on, boxy/peanut shaped bulge galaxy has an extraordinary 60.3~kpc diameter envelope, and its rotation curve has the highest peak velocity in the \citet{Bureau99} sample; $\pm 500$~km/sec. NGC\,772 is also identified as being in the blue cloud, but it is a face-on spiral with one spiral arm and 3 galaxies entrained in a stream; they all lie projected on a field of complex infrared cirrus. It is in the blue cloud by virtue of its disk, and hosts an extremely large envelope due to the ongoing interaction. NGC\,474, NGC\,772, and NGC\,5746 are anomalous in their hosting of large envelopes yet not residing on the red sequence. The remaining largest envelopes clearly reside on the red sequence and are ellipticals and S0s.

\begin{figure}
\centering
\includegraphics[width=\columnwidth]{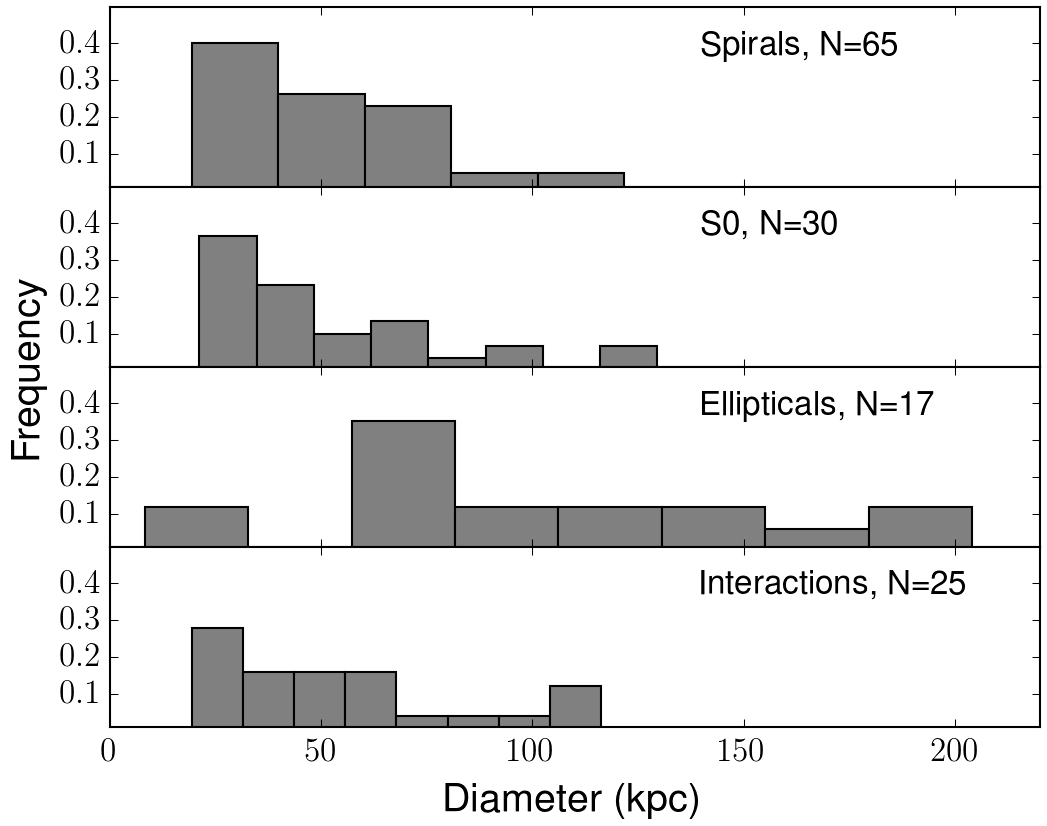}
\caption{A set of histograms displaying the number of galaxies as a function of envelope diameter (kpc), separated by Hubble type. We can see that spiral and S0 galaxies are, in general, smaller than elliptical galaxies in our sample. The ``interactions'' histogram displays the envelope diameters of galaxies in our sample that show signs of interaction; such features include streams, plumes, or shells. \label{histoenvelope}}
\end{figure}

\begin{figure}
\centering
\includegraphics[width=\columnwidth]{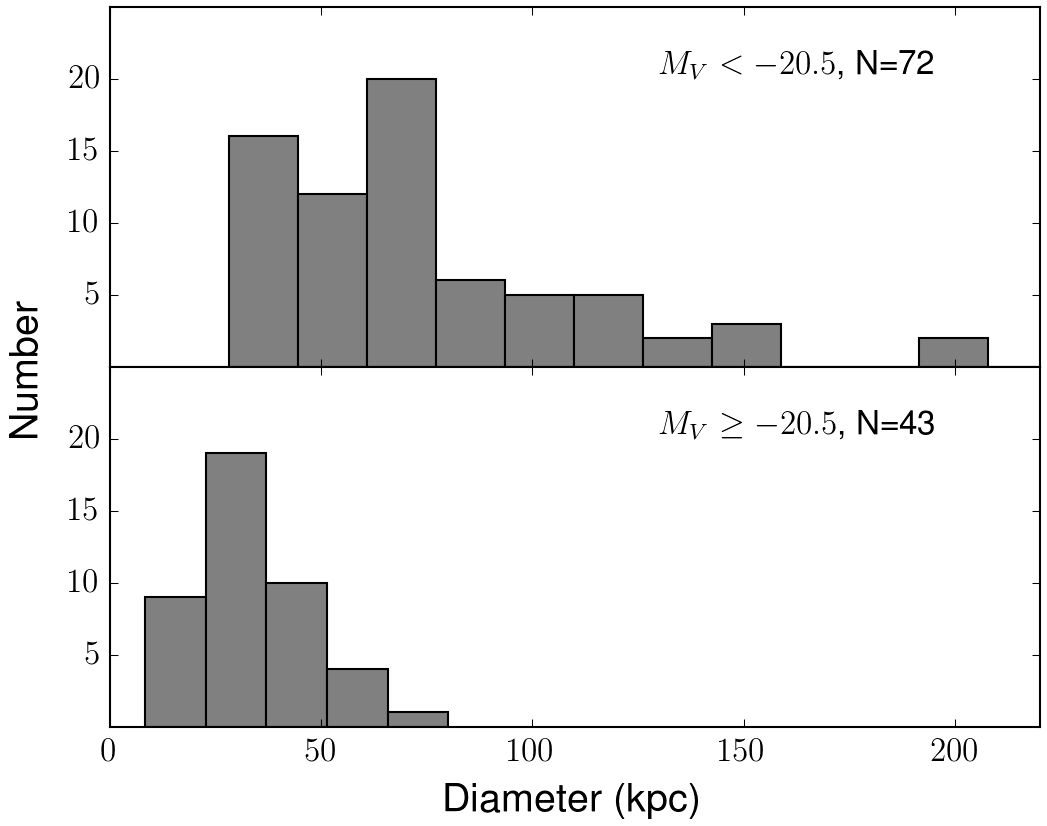}
\caption{Histograms displaying the number of galaxies as a function of envelope diameter (kpc), divided by absolute magnitude at $M_V = -20.5$. It is clear that galaxies with $M_V < -20.5$ have a greater envelope diameter. \label{histomags}}
\end{figure}

\begin{figure}
\centering
\includegraphics[width=\columnwidth]{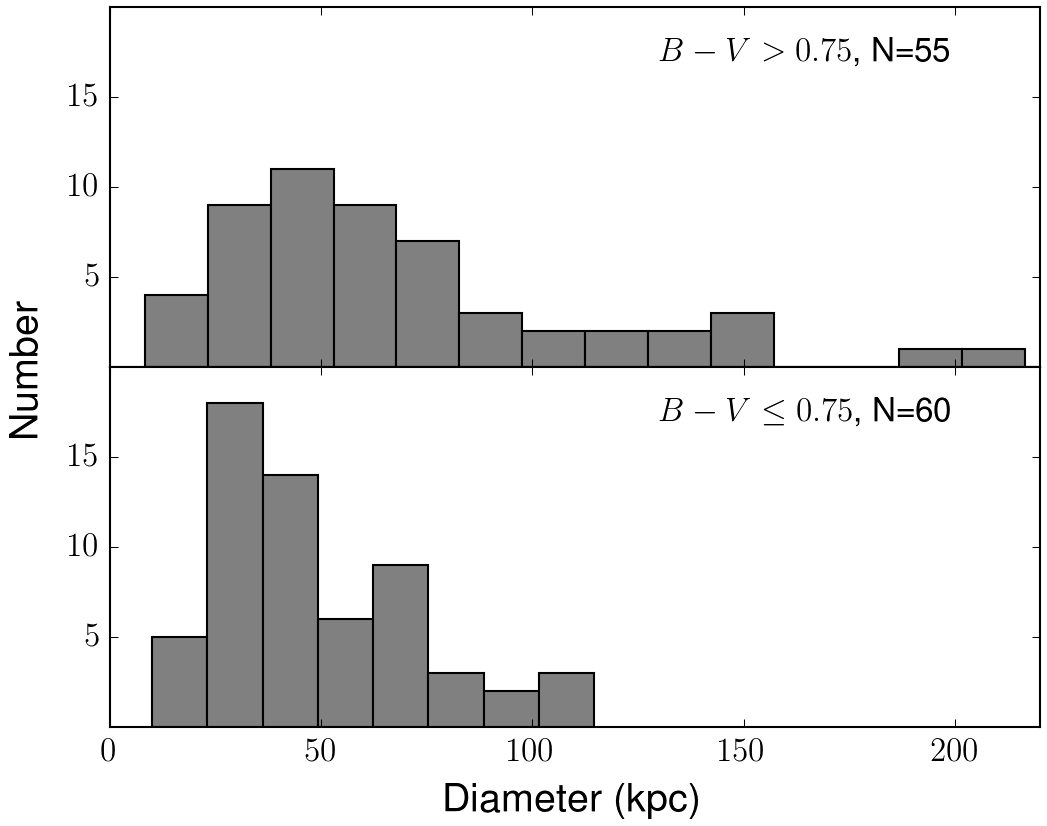}
\caption{Another illustration that the largest halos are found at the bright end
of the red sequence (Fig.~\ref{colormagsize}). Histograms displaying the number of galaxies as a function of envelope diameter (kpc), divided by colour at $B-V = 0.75$. This represents a rough colour cut between the red sequence and blue cloud.
\label{histocols}}
\end{figure}

\begin{figure}
\centering
\includegraphics[width=\columnwidth]{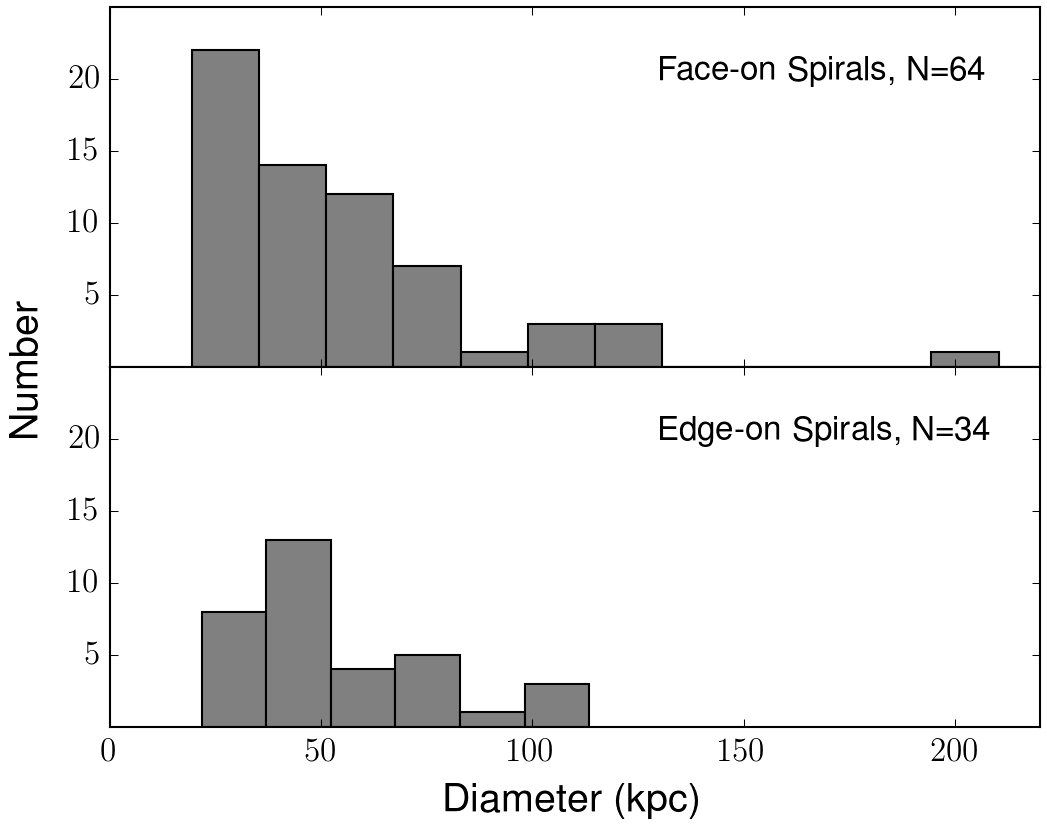}
\caption{Envelope diameter for face-on and edge-on spirals. The KS-test gives $\mathcal{D} = 0.15$ and a p-value $\mathcal{P} = 0.62$ for this result.  We do not find strong evidence for different diameters for edge-on versus face-on disks.
\label{edgeon}}
\end{figure}

In Fig.~\ref{histoenvelope}, we note that the largest envelopes are found in elliptical galaxies, but that galaxies with interactions host envelopes that span the full range of diameter. Larger samples will be required to assess whether interactions are found in specific circumstances e.g. small groups or close companions. Future work will also address low luminosity companions and their characteristics.

Fig.~\ref{histomags} shows the very clear difference in the distribution of envelope diameters when segregated by luminosity. It is clear that the bulk of galaxies with $M_V<-20.5$ have envelopes larger than those fainter, a statistic that is highly unlikely to change, even if the sample size were to increase. In  Fig.~\ref{histocols} we present the histograms that correspond to an approximate red sequence/blue cloud colour cut in Fig.~\ref{colormagsize}; this reinforces our claim that the largest halos are found on the red sequence.

\section{The nature of envelopes of disk galaxies}

In Fig.~\ref{edgeon}, we ask whether projected orientation has an impact on the measured diameter of the envelope.  We see that edge-on disks,  and find no clear difference in projected sizes. We now turn to consider the possibility that the outermost detected light in face-on disks (e.g. \citealt{Merritt16}) may arise from a disk population, and may not be a classical Population II halo.

\begin{figure}
\includegraphics[width=\columnwidth]{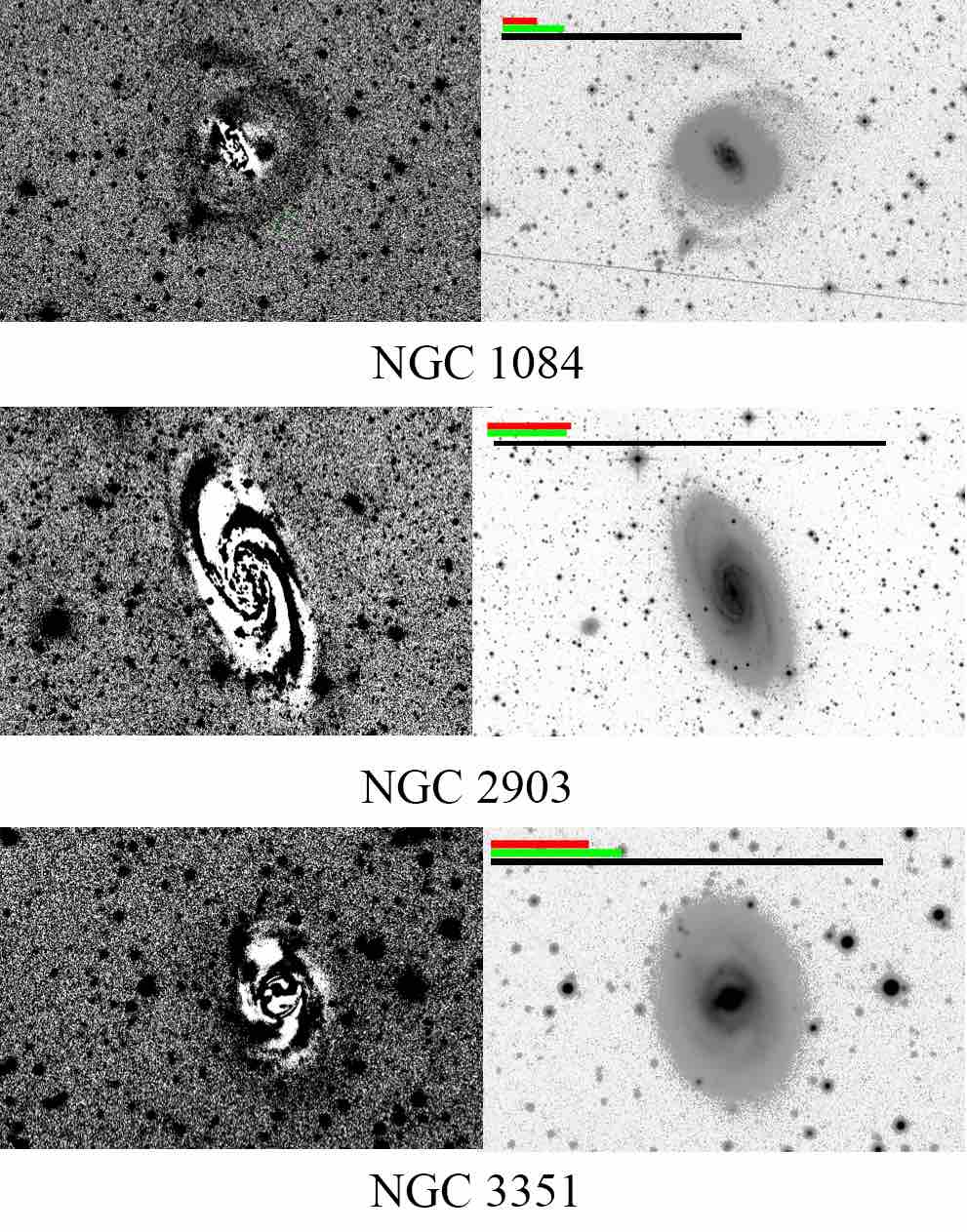}
\caption{Subtraction images next to the corresponding inverse image. The black scale bar on the inverse image panels represents the diameter at the 30th~mag/arcsec$^2$ that \citet{Merritt16} lists in their paper. The green bar represents 5' and the red bar represents 10~kpc. }
\label{subtractcompare}
\end{figure}

The outer edges of spiral galaxies are broadly observed to divide between those showing evidence of star formation and spiral or flocculent spiral structure (e.g. M\,101) and those showing a smooth outer extension (M\,83). Fig.~\ref{subtractcompare} illustrates these examples; they represent extremes, with star formation at the outer edges being the more common occurrence. M\,51 presents a smooth but unusually shaped outer envelope, whereas M\,74 and M\,101 appear to show the more commonly seen spiral structure. In the case of M\,83, the outer envelope is unusual in that it extends substantially beyond the disk, shows no spiral structure even in subtraction, and is oval and off-centred. This outermost structural feature suggests that (as is the case with M\,51) its envelope is more likely to be a flattened disk-like projection.  

Exploring further the question of which stellar population is represented in the outer parts of spirals, we present subtractions of the ELLIPSE models for galaxies that are discussed in \citet{Merritt16}. In Fig.~\ref{subtractcompare}, we can observe that the outer regions of NGC\,1084, NGC\,3351, and NGC\,2903 show clear spiral structure in subtraction. Using both imaging from the 0.7-m C28 telescope as well as verification images obtained by B. Megdal employing a single lens 8-inch refractor (to reduce scattered light issues), we do not detect envelope light outside of the extreme edge of the disturbed disk (the broad ``arm'' like structure at the top of the NGC\,1084 image in Fig.~\ref{subtractcompare}.)  The refractor observations were undertaken in order to confirm that our 0.7-m data are not compromised by scattered light. Our observations are not able to confirm the extended low surface brightness profiles reported in \citet{Merritt16}.

Finally, Fig.\ref{invim1} in the Appendix present images of galaxies included in our study, along with scale bars for apparent and physical diameter. We include several figures here, with the rest included as supplemental material. The entire set of images will also be posted on the {\sl HERON} website at IRSA.

\begin{figure}
\includegraphics[width=\columnwidth]{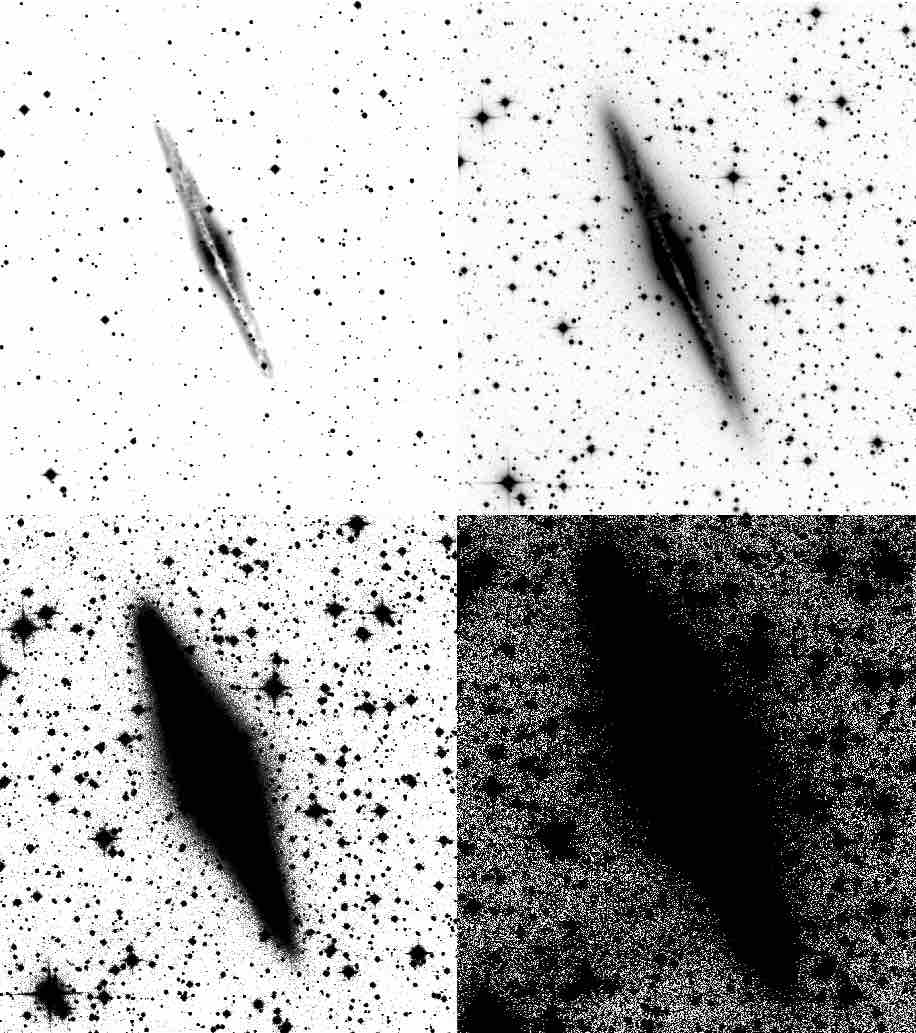}
\caption{The edge-on spiral galaxy NGC\,891 is displayed from shallow to deep stretch. Notice that the minor axis of the bulge is significantly smaller than the extent of the disk
major axis; this is typical of all edge-on disk galaxies.  At the deepest stretch,
the galaxy assumes a ``trapezoidal'' appearance due to the extent of the spheroid,
but the disk always has the greatest major axis.}
\label{891range}
\end{figure}

Noting that a significant fraction of edge-on disk galaxies exhibit disturbances (e.g. NGC\,3628 and NGC\,4216), we suspect that the faintest detectable light at the edges of face-on disks in disk galaxies should be attributed to the disk, not the halo. Fig.~\ref{891range} illustrates NGC\,891, a typical edge-on galaxy with a bulge. The deepest exposures show that the outermost isophotes are trapezoidal, with the major axis aligned with the disk. Among the most striking examples of an edge-on trapezoidal envelope is NGC\,2683. The minor axis is always aligned with the spheroidal component and perpendicular to the disk. This is characteristic of all of our edge-on disk galaxies: we have no cases where a low surface brightness envelope ever projects to a larger size than the disk, save for that M104. M104 can be considered a disk galaxy, showing a prominent dust lane and bulge, but does show one of the largest halos in our sample.

\begin{figure}
\centering
\includegraphics[width=\columnwidth]{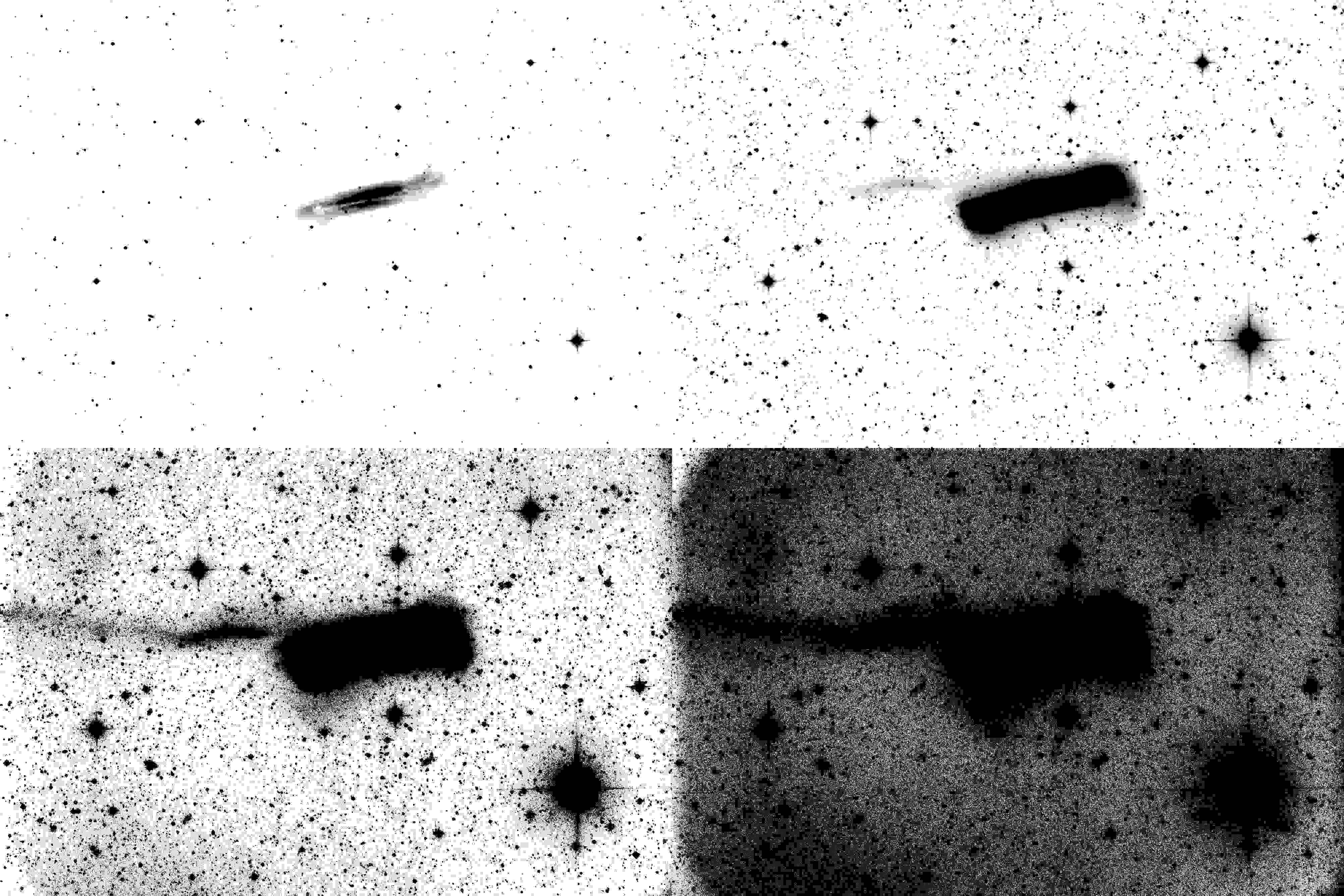}
\caption{NGC\,3628 is displayed from shallow to deep stretch. Notice that the
disk grows from roughly normal in appearance to an almost rectangular morphology.  
This kind of disturbance may give rise to ``rectangular'' envelopes seen in some S0 galaxies, like that found in NGC\,720. The thickness of the rectangular disk exceeds
50~kpc.}
\label{n3628range}
\end{figure}

\begin{figure}
\centering
\includegraphics[width=\columnwidth]{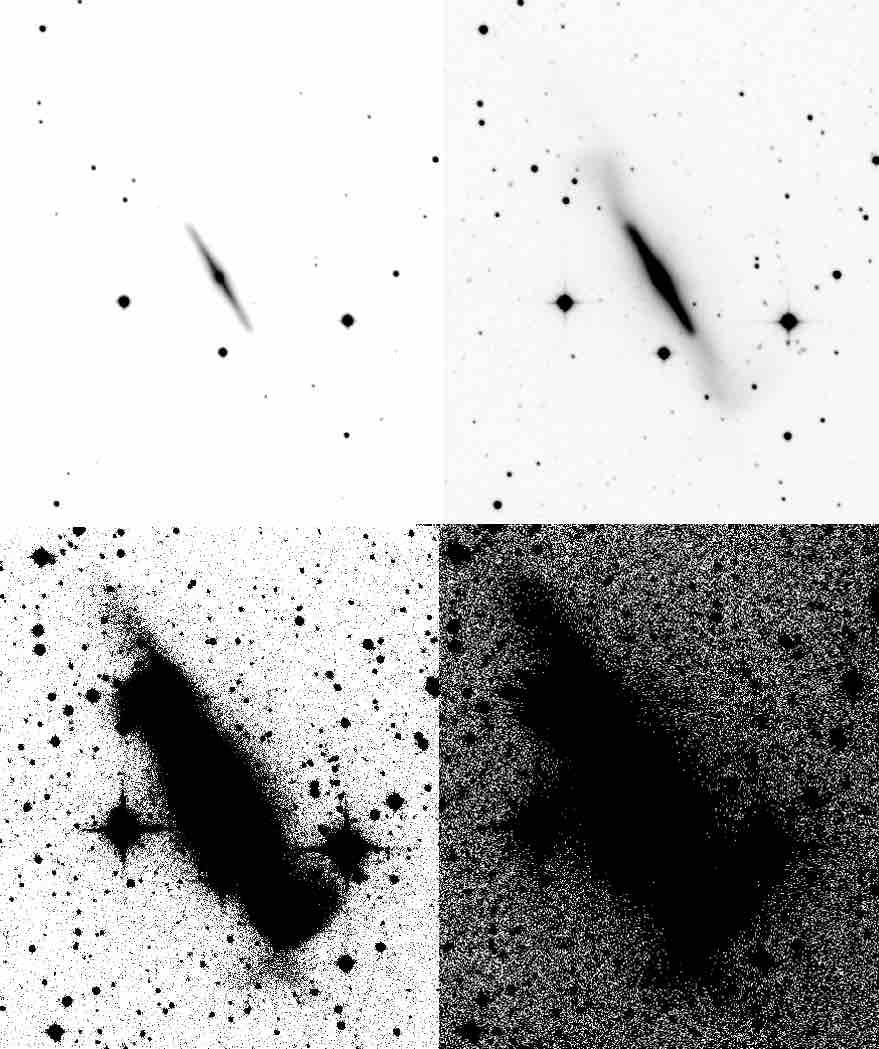}
\caption{Extreme disturbance of the S0 NGC\,4762, displayed with a range of stretch.
The shallow exposures follow those in the Hubble Atlas of Galaxies, while the deepest
stretch shows an almost ``shoe''-like structure resulting from the interaction envelope.}
\label{4762range}
\end{figure}

Fig.~\ref{n3628range} shows the unusual case of NGC\,3628. Although the tidal tail has long been noted, the buckled and disturbed disk is thick, with the deepest isophotes showing a boxy 2:1 structure. One could safely assume that all light contributing to the low surface brightness components of this galaxy belongs to the disk. Fig.~\ref{4762range} shows the disturbed edge-on S0 galaxy NGC\,4762. Deep images have previously shown the disturbed disk in the second panel, but our {\sl HERON} images show an extended ``shoe''-like structure that we suggest may consist of disk stars that were heated or disturbed during an interaction.  

We conclude that the outermost detectable envelopes of face-on disks consist of disk stars. This position is based on Fig.~\ref{edgeon}, our deep imaging of edge-on disks in Figs.~\ref{891range}-\ref{4762range}, and the edge-on disks in our sample from \ref{invim1} (including additional supplemental inverse images). Furthermore, deep H{\sc i} images (e.g. \citealt{Sancisi08} and other studies) find H{\sc i} envelopes around spiral galaxies in the disk plane, including that of M\,51. We argue that studies of face-on disk galaxies such as that of \citet{Merritt16} are in fact detecting extended disk light. Fig.~\ref{subtractcompare} illustrates our model subtractions of 3 of the galaxies in \citet{Merritt16}, and in all cases, spiral structure or disturbances dominate the outermost isophotes. Even though \citet{Merritt16} finds light outside these isophotes, we are not able extend our surface brightness measurements to such a faint level. Even so, we argue that for all disk galaxies, especially those with near face-on inclination, the lines of evidence from our stellar imaging study and that of \citet{Sancisi08} and similar H{\sc i} studies, support the outermost visible light isophotes being dominated by stars in the disk plane, not in the spheroidal old halo. These stars may owe their presence to disk flaring. Other evidence arises from studies of the extreme UV (XUV) disks e.g. \citet{Werk10}. \citet{Lemonias11} found that 4-14\% of galaxies to $z=0.05$ have XUV disks, with 7-18\% of galaxies in the green valley being candidates to transition {\it away} from the red sequence.  

The prima facie evidence of a true Population II halo would be the presence of globular clusters, but detection of globular clusters in sufficient numbers at radii $>30$~kpc would be difficult even if the spatial resolution were available to resolve them: distant clusters are rare, even in highly populated systems.

\section{Conclusion}

We report new imaging to low surface brightness for a sample of nearby galaxies predominantly from the 2MASS nearby bright galaxy catalogue, and mostly lying within the boundary of the Local Volume $\sim 50$~Mpc.  We show that our imaging using the Jeanne Rich C28 0.7-m telescope reaches $\sim 28$~mag/arcsec$^2$, and reproduces well the low surface brightness structures and surface brightness profiles reported in the literature. In $\sim$ one hour exposures, we reproduce published faint structure from  amateur exposures of tens of hours, Dragonfly, and the CFHT. We did not fail to measure, or observe, any low surface brightness features reported by others in the literature.  

We measure the diameters of the envelopes not including transient structures such as streams, arcs, and interaction filaments. We find a strong primary correlation between envelope diameter and $M_V$, after carefully checking for spurious correlations between envelope diameter and apparent surface brightness, and distance. We find that the largest envelopes are hosted by the most luminous elliptical galaxies. However, very large envelopes are found spanning the full range of morphological types, in the most luminous galaxies.  

We consider our sample in the colour--magnitude diagram. While the largest envelopes are found in all parts of the CMD, the envelopes with $D > 100$~kpc are almost always found on the bright end of the red sequence, with $M_V < -21$ in E/S0 galaxies. The largest envelopes, those with $D>40$~kpc, are only found in galaxies with $M_V<-20$; however 80\% of the envelopes with $D > 60$~kpc are on the bright end of the red sequence. We find that interactions can occur with equal likelihood across the CMD, even on the red sequence. Although we can observe signs of interactions in the last 1-2 Gyr, these are not necessarily playing a role in quenching of star formation. However, this question deserves more exploration, potentially in a future {\sl HERON} project.

It is noteworthy to emphasize that it is mostly the total intrinsic luminosity, and {\it not} presence on the red sequence, that determines envelope diameter. Galaxies at the faint ends $(M_V>-21)$ of the blue and red sequences have the same distribution of envelope sizes. While presence on the red sequence may have resulted from an early interaction history, the present-day absolute luminosity appears to be the critical factor that determines the size of the low surface brightness envelope.

We consider disk galaxies, and find that edge-on spirals have larger diameters than face-on spirals. We develop several lines of argument that the envelopes of disk galaxies are dominated by stars on the disk plane. We show that in the sample of \citet{Merritt16} that the outermost portions of disks are dominated by spiral structure. We also illustrate two cases, NGC\,4762 and NGC\,3628, where interactions have resulted in the disk outskirts being strongly disturbed and thickened. Appealing to the H{\sc i} imaging of \citet{Sancisi08} 
and studies of XUV disks \citep{Lemonias11}, we argue that the envelopes of disks are dominated by disk stars, not by the classical halo spheroid. The low surface brightness structures of all edge-on galaxies are dominated by their disks; there are no cases where the greatest diameter at low surface brightness arises from a classical spheroidal structure.

Future {\sl HERON} work amongst an international team of observational and theoretical collaborators will report the quantitative analysis of surface brightness profiles, discuss outer envelope morphologies, and other properties including comparisons of extended structures in multiple wavelengths. We will also report and catalogue all low surface brightness companions detected in our survey, listing luminosities, diameters, and coordinates among other details. Finally, we will upload our complete datasets and imaging to the {\sl HERON} archive at the IRSA/IPAC database.

\newpage
\section*{Acknowledgements}
The authors express thanks to Eija Laurikannen and Sebastien Comeron for valuable comments on the project. 
We would like to thank UCLA alumnus Kyra Mitchell for her valuable help creating our catalogue of inverse image grids, several of which are displayed in the Appendix with the rest included as supplemental materials (Fig.~\ref{invim1}). We acknowledge UCLA alumnus Dylan Schaul, who worked on data reduction for the early images.  We also acknowledge Nanjing University (Jiangsu Province, China) students Xu Zizheng and Weigung Cao for their help creating an image reduction pipeline which will be used in our future work.   We also thank David Gedalia for his assistance in improvements of the 0.7-m Lockwood Valley telescope, and B. Megdal for his assistance in obtaining images using his 8-inch refractor. We also acknowledge Ian Kearns-Brown for technical and IT support of the Jeanne Rich telescope.
We also thank the membership of the Polaris Observatory Association for their maintenance of the observatory infrastructure.
Aleksandr Mosenkov expresses gratitude for the grant of the Russian Foundation for Basic Researches number mol\_a 18-32-00194. \\

This research has made use of the NASA/IPAC Infrared Science Archive (IRSA; \url{http://irsa.ipac.caltech.edu/frontpage/}), and the NASA/IPAC Extragalactic Database (NED; \url{https://ned.ipac.caltech.edu/}), both of which are operated by the Jet Propulsion Laboratory, California Institute of Technology, under contract with the National Aeronautics and Space Administration.  This research has made use of the HyperLEDA database (\url{http://leda.univ-lyon1.fr/}; \citealp{2014A&A...570A..13M}).

\noindent Software: IRAF, IMSURFIT, Ellipse, DS9, Python, Veusz Graphing, Microsoft Excel. 








\appendix

\section{{\sl HERON} images}
\label{Appendix:figs}

\begin{figure*}
\centering
\includegraphics[width=\textwidth]{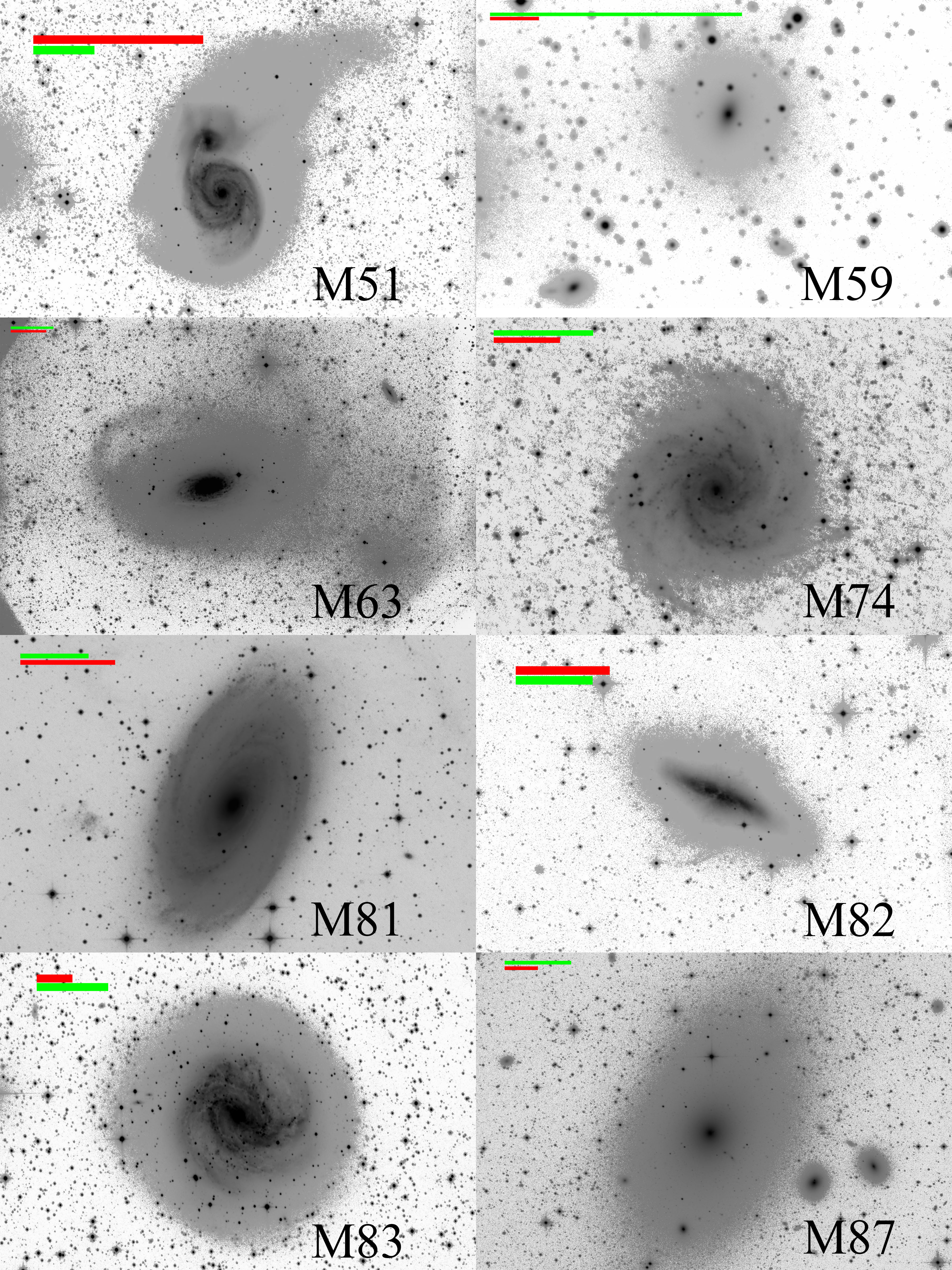}
\caption{Inverse images of some galaxies from our sample. The green scale bar is 5'; the red scale
bar is 10~kpc at the distance of the galaxy as listed in Table~\ref{tab:table2}.} \label{invim1}
\end{figure*}

\addtocounter{figure}{-1}
\begin{figure*}
\centering
\includegraphics[width=\textwidth]{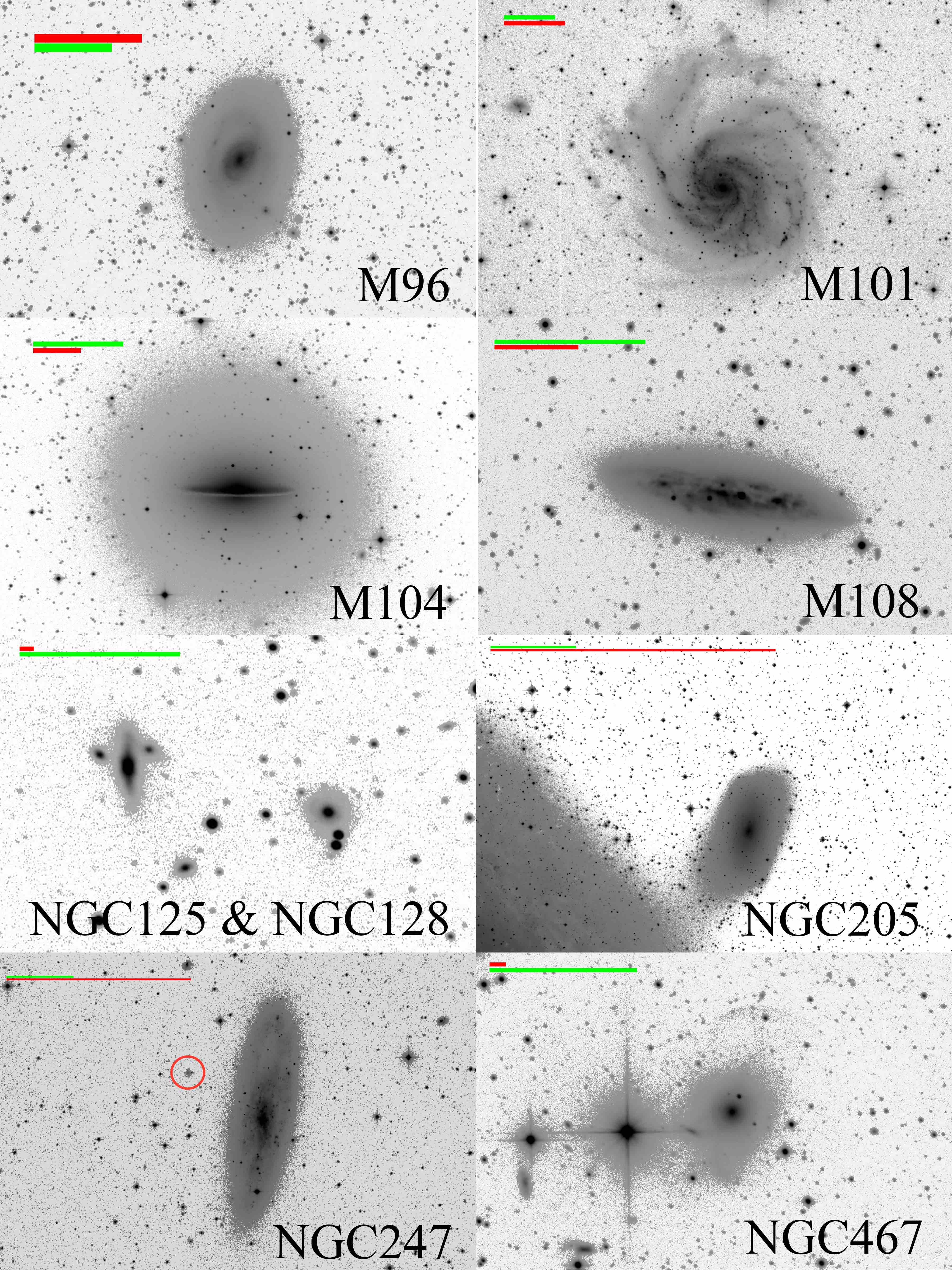}
\caption{(continued)}
\end{figure*}

\addtocounter{figure}{-1}
\begin{figure*}
\centering
\includegraphics[width=\textwidth]{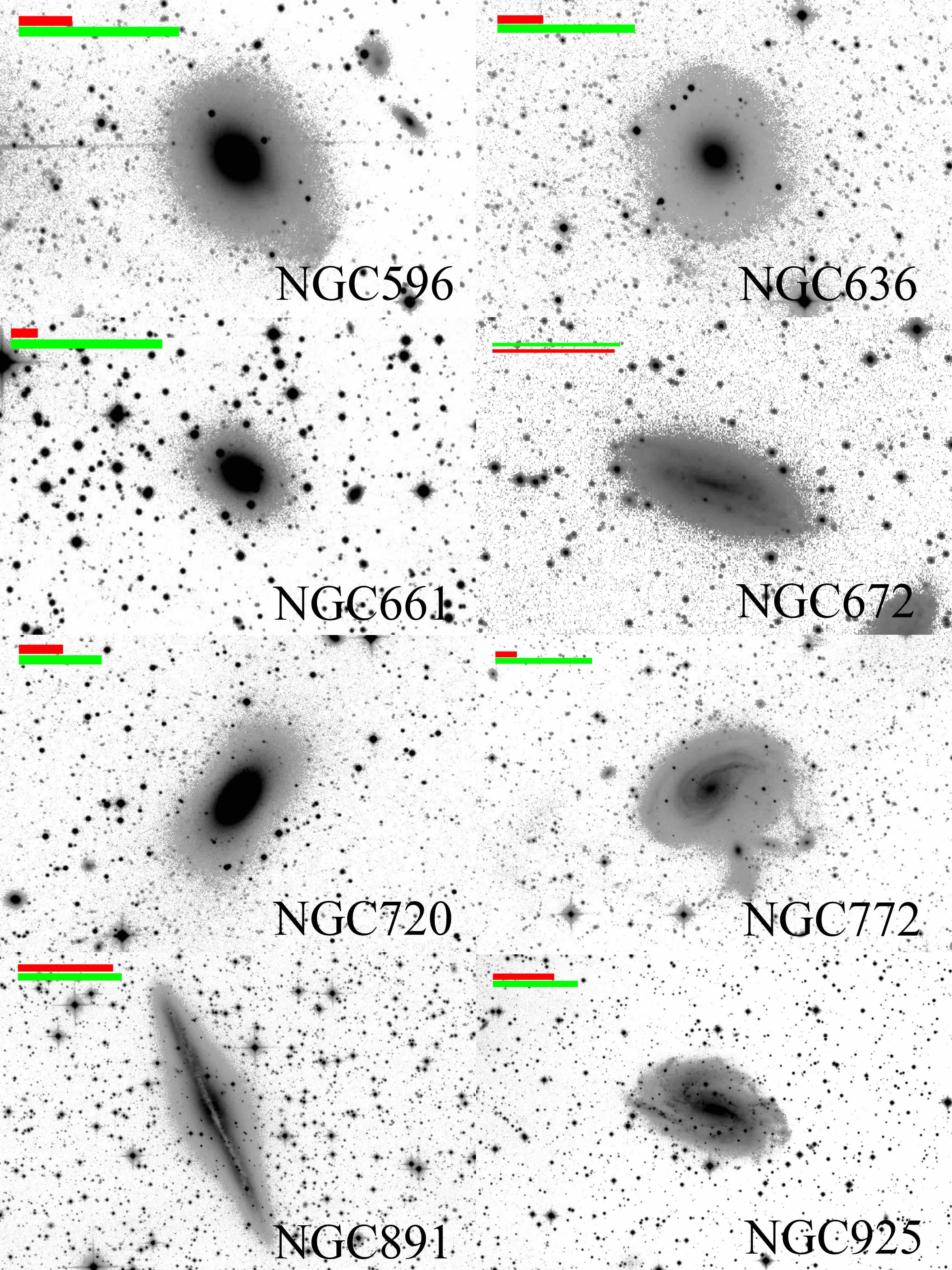}
\caption{(continued)}
\end{figure*}

\addtocounter{figure}{-1}
\begin{figure*}
\centering
\includegraphics[width=\textwidth]{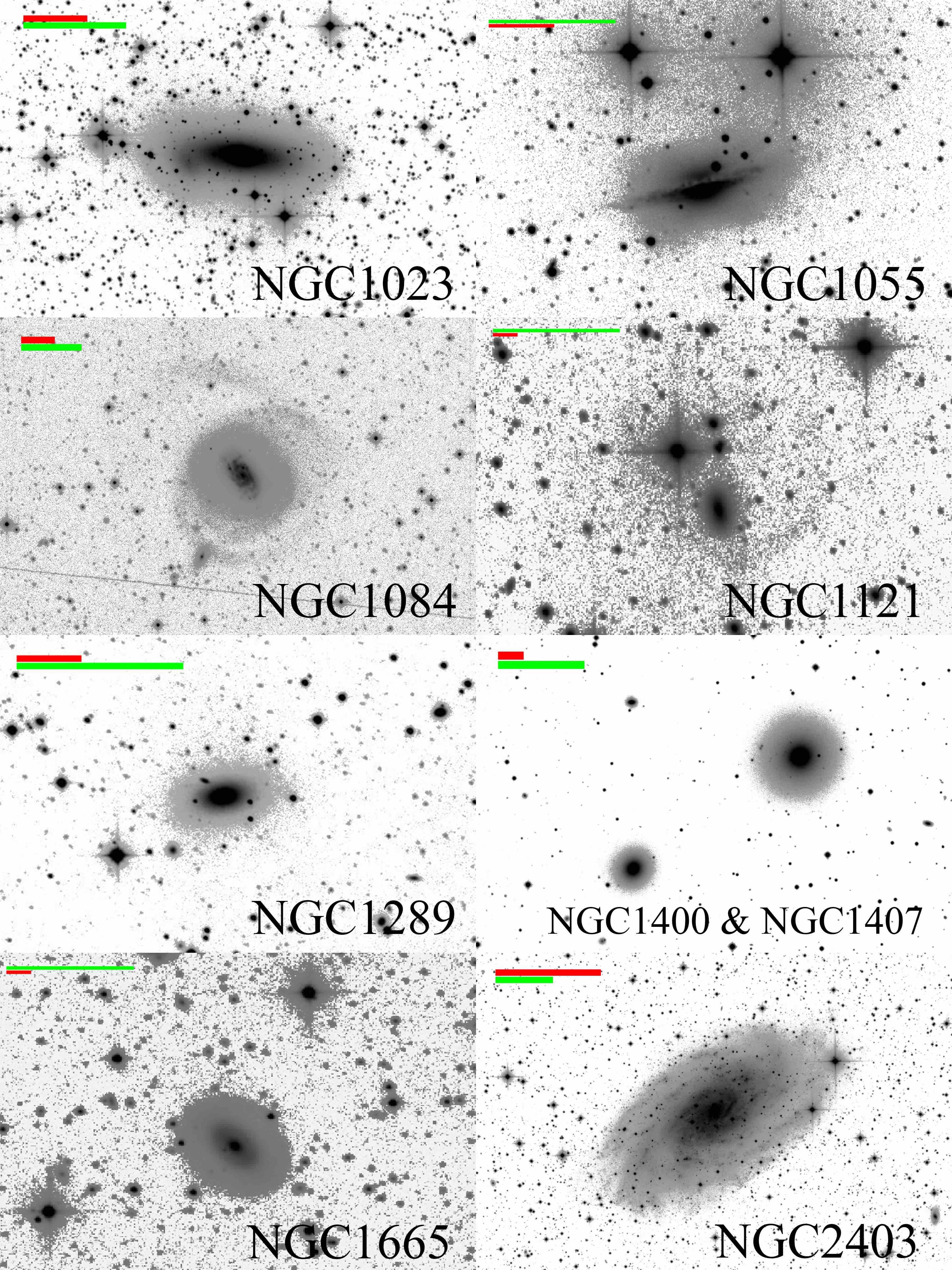}
\caption{(continued)}
\end{figure*}

\addtocounter{figure}{-1}
\begin{figure*}
\centering
\includegraphics[width=\textwidth]{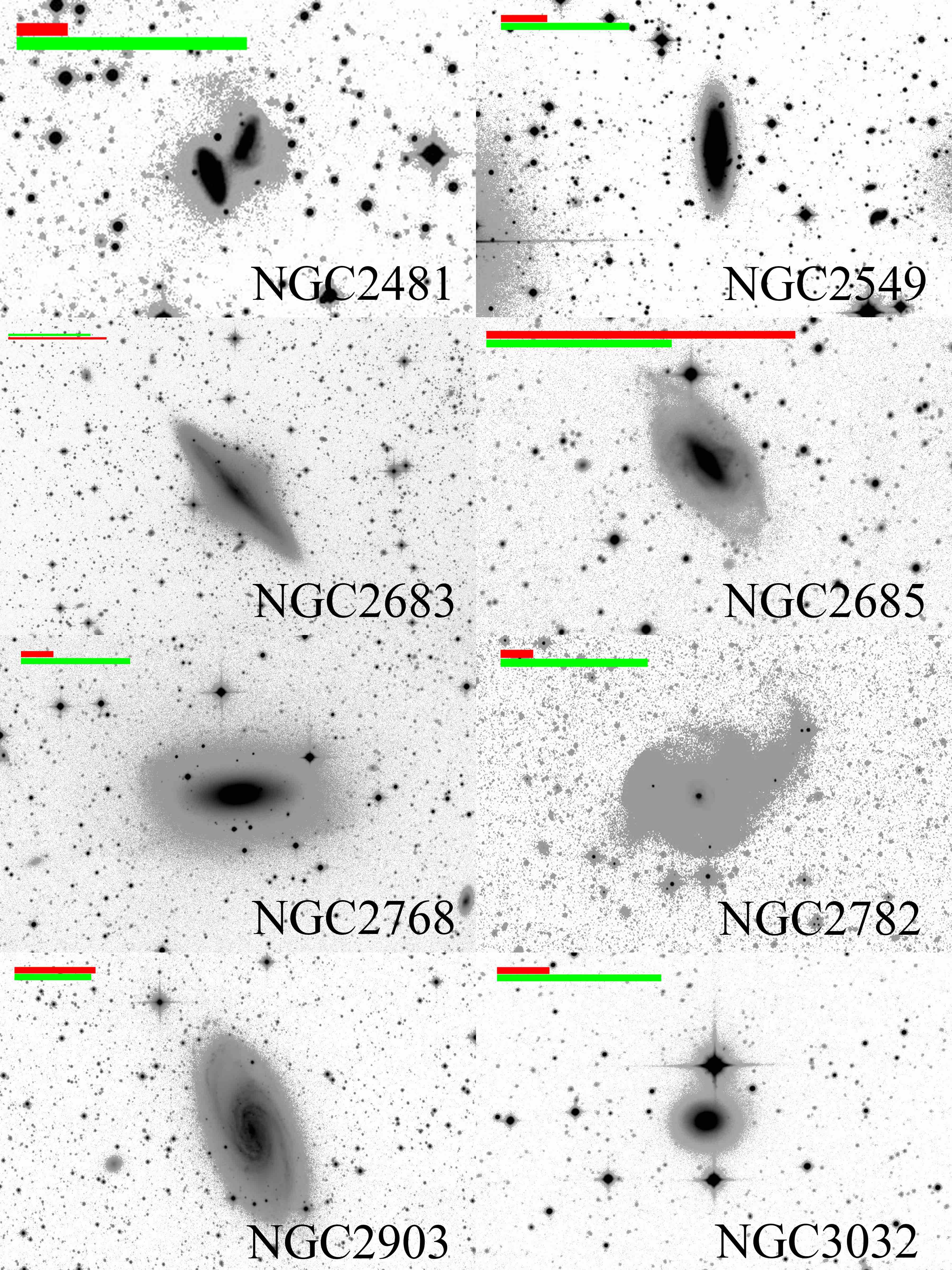}
\caption{(continued)}
\end{figure*}

\addtocounter{figure}{-1}
\begin{figure*}
\centering
\includegraphics[width=\textwidth]{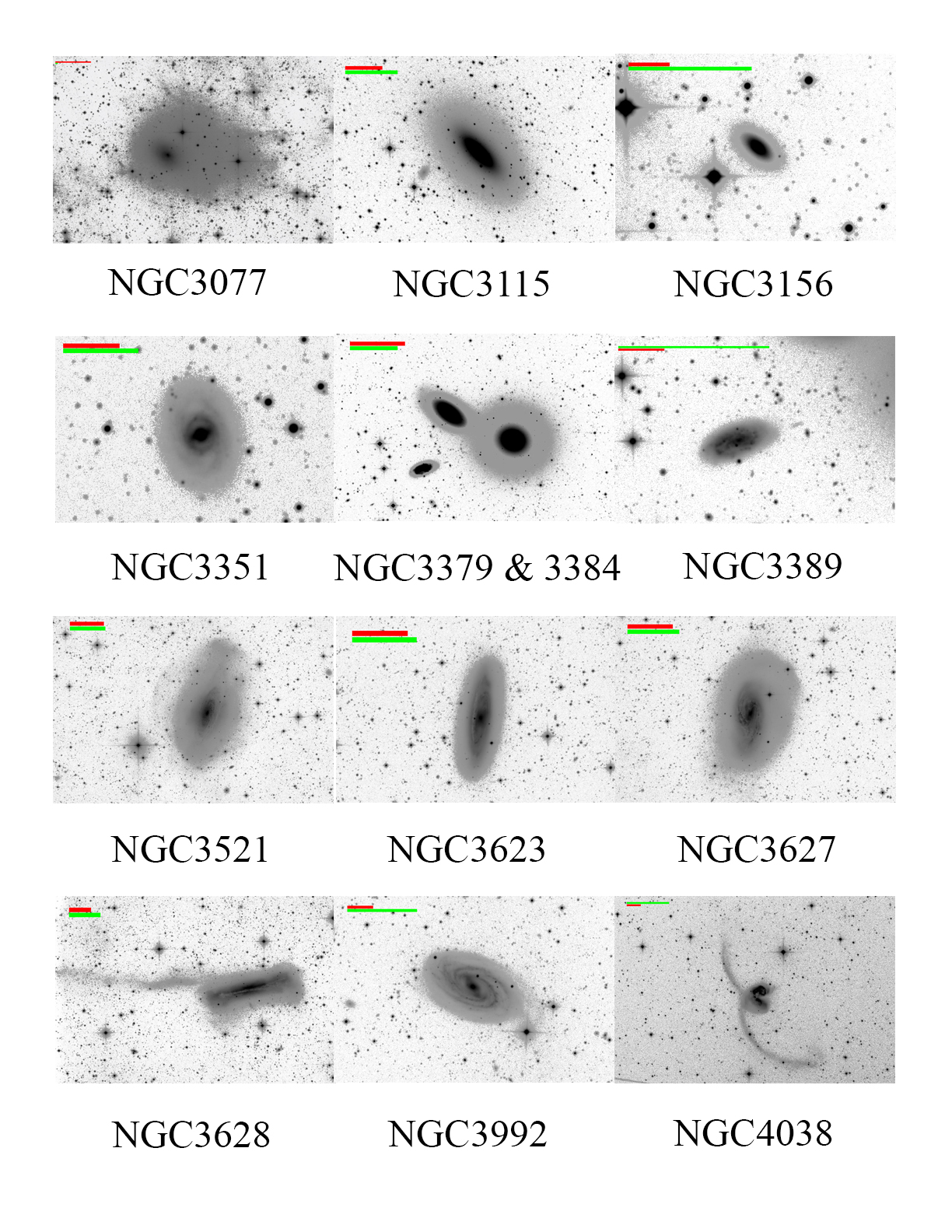}
\caption{(continued)}
\end{figure*}

\addtocounter{figure}{-1}
\begin{figure*}
\centering
\includegraphics[width=\textwidth]{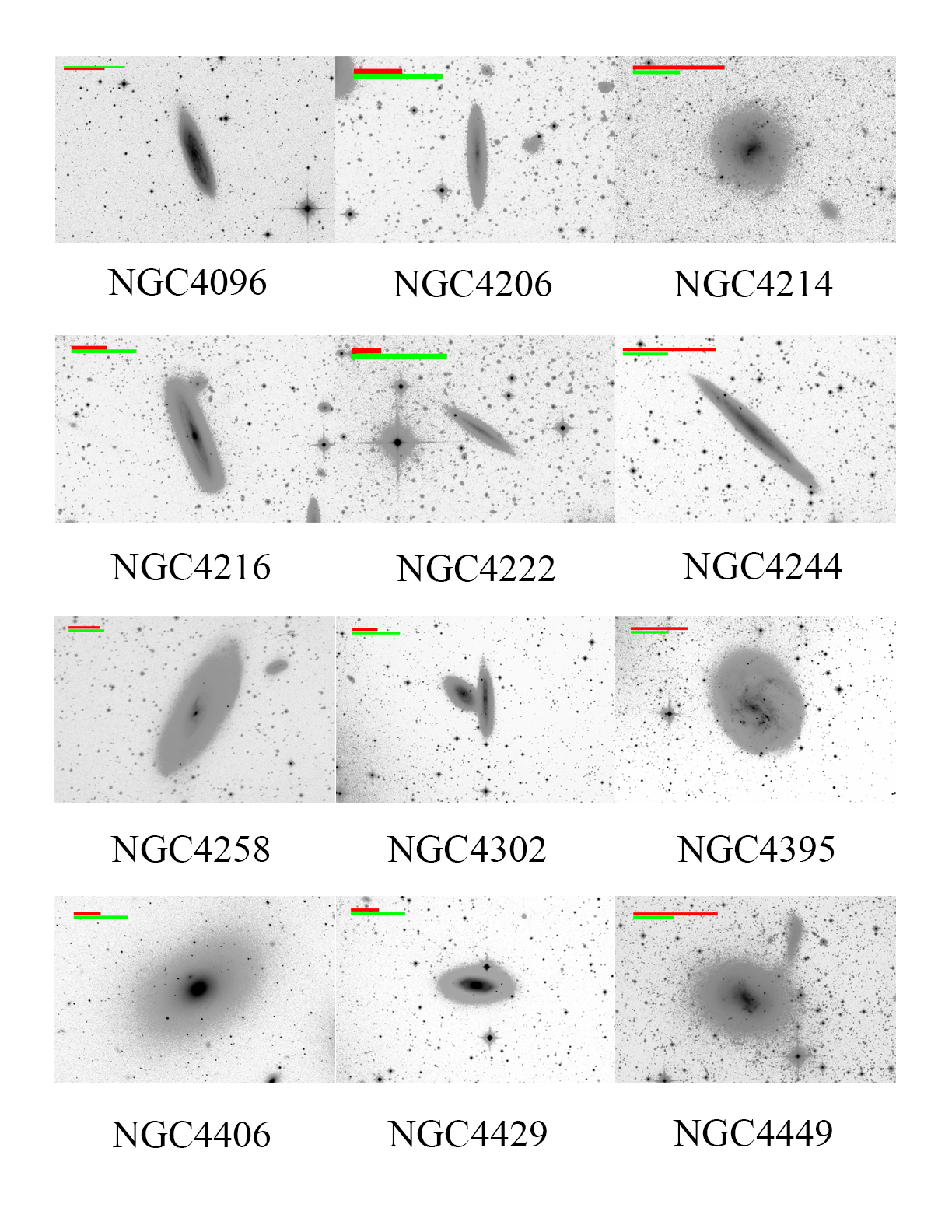}
\caption{(continued)}
\end{figure*}

\addtocounter{figure}{-1}
\begin{figure*}
\centering
\includegraphics[width=\textwidth]{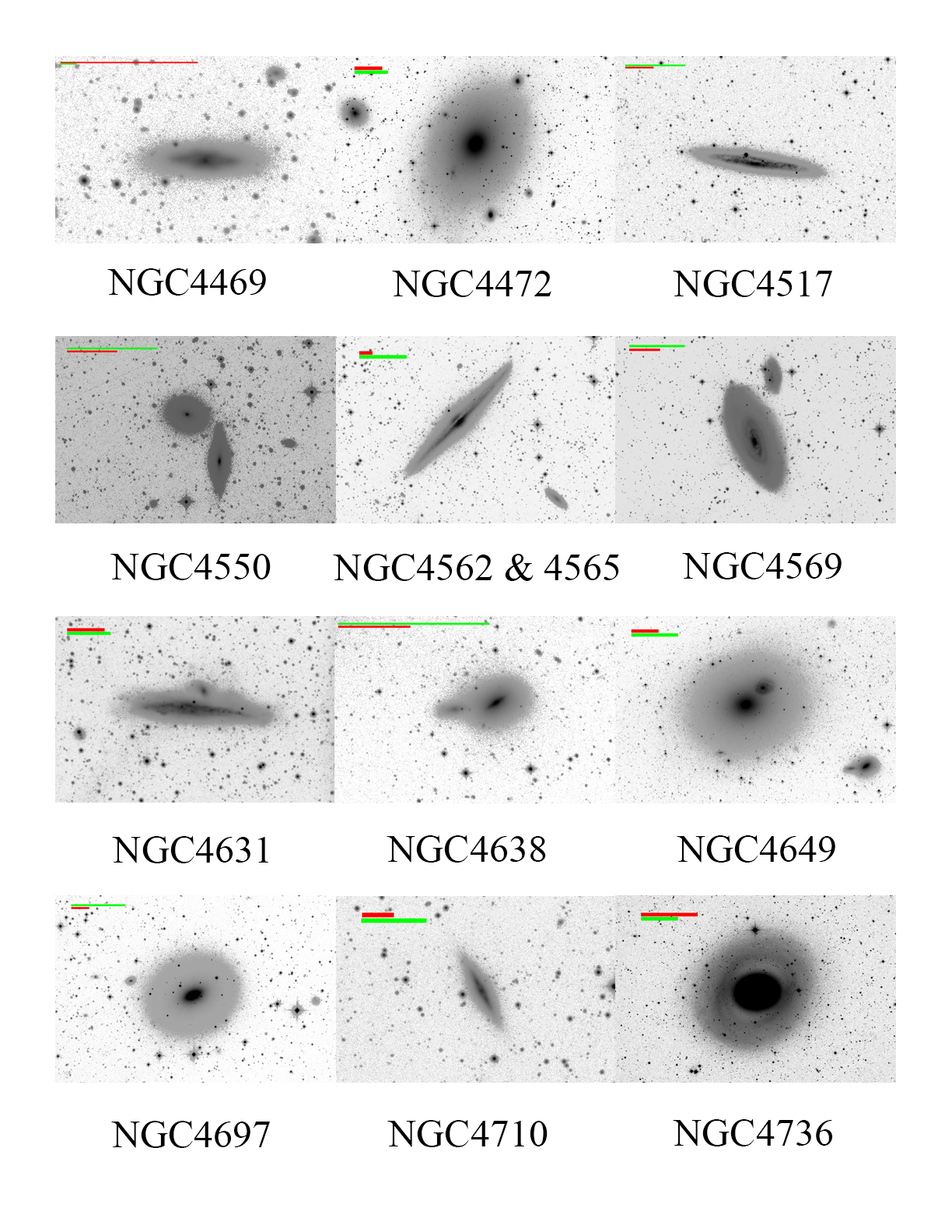}
\caption{(continued)}
\end{figure*}

\addtocounter{figure}{-1}
\begin{figure*}
\centering
\includegraphics[width=\textwidth]{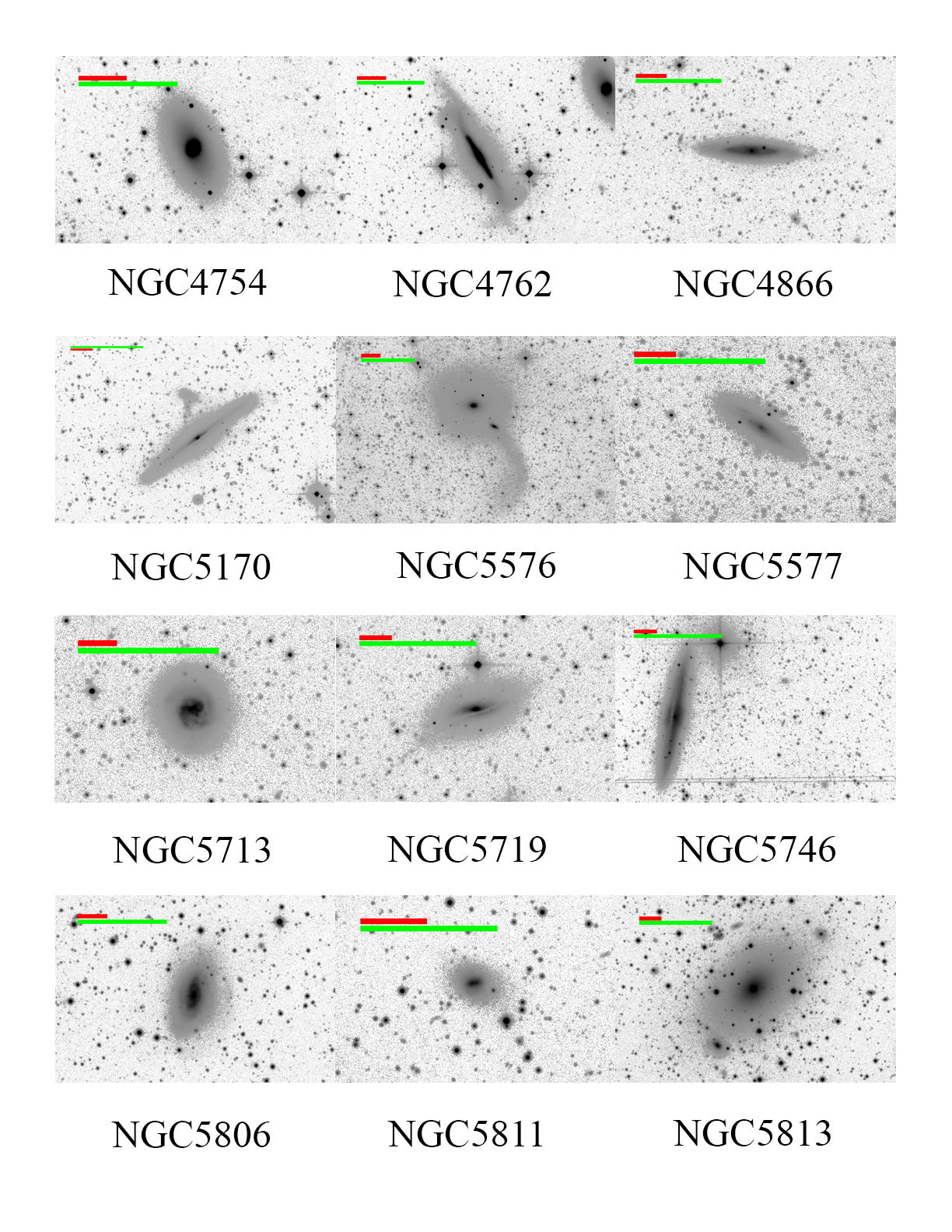}
\caption{(continued)}
\end{figure*}

\addtocounter{figure}{-1}
\begin{figure*}
\centering
\includegraphics[width=\textwidth]{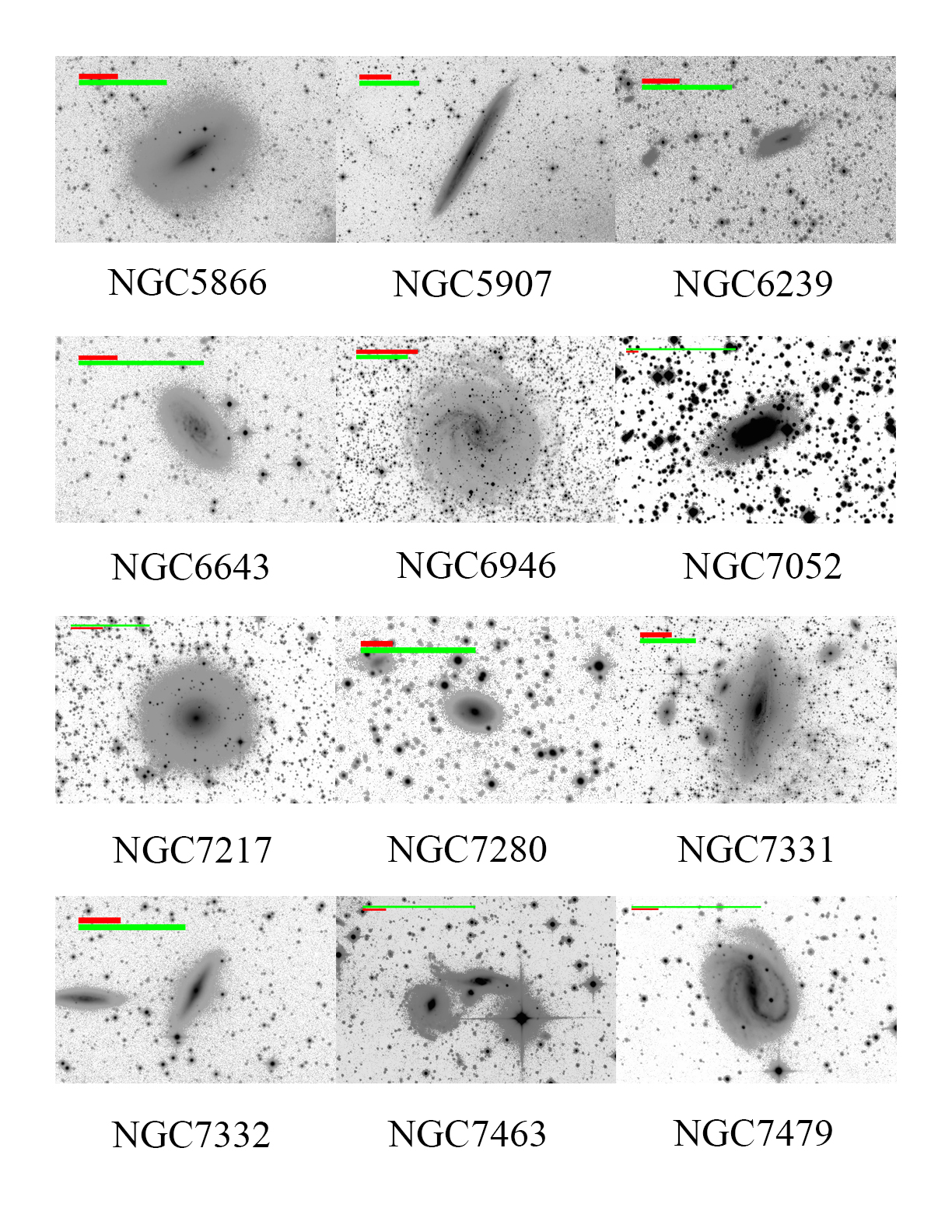}
\caption{(continued)}
\end{figure*}

\addtocounter{figure}{-1}
\begin{figure*}
\centering
\includegraphics[width=\textwidth]{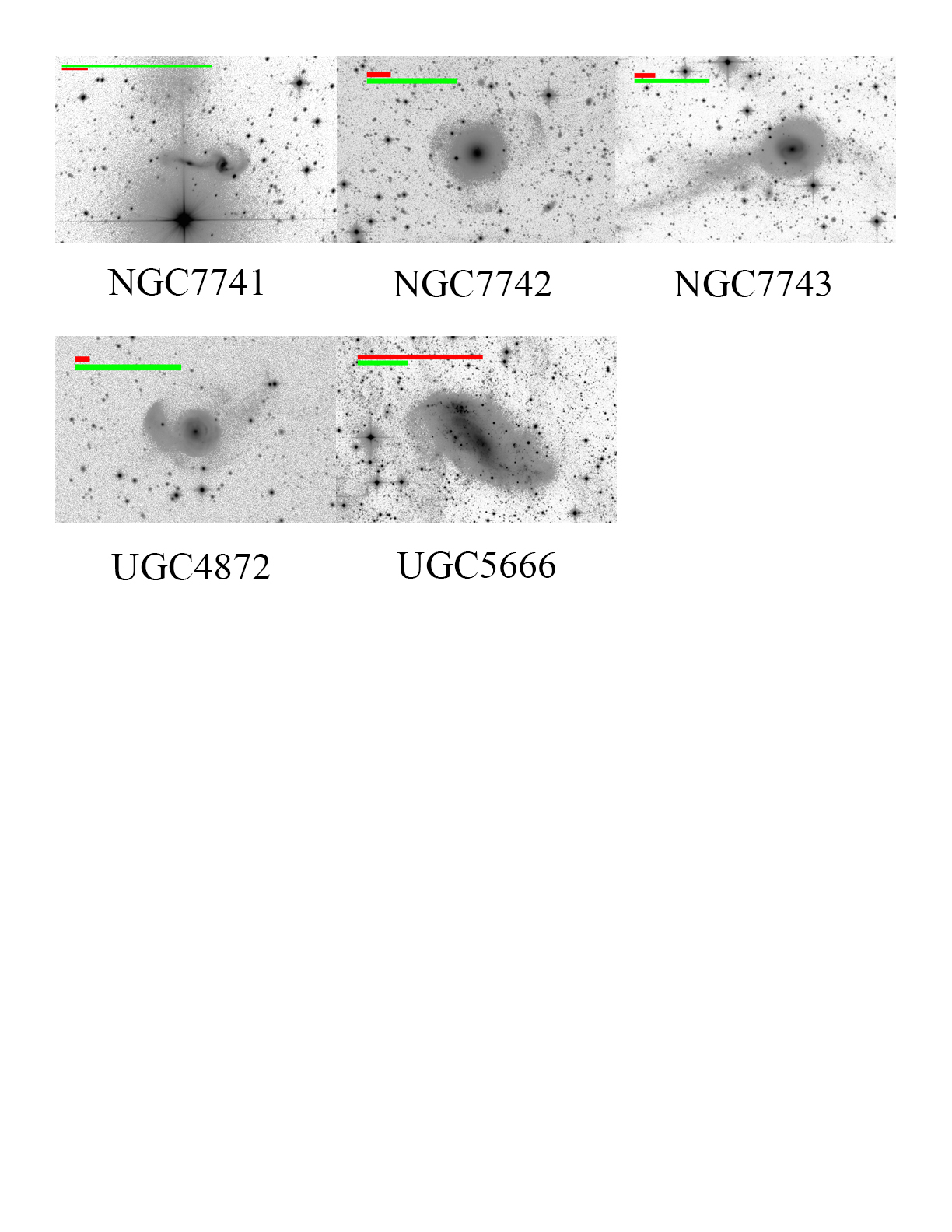}
\caption{(continued)}
\end{figure*}

\section{{\sl HERON} tables}
\label{Appendix:tabs}

\begin{table*}
	\centering
	\label{tab:table2}
    \tabcolsep=0.2cm
    \footnotesize
    \caption{Lists the coordinates (from HyperLeda), distance (see the text), radial velocity (from HyperLeda), type (from NED), apparent and absolute magnitude in the $V$ band and the colour $B-V$ (using the HyperLeda values {\it btc} and {\it bvtc}). Also, in the column Features we note if a galaxy is viewed edge-on (`E') or interacting (`I').}
\begin{tabular}{ccccccccc}
\hline
NGC & RA, Dec  & $D$ & $v$  & Type & Features & $m_V$ & $M_V$ & $B-V$ \\
    & hrs,degs & Mpc & km/s &      &     &  mag  &  mag  &  mag  \\
\hline
M49(4472) & 12h29m46s +07h59m59s & 16.63 & 978 & E2/S0 & ... & 8.32 & -22.79 & 0.93\\
M51(5194/5195) & 13h29m52s +47h11m42s & 8.58 & 460 & SA(s)bc & I & 7.81 & -21.86 & 0.55\\
M59(4621) & 12h42m02s +11h38m50s & 14.74 & 438 & E5 & ... & 9.54 & -21.31 & 0.91\\
M60(4649) & 12h43m39s +11h33m09s & 17.05 & 1107 & E2 & I & 8.74 & -22.41 & 0.94\\
M63(5055) & 13h15m49s +42h01m45s & 9.04 & 507 & SA(rs)bc & ... & 8.24 & -21.54 & 0.66\\
M65(3623) & 11h18m55s +13h05m32s & 12.87 & 801 & SAB(rs)a & ... & 8.81 & -21.74 & 0.78\\
M66(3627) & 11h20m15s +12h59m29s & 11.46 & 718 & SAB(s)b & ... & 8.47 & -21.82 & 0.63\\
M74(628) & 01h36m41s +15h47m00s & 10.14 & 658 & SA(s)c & ... & 8.91 & -21.12 & 0.49\\
M81(3031) & 09h55m33s +69h03m54s & 3.60 & -37 & SA(s)ab & ... & 6.29 & -21.49 & 0.83\\
M82(3034) & 09h55m52s +69h40m47s & 3.62 & 231 & I0 & E & 7.07 & -20.72 & 0.69\\
M83(5236) & 13h37m00s -29h51m56s & 4.90 & 508 & SAB(s)c & ... & 6.95 & -21.50 & 0.60\\
M86(4406) & 12h26m11s +12h56m44s & 17.39 & -291 & S0(3)/E3 & ... & 8.72 & -22.48 & 0.91\\
M87(4486) & 12h30m49s +12h23m25s & 16.78 & 1283 & E+ & ... & 8.62 & -22.50 & 0.93\\
M90(4569) & 12h36m49s +13h09m48s & 11.86 & -220 & SAB(rs)ab & ... & 8.91 & -21.46 & 0.61\\
M94(4736) & 12h50m53s +41h07m13s & 4.39 & 313 & (R)SA(r)ab & ... & 7.83 & -20.38 & 0.72\\
M95(3351) & 10h43m57s +11h42m13s & 9.93 & 777 & SB(r)b & ... & 9.43 & -20.55 & 0.73\\
M96(3368) & 10h46m45s +11h49m11s & 9.79 & 892 & SAB(rs)ab & ... & 8.96 & -21.00 & 0.80\\
M101(5457) & 14h03m12s +54h20m56s & 7.11 & 236 & SAB(rs)cd & ... & 7.86 & -21.40 & 0.44\\
M104(4594) & 12h39m59s -11h37m22s & 8.60 & 1087 & SA(s)a & E & 7.85 & -21.82 & 0.88\\
M105(3379) & 10h47m49s +12h34m53s & 11.32 & 918 & E1 & I & 9.19 & -21.08 & 0.93\\
M106(4258) & 12h18m57s +47h18m13s & 7.61 & 454 & SAB(s)bc & ... & 7.88 & -21.53 & 0.60\\
M108(3556) & 11h11m30s +55h40m27s & 9.83 & 697 & SB(s)cd & E & 9.41 & -20.56 & 0.57\\
M109(3992) & 11h57m35s +53h22m28s & 25.27 & 1047 & SB(rs)bc & ... & 9.40 & -22.61 & 0.71\\
M110(205) & 00h40m22s +41h41m07s & 0.80 & -241 & E5 & ... & 7.87 & -16.65 & 0.80\\
125 & 00h28m50s +02h50m20s & 63.68 & 5273 & (R)SA0+ & ... & 12.24 & -21.78 & 0.89\\
128 & 00h29m15s +02h51m50s & 44.87 & 4129 & S0 & EI & 11.58 & -21.68 & 0.90\\
247 & 00h47m08s -20h45m37s & 3.71 & 155 & SAB(s)d & ... & 8.18 & -19.67 & 0.44\\
278 & 00h52m04s +47h33m01s & 17.78 & 639 & SAB(rs)b & ... & 10.44 & -20.81 & 0.51\\
467 & 01h19m10s +03h18m02s & 58.61 & 5568 & SA(s)0 & I & 11.98 & -21.86 & 0.96\\
470 & 01h19m44s +03h24m35s & 48.75 & 2373 & SA(rs)b & I & 11.38 & -22.06 & 0.66\\
474 & 01h20m06s +03h24m55s & 27.54 & 2342 & (R')SA(s)0 & I & 11.42 & -20.78 & 0.81\\
509 & 01h23m24s +09h26m00s & 32.89 & 2273 & S0 & ... & 13.67 & -18.91 & 0.74\\
518 & 01h24m17s +09h19m51s & 38.73 & 2685 & Sa & E & 12.89 & -20.05 & 0.74\\
520 & 01h24m35s +03h47m32s & 17.86 & 2161 & Sa & I & 10.99 & -20.27 & 0.70\\
524 & 01h24m47s +09h32m19s & 32.58 & 2422 & SA(rs)0+ & ... & 10.11 & -22.45 & 0.96\\
525 & 01h24m52s +09h42m11s & 23.58 & 1624 & S0 & ... & 13.44 & -18.42 & 0.75\\
530 & 01h24m41s -01h35m13s & 71.81 & 5024 & SB0+ & E & 12.61 & -21.68 & 0.91\\
532 & 01h25m17s +09h15m50s & 34.66 & 2369 & SAb & E & 12.30 & -20.40 & 0.86\\
596 & 01h32m51s -07h01m53s & 21.55 & 1903 & E+ & ... & 10.83 & -20.83 & 0.85\\
636 & 01h39m06s -07h30m45s & 29.47 & 1854 & E3 & ... & 11.32 & -21.03 & 0.91\\
661 & 01h44m14s +28h42m21s & 37.15 & 3817 & E+ & ... & 11.99 & -20.86 & 0.88\\
672 & 01h47m53s +27h25m55s & 7.21 & 426 & SB(s)cd & ... & 9.97 & -19.32 & 0.43\\
720 & 01h53m00s -13h44m18s & 27.38 & 1717 & E5 & ... & 10.10 & -22.08 & 0.95\\
772 & 01h59m19s +19h00m27s & 25.47 & 2460 & SA(s)b & I & 9.39 & -22.64 & 0.65\\
891 & 02h22m33s +42h20m51s & 9.85 & 527 & SA(s)b & E & 9.04 & -20.93 & 0.71\\
925 & 02h27m16s +33h34m44s & 9.20 & 553 & SAB(s)d & ... & 9.38 & -20.44 & 0.45\\
1023 & 02h40m23s +39h03m47s & 9.56 & 645 & SB(rs)0- & I & 8.42 & -21.49 & 0.92\\
1055 & 02h41m45s +00h26m36s & 19.34 & 993 & SBb & E & 10.14 & -21.30 & 0.72\\
1084 & 02h45m59s -07h34m42s & 15.46 & 1422 & SA(s)c & I & 10.26 & -20.69 & 0.51\\
1289 & 03h18m49s -01h58m23s & 29.92 & 2835 & SB(rs)0 & ... & 12.34 & -20.04 & 0.80\\
1400 & 03h39m30s -18h41m17s & 26.06 & 589 & SA0- & ... & 10.79 & -21.29 & 0.90\\
1407 & 03h40m11s -18h34m49s & 28.54 & 1791 & E0 & ... & 9.46 & -22.82 & 0.95\\
2403 & 07h36m51s +65h36m08s & 3.20 & 140 & SAB(s)cd & ... & 7.78 & -19.75 & 0.38\\
2481 & 07h57m13s +23h46m03s & 31.72 & 2180 & S0/a & I & 12.32 & -20.19 & 0.67\\
2549 & 08h18m58s +57h48m11s & 12.51 & 1075 & SA(r)0 & E & 11.01 & -19.48 & 0.85\\
2683 & 08h52m41s +33h25m20s & 9.38 & 420 & SA(rs)b & E & 8.37 & -21.50 & 0.76\\
2685 & 08h55m34s +58h44m03s & 12.65 & 877 & (R)SB0+ & I & 11.16 & -19.35 & 0.76\\
2768 & 09h11m37s +60h02m14s & 22.15 & 1397 & S0 & ... & 9.71 & -22.01 & 0.92\\
2782 & 09h14m05s +40h06m49s & 16.98 & 2623 & SAB(rs)a & I & 11.50 & -19.65 & 0.61\\
2903 & 09h32m10s +21h30m05s & 9.32 & 555 & SAB(rs)bc & ... & 8.28 & -21.57 & 0.57\\
3032 & 09h52m08s +29h14m11s & 20.00 & 1540 & SAB(r)0 & ... & 12.34 & -19.16 & 0.64\\
\hline
\end{tabular}
\end{table*}

\begin{table*}
\contcaption{}
\label{tab:tab2:continued}
\begin{tabular}{ccccccccc}
\hline
NGC & RA, Dec  & $D$ & $v$  & Type & Features & $m_V$ & $M_V$ & $B-V$ \\
    & hrs,degs & Mpc & km/s &      &     &  mag  &  mag  &  mag  \\
\hline
3079 & 10h01m58s +55h40m47s & 16.52 & 1163 & SB(s)c & E & 9.45 & -21.64 & 0.53\\
3115 & 10h05m13s -07h43m06s & 9.65 & 648 & S0- & E & 9.00 & -20.92 & 0.90\\
3156 & 10h12m41s +03h07m45s & 22.15 & 1346 & S0 & ... & 12.21 & -19.51 & 0.71\\
3384 & 10h48m16s +12h37m45s & 9.42 & 563 & SB(s)0- & I & 9.90 & -19.97 & 0.88\\
3389 & 10h48m27s +12h31m59s & 19.32 & 1301 & SA(s)c & ... & 11.42 & -20.01 & 0.36\\
3521 & 11h05m48s +00h02m05s & 12.39 & 801 & SAB(rs)bc & ... & 8.55 & -21.92 & 0.71\\
3628 & 11h20m16s +13h35m22s & 10.82 & 845 & SAb & E & 8.49 & -21.68 & 0.68\\
4038 & 12h01m53s -18h52m05s & 24.49 & 1634 & SB(s)m & I & 9.82 & -22.12 & 0.57\\
4096 & 12h06m01s +47h28m42s & 11.99 & 563 & SAB(rs)c & ... & 9.53 & -20.87 & 0.50\\
4206 & 12h15m16s +13h01m26s & 18.87 & 702 & SA(s)bc & E & 11.16 & -20.22 & 0.50\\
4214 & 12h15m39s +36h19m35s & 2.98 & 292 & IAB(s)m & ... & 9.51 & -17.86 & 0.41\\
4216 & 12h15m54s +13h08m57s & 13.80 & 134 & SAB(s)b & E & 9.09 & -21.61 & 0.84\\
4222 & 12h16m22s +13h18m26s & 19.73 & 229 & Sc & E & 11.57 & -19.90 & 0.67\\
4244 & 12h17m29s +37h48m28s & 4.35 & 245 & SA(s)cd & E & 9.33 & -18.86 & 0.41\\
4302 & 12h21m42s +14h35m54s & 14.32 & 1129 & Sc & EI & 10.06 & -20.72 & 0.74\\
4395 & 12h25m48s +33h32m48s & 4.76 & 317 & SA(s)m & ... & 9.55 & -18.84 & 0.34\\
4429 & 12h27m26s +11h06m27s & 13.00 & 992 & SA(r)0+ & ... & 9.85 & -20.72 & 0.89\\
4449 & 12h28m11s +44h05m37s & 4.27 & 203 & IBm & ... & 8.66 & -19.49 & 0.33\\
4469 & 12h29m28s +08h45m00s & 16.75 & 582 & SB(s)0/a & E & 10.85 & -20.27 & 0.86\\
4517 & 12h32m45s +00h06m52s & 8.39 & 1127 & SA(s)cd & E & 9.20 & -20.42 & 0.53\\
4550 & 12h35m30s +12h13m15s & 15.28 & 410 & SB0 & EI & 11.52 & -19.40 & 0.78\\
4551 & 12h35m37s +12h15m50s & 16.06 & 1200 & E2 & I & 11.79 & -19.24 & 0.90\\
4565 & 12h36m20s +25h59m15s & 12.07 & 1226 & SA(s)b & E & 8.30 & -22.11 & 0.68\\
4631 & 12h42m07s +32h32m33s & 7.35 & 615 & SB(s)d & EI & 7.61 & -21.72 & 0.39\\
4638 & 12h42m47s +11h26m32s & 17.21 & 1163 & S0- & EI & 11.06 & -20.12 & 0.89\\
4697 & 12h48m35s -05h48m02s & 12.23 & 1240 & E6 & ... & 9.25 & -21.19 & 0.87\\
4710 & 12h49m38s +15h09m53s & 16.83 & 1103 & SA(r)0+ & E & 10.71 & -20.42 & 0.77\\
4754 & 12h52m17s +11h18m50s & 16.02 & 1317 & SB(r)0- & ... & 10.41 & -20.61 & 0.86\\
4762 & 12h52m55s +11h13m51s & 10.81 & 973 & SB(r)0 & E & 10.14 & -20.03 & 0.79\\
4866 & 12h59m27s +14h10m15s & 29.65 & 1984 & SB(rs)bc & E & 10.93 & -21.43 & 0.74\\
5170 & 13h29m48s -17h57m59s & 29.30 & 1502 & SA(s)c & E & 9.85 & -22.48 & 0.72\\
5576 & 14h21m03s +03h16m15s & 25.20 & 1422 & E3 & I & 10.80 & -21.20 & 0.85\\
5577 & 14h21m13s +03h26m09s & 19.23 & 1489 & SA(rs)bc & ... & 11.66 & -19.76 & 0.75\\
5713 & 14h40m11s +00h17m20s & 20.32 & 1948 & SAB(rs)bc & ... & 10.64 & -20.90 & 0.56\\
5719 & 14h40m56s +00h19m05s & 26.47 & 1726 & SAB(s)ab & ... & 11.78 & -20.32 & 0.95\\
5746 & 14h44m55s +01h57m17s & 27.04 & 1711 & SAB(rs)b & E & 9.38 & -22.78 & 0.77\\
5806 & 15h00m00s +01h53m28s & 20.69 & 1348 & SAB(s)b & ... & 11.16 & -20.42 & 0.60\\
5811 & 15h00m27s +01h37m24s & 24.17 & 1523 & SB(s)m & ... & 13.58 & -18.34 & 0.74\\
5813 & 15h01m11s +01h42m07s & 31.87 & 1955 & E1-2 & ... & 10.36 & -22.16 & 0.92\\
5814 & 15h01m21s +01h38m13s & 156.31 & 10526 & (R')Sab & ... & 13.21 & -22.77 & 0.79\\
5866 & 15h06m29s +55h45m47s & 14.42 & 674 & S0\_3 & E & 9.89 & -20.90 & 0.79\\
5907 & 15h15m53s +56h19m44s & 17.23 & 667 & SA(s)c & E & 9.10 & -22.08 & 0.62\\
6239 & 16h50m04s +42h44m23s & 24.30 & 926 & SB(s)b & I & 11.91 & -20.01 & 0.41\\
6340 & 17h10m24s +72h18m15s & 19.41 & 1215 & SA(s)0/a & ... & 10.95 & -20.49 & 0.80\\
6643 & 18h19m46s +74h34m06s & 20.25 & 1484 & SA(rs)c & ... & 10.48 & -21.05 & 0.52\\
6946 & 20h34m52s +60h09m12s & 6.72 & 45 & SAB(rs)cd & ... & 8.02 & -21.12 & 0.45\\
7052 & 21h18m33s +26h26m47s & 51.29 & 4700 & E & ... & 11.72 & -21.83 & 1.71\\
7217 & 22h07m52s +31h21m33s & 18.28 & 951 & (R)SA(r)ab & ... & 9.80 & -21.51 & 0.80\\
7280 & 22h26m27s +16h08m53s & 21.48 & 1862 & SAB(r)0+ & ... & 12.03 & -19.64 & 0.83\\
7331 & 22h37m04s +34h24m56s & 14.53 & 815 & SA(s)b & ... & 8.53 & -22.28 & 0.71\\
7332 & 22h37m24s +23h47m53s & 22.77 & 1180 & S0 & E & 11.00 & -20.78 & 0.81\\
7463 & 23h01m52s +15h58m54s & 22.82 & 2262 & SABb & I & 12.33 & -19.46 & 0.39\\
7465 & 23h02m00s +15h57m53s & 27.29 & 1948 & (R')SB(s)0 & I & 12.40 & -19.78 & 0.64\\
7479 & 23h04m56s +12h19m22s & 30.20 & 2380 & SB(s)c & ... & 10.48 & -21.92 & 0.61\\
7742 & 23h44m15s +10h46m01s & 21.88 & 1659 & SA(r)b & ... & 11.40 & -20.30 & 0.65\\
7743 & 23h44m21s +09h56m02s & 18.69 & 1684 & (R)SB(s)0+ & ... & 11.31 & -20.05 & 0.81\\
UGC4872 & 09h15m01s +40h02m11s & 112.20 & 8211 & SBb & E & 13.73 & -21.52 & 0.74\\
UGC5666 & 10h28m23s +68h24m43s & 4.00 & 46 & SAB(s)m & I & 9.69 & -18.32 & 0.33\\
\hline
\end{tabular}
\end{table*}

\begin{table}
	\label{tab:table3}
    \caption{Lists exposure dates, exposure lengths, and the limiting surface brightness level which we reach in our observations.}
\begin{tabular}{cccc}
\hline
NGC & Exp.date  & Exp.length & SBmin \\
    & mm/dd/yy  &   sec x n  &  mag/arcsec$^{2}$ \\
\hline
M49(4472) & 13/03/13 & 300$\times$11 & 29.3\\
M51(5194/5195) & 21/04/12 & 300$\times$3 & 29.9\\
M59(4621) & 24/05/12 & 300$\times$10 & 29.1\\
M60(4649) & 13/03/13 & 300$\times$12 & 28.6\\
M63(5055) & 29/01/11 & 300$\times$16 & 29.5\\
M65(3623) & 12/02/13 & 300$\times$12 & 29.1\\
M66(3627) & 13/03/13 & 300$\times$12 & 29.6\\
M74(628) & 22/10/11 & 300$\times$31 & 29.7\\
M81(3031) & 04/12/11 & 300$\times$6 & 29.2\\
M82(3034) & 18/10/12 & 300$\times$4 & 28.7\\
M83(5236) & 20/05/12 & 300$\times$12 & 28.2\\
M86(4406) & 09/06/13 & 300$\times$5 & 28.1\\
M87(4486) & 13/03/13 & 300$\times$11 & 29.0\\
M90(4569) & 22/04/14 & 300$\times$17 & 28.9\\
M94(4736) & 13/03/13 & 300$\times$7 & 28.9\\
M95(3351) & 08/02/11 & 300$\times$5 & 28.6\\
M96(3368) & 16/01/13 & 300$\times$10 & 29.9\\
M101(5457) & 11/06/13 & 300$\times$8 & 29.7\\
M104(4594) & 07/02/13 & 300$\times$14 & 29.2\\
M105(3379) & 07/02/13 & 300$\times$12 & 29.1\\
M106(4258) & 03/05/11 & 300$\times$9 & 27.0\\
M108(3556) & 27/05/14 & 300$\times$6 & 28.6\\
M109(3992) & 09/06/13 & 300$\times$11 & 28.8\\
M110(205) & 20/10/11 & 300$\times$8 & 28.4\\
125 & 22/10/11 & 300$\times$3 & 28.0\\
128 & 22/10/11 & 300$\times$9 & 28.1\\
247 & 08/10/12 & 300$\times$9 & 28.9\\
278 & 07/10/16 & 100$\times$6 & 28.4\\
467 & 16/10/12 & 300$\times$12 & 28.0\\
470 & 16/10/12 & 300$\times$12 & 28.3\\
474 & 16/10/12 & 300$\times$12 & 28.9\\
509 & 19/10/12 & 300$\times$11 & 28.6\\
518 & 30/11/11 & 300$\times$4 & 28.5\\
520 & 27/09/16 & 300$\times$10 & 28.4\\
524 & 30/11/11 & 300$\times$4 & 28.3\\
525 & 30/11/11 & 300$\times$4 & 27.6\\
530 & 18/10/12 & 300$\times$11 & 28.5\\
532 & 30/11/11 & 300$\times$4 & 28.2\\
596 & 11/12/12 & 300$\times$11 & 28.9\\
636 & 11/12/12 & 300$\times$10 & 29.6\\
661 & 19/10/12 & 300$\times$12 & 29.1\\
672 & 11/12/12 & 300$\times$13 & 29.2\\
720 & 27/10/11 & 300$\times$13 & 29.9\\
772 & 18/10/12 & 300$\times$12 & 29.0\\
891 & 04/09/11 & 300$\times$11 & 28.2\\
925 & 07/10/16 & 300$\times$12 & 28.8\\
1023 & 18/10/12 & 300$\times$6 & 29.4\\
1055 & 20/12/12 & 300$\times$10 & 28.3\\
1084 & 14/09/12 & 600$\times$3 & 28.6\\
1289 & 20/11/12 & 300$\times$9 & 28.7\\
1400 & 10/10/16 & 300$\times$12 & 28.4\\
1407 & 10/10/16 & 300$\times$12 & 28.2\\
2403 & 11/12/12 & 300$\times$13 & 30.1\\
2481 & 16/01/13 & 300$\times$12 & 28.4\\
2549 & 16/01/13 & 300$\times$12 & 28.6\\
2683 & 01/01/14 & 300$\times$12 & 29.3\\
2685 & 30/11/11 & 300$\times$7 & 28.3\\
2768 & 30/11/11 & 300$\times$6 & 28.9\\
2782 & 20/11/12 & 300$\times$7 & 28.2\\
2903 & 20/12/12 & 300$\times$13 & 29.4\\
3032 & 30/11/11 & 300$\times$10 & 28.5\\
\hline
\end{tabular}
\end{table}

\begin{table}
\contcaption{}
\label{tab:table3:continued}
\begin{tabular}{cccc}
\hline
NGC & Exp.date  & Exp.length & SBmin \\
    & mm/dd/yy  &   sec x n  &  mag/arcsec$^{2}$ \\
\hline
3079 & 29/03/14 & 300$\times$196 & 28.8\\
3115 & 20/05/12 & 300$\times$7 & 29.4\\
3156 & 04/12/11 & 300$\times$11 & 28.1\\
3384 & 03/05/11 & 300$\times$18 & 28.9\\
3389 & 03/05/11 & 300$\times$18 & 28.6\\
3521 & 07/02/13 & 300$\times$12 & 29.7\\
3628 & 11/12/12 & 300$\times$11 & 29.6\\
4038 & 24/02/14 & 300$\times$49 & 28.5\\
4096 & 29/03/14 & 300$\times$13 & 29.2\\
4206 & 22/04/12 & 300$\times$11 & 28.5\\
4214 & 22/03/12 & 300$\times$18 & 29.1\\
4216 & 22/04/12 & 300$\times$11 & 29.1\\
4222 & 22/04/12 & 300$\times$11 & 28.6\\
4244 & 14/04/13 & 300$\times$12 & 28.9\\
4302 & 14/04/15 & 300$\times$5 & 28.7\\
4395 & 06/06/15 & 300$\times$17 & 29.3\\
4429 & 13/04/15 & 300$\times$3 & 28.8\\
4449 & 31/05/11 & 300$\times$35 & 28.9\\
4469 & 11/04/13 & 300$\times$14 & 28.4\\
4517 & 29/05/14 & 300$\times$16 & 28.5\\
4550 & 03/06/14 & 300$\times$15 & 28.7\\
4551 & 03/06/14 & 300$\times$15 & 28.7\\
4565 & 14/05/12 & 300$\times$18 & 30.0\\
4631 & 14/04/13 & 300$\times$12 & 29.8\\
4638 & 24/05/12 & 300$\times$10 & 29.1\\
4697 & 22/02/15 & 300$\times$2 & 29.1\\
4710 & 11/04/13 & 300$\times$15 & 28.6\\
4754 & 18/05/12 & 300$\times$7 & 28.1\\
4762 & 18/05/12 & 300$\times$7 & 28.9\\
4866 & 12/02/13 & 300$\times$12 & 29.2\\
5170 & 25/02/14 & 300$\times$22 & 29.5\\
5576 & 20/04/12 & 300$\times$21 & 29.4\\
5577 & 20/04/12 & 300$\times$21 & 28.2\\
5713 & 03/06/11 & 300$\times$10 & 28.5\\
5719 & 03/06/11 & 300$\times$10 & 28.3\\
5746 & 20/05/12 & 300$\times$9 & 28.9\\
5806 & 11/04/13 & 300$\times$10 & 28.5\\
5811 & 11/04/13 & 300$\times$10 & 28.2\\
5813 & 11/04/13 & 300$\times$10 & 28.4\\
5814 & 11/04/13 & 300$\times$10 & 28.4\\
5866 & 20/05/12 & 300$\times$7 & 29.0\\
5907 & 09/06/13 & 300$\times$13 & 29.0\\
6239 & 08/10/12 & 300$\times$4 & 28.4\\
6340 & 11/06/13 & 300$\times$9 & 28.1\\
6643 & 20/05/12 & 300$\times$5 & 28.0\\
6946 & 21/09/11 & 300$\times$14 & 28.4\\
7052 & 06/09/16 & 300$\times$6 & 28.5\\
7217 & 29/08/16 & 300$\times$12 & 28.6\\
7280 & 19/10/12 & 300$\times$8 & 28.6\\
7331 & 21/08/12 & 600$\times$8 & 28.4\\
7332 & 11/06/13 & 300$\times$8 & 28.1\\
7463 & 14/09/12 & 600$\times$6 & 28.1\\
7465 & 14/09/12 & 600$\times$6 & 27.7\\
7479 & 17/09/12 & 600$\times$6 & 28.8\\
7742 & 14/09/12 & 600$\times$7 & 29.2\\
7743 & 14/09/12 & 600$\times$7 & 28.3\\
UGC4872 & 16/10/12 & 300$\times$9 & 26.9\\
UGC5666 & 16/01/13 & 300$\times$12 & 28.7\\
\hline
\end{tabular}
\end{table}

\begin{table}
 \begin{minipage}{90mm}
	\label{tab:table4}
    \caption{Lists our envelope diameter measurements (at the 28~mag/arcsec$^{2}$ isophote) for each galaxy in arc-minutes and kpc, along with the envelope shape for each envelope.}
\begin{tabular}{cccc}
\hline
NGC & Diameter  & Diameter & Envelope shape \\
    &  arcmin   &    kpc   &                \\
\hline
M49(4472) & $41.5\pm1.7$ & $200.8\pm8.4$ & Round\\
M51(5194/5195) & $26.8\pm0.9$ & $66.8\pm2.1$ & Disturbed\\
M59(4621) & $17.0\pm0.9$ & $72.9\pm3.9$ & Round\\
M60(4649) & $14.6\pm0.5$ & $72.6\pm2.5$ & Round\\
M63(5055) & $38.1\pm1.2$ & $100.2\pm3.0$ & Oval\\
M65(3623) & $12.0\pm0.2$ & $44.9\pm0.9$ & Slightly boxy\\
M66(3627) & $22.4\pm0.5$ & $74.8\pm1.6$ & Oval\\
M74(628) & $13.9\pm0.1$ & $40.9\pm0.3$ & Round\\
M81(3031) & $33.6\pm0.8$ & $35.2\pm0.8$ & Oval\\
M82(3034) & $41.3\pm1.6$ & $43.5\pm1.7$ & Slightly oval\\
M83(5236) & $22.9\pm0.5$ & $32.7\pm0.7$ & Round\\
M86(4406) & $25.8\pm1.1$ & $130.4\pm5.5$ & Round\\
M87(4486) & $41.8\pm2.4$ & $204.0\pm11.5$ & Oval\\
M90(4569) & $16.7\pm0.7$ & $57.7\pm2.4$ & Oval\\
M94(4736) & $23.5\pm0.3$ & $30.0\pm0.4$ & Round\\
M95(3351) & $11.7\pm0.6$ & $33.8\pm1.7$ & Oval\\
M96(3368) & $14.2\pm0.1$ & $40.4\pm0.2$ & Oval\\
M101(5457) & $32.7\pm0.6$ & $67.6\pm1.2$ & Round\\
M104(4594) & $30.8\pm1.2$ & $77.0\pm3.0$ & Round\\
M105(3379) & $19.5\pm0.7$ & $64.1\pm2.4$ & Round\\
M106(4258) & --- & --- & Slightly diamond\\
M108(3556) & $22.3\pm0.5$ & $63.6\pm1.3$ & Oval\\
M109(3992) & $10.7\pm0.4$ & $79.0\pm2.6$ & Oval\\
M110(205) & $36.0\pm2.3$ & $8.4\pm0.5$ & Slight parallelogram\\
125 & $3.6\pm0.3$ & $66.3\pm6.2$ & Circular\\
128 & $7.6\pm0.7$ & $99.6\pm9.3$ & Oval\\
247 & $28.3\pm0.7$ & $30.5\pm0.8$ & Disky\\
278 & $7.0\pm0.3$ & $36.1\pm1.4$ & Circular\\
467 & $6.8\pm0.2$ & $116.3\pm3.1$ & Circular disturbed\\
470 & $4.2\pm0.2$ & $59.3\pm2.9$ & Oval\\
474 & $7.9\pm0.6$ & $62.9\pm5.0$ & Circular disturbed\\
509 & $2.3\pm0.1$ & $21.8\pm1.2$ & Oval\\
518 & $3.4\pm0.2$ & $38.6\pm1.8$ & Oval\\
520 & $6.0\pm0.2$ & $31.0\pm0.8$ & Disturbed\\
524 & $13.7\pm0.4$ & $129.6\pm3.7$ & Circular\\
525 & --- & --- & Diamond\\
530 & $2.0\pm0.1$ & $41.8\pm2.4$ & Oval\\
532 & $8.0\pm0.4$ & $80.2\pm4.0$ & Oval\\
596 & $15.7\pm0.7$ & $98.2\pm4.1$ & Circular\\
636 & $9.2\pm0.4$ & $79.1\pm3.3$ & Circular\\
661 & $5.9\pm0.3$ & $64.0\pm3.2$ & Round\\
672 & $10.2\pm0.5$ & $21.4\pm1.0$ & Oval\\
720 & $19.7\pm0.9$ & $156.8\pm6.8$ & Boxy\\
772 & $14.8\pm0.7$ & $109.9\pm5.2$ & Oval disturbed\\
891 & $13.1\pm0.2$ & $37.4\pm0.6$ & Diamond\\
925 & $12.3\pm0.3$ & $32.8\pm0.8$ & Oval\\
1023 & $18.0\pm0.5$ & $50.1\pm1.4$ & Boxy hexagon\\
1055 & $20.2\pm0.9$ & $113.4\pm5.0$ & Boxy hexagon\\
1084 & $12.1\pm0.7$ & $54.3\pm3.3$ & Round\\
1289 & $6.4\pm0.5$ & $56.0\pm4.4$ & Round\\
1400 & $10.1\pm0.7$ & $76.9\pm5.0$ & Round\\
1407 & $17.6\pm1.0$ & $146.0\pm8.6$ & Round\\
2403 & $26.4\pm0.5$ & $24.6\pm0.5$ & Oval\\
2481 & $3.2\pm0.1$ & $29.1\pm1.1$ & Oval\\
2549 & $6.2\pm0.1$ & $22.5\pm0.3$ & Oval\\
2683 & $11.4\pm0.4$ & $31.2\pm1.0$ & Diamond\\
2685 & $7.1\pm0.3$ & $26.1\pm1.1$ & Slightly diamond\\
2768 & $14.9\pm0.5$ & $95.8\pm3.5$ & Round\\
2782 & $7.3\pm0.2$ & $36.3\pm1.2$ & Round disturbed\\
2903 & $24.1\pm0.6$ & $65.3\pm1.7$ & Disky\\
3032 & $4.3\pm0.2$ & $24.8\pm1.0$ & Round\\
\hline
\end{tabular}
\end{minipage}
\end{table}

\begin{table}
 \begin{minipage}{90mm}
\contcaption{}
\label{tab:table4:continued}
\begin{tabular}{cccc}
\hline
NGC & Diameter  & Diameter & Envelope shape \\
    &  arcmin   &    kpc   &                \\
\hline
3079 & $11.2\pm0.5$ & $53.8\pm2.2$ & Oval\\
3115 & $24.8\pm1.0$ & $69.7\pm2.9$ & Oval\\
3156 & $5.0\pm0.3$ & $32.2\pm1.8$ & Oval\\
3384 & $12.3\pm0.1$ & $33.8\pm0.2$ & Oval\\
3389 & $4.7\pm0.1$ & $26.6\pm0.8$ & Oval\\
3521 & $22.4\pm0.4$ & $80.6\pm1.4$ & Oval/diamond\\
3628 & $20.7\pm0.2$ & $65.1\pm0.5$ & Boxy\\
4038 & $16.1\pm1.6$ & $114.7\pm11.6$ & Disturbed with antennae\\
4096 & $13.1\pm0.3$ & $45.7\pm1.1$ & Disky\\
4206 & $6.8\pm0.2$ & $37.3\pm1.0$ & Oval\\
4214 & $11.6\pm0.2$ & $10.0\pm0.1$ & Round\\
4216 & $10.7\pm0.1$ & $42.8\pm0.3$ & Boxy\\
4222 & $6.9\pm0.3$ & $39.8\pm1.6$ & Disky\\
4244 & $22.1\pm0.5$ & $27.9\pm0.7$ & Disky\\
4302 & $11.0\pm0.2$ & $45.8\pm0.9$ & Oval\\
4395 & $16.9\pm0.3$ & $23.4\pm0.4$ & Round\\
4429 & $12.1\pm0.4$ & $45.8\pm1.4$ & Oval\\
4449 & $21.3\pm0.7$ & $26.4\pm0.9$ & Round\\
4469 & $10.8\pm0.5$ & $52.8\pm2.2$ & Oval\\
4517 & $18.5\pm0.9$ & $45.1\pm2.1$ & Oval\\
4550 & $6.3\pm0.3$ & $27.8\pm1.3$ & Diamond\\
4551 & $4.5\pm0.3$ & $21.2\pm1.2$ & Round\\
4565 & $21.1\pm0.6$ & $74.1\pm2.1$ & Diamond\\
4631 & $21.5\pm0.4$ & $46.0\pm0.9$ & Boxy\\
4638 & $5.3\pm0.3$ & $26.7\pm1.7$ & Round\\
4697 & $17.5\pm0.7$ & $62.1\pm2.6$ & Round\\
4710 & $9.4\pm0.4$ & $46.0\pm1.9$ & Diamond\\
4754 & $7.9\pm0.6$ & $36.7\pm2.7$ & Oval\\
4762 & $14.9\pm0.4$ & $47.0\pm1.2$ & Boxy\\
4866 & $8.0\pm0.2$ & $68.8\pm1.8$ & Oval\\
5170 & $11.6\pm0.2$ & $98.7\pm1.6$ & Boxy/diamond\\
5576 & $11.5\pm0.4$ & $84.0\pm2.8$ & Round\\
5577 & $4.4\pm0.2$ & $24.9\pm0.9$ & Disky\\
5713 & $4.8\pm0.1$ & $28.2\pm0.8$ & Round\\
5719 & $7.0\pm0.2$ & $53.7\pm1.6$ & Slightly diamond\\
5746 & $11.1\pm0.4$ & $87.0\pm3.2$ & Oval\\
5806 & $7.3\pm0.2$ & $43.6\pm1.2$ & Oval\\
5811 & $3.2\pm0.1$ & $22.7\pm0.5$ & Oval\\
5813 & $16.5\pm0.7$ & $153.1\pm6.5$ & Oval\\
5814 & $6.7\pm0.4$ & $305.8\pm17.2$ & Oval\\
5866 & $9.1\pm0.3$ & $38.4\pm1.1$ & Round/boxy\\
5907 & $15.5\pm0.3$ & $77.9\pm1.7$ & Oval\\
6239 & $5.3\pm0.3$ & $37.6\pm1.9$ & Disturbed\\
6340 & $11.5\pm0.8$ & $65.1\pm4.4$ & Round\\
6643 & $5.5\pm0.2$ & $32.4\pm1.4$ & Oval\\
6946 & $25.9\pm0.5$ & $50.7\pm1.0$ & Round\\
7052 & $7.8\pm0.4$ & $116.2\pm6.2$ & Boxy\\
7217 & $11.0\pm0.4$ & $58.5\pm1.9$ & Round\\
7280 & $3.4\pm0.1$ & $21.3\pm0.8$ & Round\\
7331 & $14.7\pm0.5$ & $62.3\pm1.9$ & Diamond\\
7332 & $5.2\pm0.1$ & $34.3\pm1.0$ & Diamond\\
7463 & $6.4\pm0.3$ & $42.8\pm1.9$ & Disturbed\\
7465 & --- & --- & Round\\
7479 & $8.2\pm0.4$ & $71.8\pm3.1$ & Oval\\
7742 & $4.7\pm0.1$ & $29.9\pm0.6$ & Round\\
7743 & $8.5\pm0.4$ & $46.4\pm2.0$ & Oval\\
UGC4872 & --- & --- & Oval\\
UGC5666 & $16.7\pm0.6$ & $19.4\pm0.7$ & Oval\\
\hline
\end{tabular}
\end{minipage}
\end{table}


\bsp	
\label{lastpage}
\end{document}